\documentclass[aps,prb,reprint,,superscriptaddress,twocolumn]{revtex4-2}

\usepackage{amsmath,amssymb,bm,braket,mathrsfs,mathtools}
\usepackage{graphicx}
\usepackage{color}
\usepackage{multirow}
\usepackage{hyperref}

\DeclareMathOperator{\trace}{Tr}
\DeclareMathOperator{\bz}{BZ}
\DeclareMathOperator{\sgn}{sgn}
\DeclareMathOperator{\Pf}{Pf}
\graphicspath{{./figure/}}

\begin{document}
\title{Possible topological phases in quantum anomalous Hall insulator/unconventional superconductor hybrid systems}
\author{Ryoi Ohashi}
\affiliation{Department of Applied Physics, Nagoya University, Nagoya 464-8603, Japan}
\author {Shingo Kobayashi}
\affiliation{RIKEN Center for Emergent Matter Science, Wako, Saitama, 351-0198, Japan}
\author{Yukio Tanaka}
\affiliation{Department of Applied Physics, Nagoya University, Nagoya 464-8603, Japan}

\begin{abstract}
 Quantum anomalous Hall insulator (QAH)/$s$-wave superconductor (SC) hybrid systems are known to be an ideal platform for realizing two-dimensional topological superconductors with chiral Majorana edge modes. 
  In this paper we study QAH/unconventional SC hybrid systems whose pairing symmetry is $p$-wave, $d$-wave, chiral $p$-wave, or chiral $d$-wave. The hybrid systems are a generalization of the QAH/$s$-wave SC hybrid system. In view of symmetries of the QAH and pairings, we introduce three topological numbers to classify topological phases of the hybrid systems. One is the Chern number that characterizes chiral Majorana edge modes and the others are topological numbers associated with crystalline symmetries. We numerically calculate the topological numbers and associated surface states for three characteristic regimes that feature an influence of unconventional SCs on QAHs.  Our calculation shows a rich variety of topological phases and unveils the following topological phases that are no counterpart of the $s$-wave case: crystalline symmetry-protected helical Majorana edge modes, a line node phase (crystalline-symmetry-protected Bogoliubov Fermi surface), and multiple chiral Majorana edge modes. The phenomena result from a nontrivial topological interplay between the QAH and unconventional SCs.
  Finally, we discuss tunnel conductance in a junction between a normal metal and the hybrid systems, and show that the chiral and helical Majorana edge modes are distinguishable in terms of the presence/absence of zero-bias conductance peak. 
\end{abstract}
\maketitle

\section{Introduction}

Unconventional superconductors (SCs) with nonzero topology, dubbed topological SCs, have been the subject of intense study~\cite{qi_2011,tanaka_2012,alicea_2012,elliott_2015,sato_2016,mizushima_2016,sato_2017,chiu_2016}, since they host Majorana fermions on their surface as surface zero-energy Andreev bound states. The Majorana fermions follow non-Abelian statistics~\cite{moore_1991,read_2000,ivanov_2001,stone_2006,fujimoto2008,alicea_2011} 
and are proposed as a topological qubit to implement fault-tolerant topological quantum computation~\cite{nayak_2008}. To realize emergent Majorana fermions, a lot of efforts have been devoted to one-dimensional systems, such as semiconductor/SC hybrid systems~\cite{kitaev_2001,sato_2009,sau_2010,alicea_2010,oreg_2010,cook_2011,mourik_2012,das_2012,rokhinson_2012,deng_2012,finck_2013,Deng_2016}
 and a magnetic atomic chain on superconductors~\cite{choy_2011,braunecker_2013,klinovaja_2013,nadj_2013,vazifeh_2013,nadj_2014,ruby_2015,pawlak_2016,jeon_2017,kim_2018}.

\begin{figure}[!tb]
  \centering
  \includegraphics[width=0.8\hsize]
  {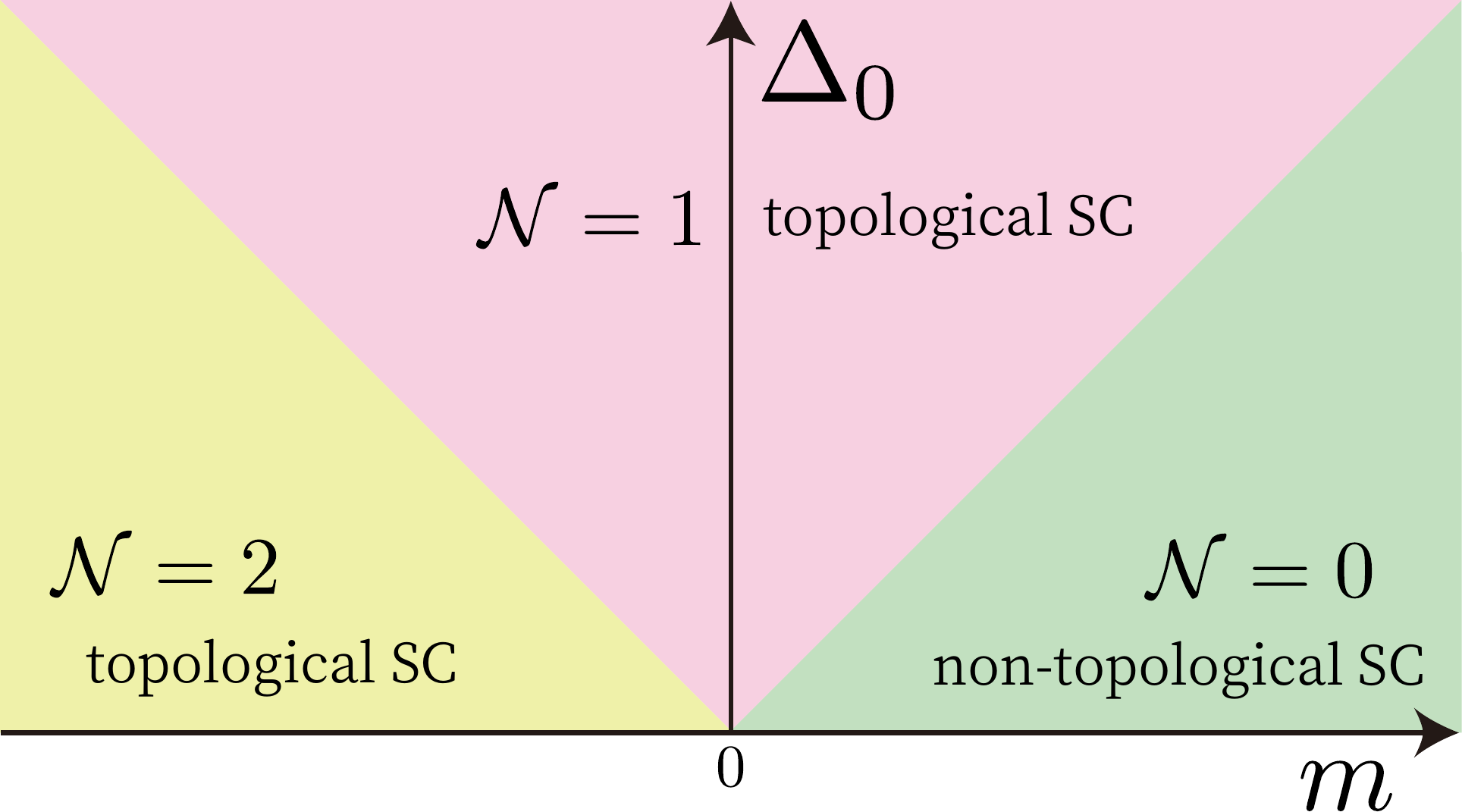}
  \caption{Phase diagram of QAH/$s$-wave SC hybrid system~\cite{he_2014,qi_2010}. Here, $m$ and $\Delta_0$ represent the mass gap and the amplitude of the $s$-wave pair potential. $\mathcal{N}$ indicates the Chern number for each phase. }
  \label{fg:s-wave_phase}
\end{figure}

Alongside those efforts, quantum anomalous Hall insulator (QAH)/SC hybrid systems have attracted a lot of interest as another platform of topological SCs. QAHs are two-dimensional (2D) topological insulators that exhibit quantized Hall conductance without an external magnetic field and host chiral fermions on the edge~\cite{haldane_1988,yu_2010,chang_2013}.  
The QAHs in proximity to an $s$-wave SC become 2D topological SCs~\cite{qi_2010,ii_2011,ii_2012,he_2014}. 
Due to the breaking of time-reversal symmetry (TRS), the superconducting states belong to class D of the Altland-Zirnbauer symmetry classes~\cite{altland_1997,zirnbauer_1996,schnyder_2008,kitaev_2009,schnyder_2009,ryu_2010}. The topological phases are characterized by the Chern number $\mathcal{N}$. Figure~\ref{fg:s-wave_phase} shows the phase diagram of QAH/$s$-wave SC hybrid systems, where $m$ is a mass parameter and $\Delta_0$ is the amplitude of the gap function. The phases with $\mathcal{N} = 2$, $1$, and $0$ indicate the presence of two, one and zero chiral Majorana edge modes, respectively.  Recent experiments on QAH/topological SC/QAH junction have observed a half-quantized conductance~\cite{he_2017}, which signals the $\mathcal{N} = 1$ phase~\cite{chung_2011,wang_2015,lian_2018}, i.e., a single Majorana chiral edge mode, but the existence is still open to debate~\cite{yasuda_2017,ji_2018,huang_2018,kayyalha_2020,shen_2020}.

Recently, QAH/SC hybrid systems have been generalized to hybrid systems with antiferromagnetic topological insulators~\cite{huang_2018,lado_2018}, $s$-wave SC with nontrivial topology~\cite{wang_2018}, or $d$-wave SCs~\cite{he_2019}. Interestingly, such extended hybrid systems bring other 2D topological phases showing helical Majorana edge modes protected by a pseudo TRS~\cite{huang_2018} and multiple chiral Majorana edge modes~\cite{wang_2018,lado_2018,he_2019}. These exotic topological surface states have provided another route for probing Majorana edge modes~\cite{wang_2018,he_2019}. In addition, superconducting doped topological materials, including topological insulators~\cite{Fu2010,Sato2010,Yamakage_2012,Hashimoto_2015} and Dirac/Weyl semimetals~\cite{Lu_2015,Kobayashi2015,Hashimoto_2016,Oudah2016,Kawakami2018,Kawakami2019}, are a promising platform to explore topological superconductivity induced by topologically nontrivial electron states.

In this paper, we study topological phases of QAH/unconventional SC hybrid systems as a generalization of the QAH/$s$-wave SC hybrid system. Unconventional SCs, which have non-$s$-wave pairing, can be realized in cuprate~\cite{bednorz_1986,scalapino_1995,tsuei_2000}, Sr$_2$RuO$_4$~\cite{maeno_1994,rice_1995,mackenzie_2003,maeno_2012,pustogow_2019,Suh2020,Gingras2019,Roising2019,Ghosh2021,Grinenko2021,Benhabib2021}, SrPtAs~\cite{nishikubo_2011,biswas_2013,fischer_2014}, heavy fermion materials~\cite{sauls_1994,joynt_2002,strand_2010,schemm_2014,kasahara_2007,shibauchi_2014,Schemm_2015,ran_2019,jiao_2020}, and so on.  We consider $p$-wave, $d$-wave, chiral $p$-wave, and chiral $d$-wave pairings as possible non-$s$-wave pairings in the SC side and examine a superconducting state realized by a topological interplay between these pairings and the QAH. 

Numerically calculating surface states and topological numbers peculiar to QAH/unconventional SC hybrid systems, we show that the hybrid systems exhibit versatile topological phases with no counterpart of the QAH/s-wave SC hybrid system; these topological phases are attributed to nontrivial topology or spin structures in unconventional Cooper pairs. For instance, chiral SCs manifest itself a nontrivial Chern number ($\mathcal{N}\neq 0$). Thus, a nontrivial topological interplay may emerge in the hybrid systems.  We discuss possible topological phases for three characteristic parameter regimes: (i) $E_{\rm g} > 2|\Delta_0|$, (ii) $E_{\rm g} < 2|\Delta_0|$, and (iii) metallic state, where $E_{\rm g}$ is the amplitude of the insulating gap in the QAH and the metallic state is the situation where the chemical potential lies above or below the insulating gap. The topological properties of hybrid systems for each parameter regime are briefly summarized as follows.  Hereafter, we assume the topologically nontrivial states in the underlying QAHs. 
\begin{itemize}
\item[(i)]
Only a chiral edge mode is contributed to the superconducting state for $E_{\rm g} > 2|\Delta_0|$, whereby leading to two chiral Majorana edge modes irrespective to pairing symmetries. 
\item[(ii)]
 Different topological phases appear for some pairing symmetries when the gap function satisfies $E_{\rm g} < 2|\Delta_0|$. We find a crystalline-symmetry-protected helical Majorana edge mode when the pairing symmetry is a $p$-wave pairing. Moreover, a gapless phase appears when there occurs an inconsistency between the spin structures of QAH and spin-triplet pairing. In view of a topological invariant, the gapless phase is regarded as a crystalline symmetry-protected Bogoliubov Fermi surface.
\item[(iii)]
Topological phases inherent to unconventional pairings appear when the chemical potential lies at the bulk bands; if the pair potential has nodes, zero-energy flat bands or sine-curved states are realized on the surface. On the other hand, if it has a full gap, we find multiple chiral Majorana edge modes, the number of which is determined from a sum rule of the Chern number.  In addition, the gapless phase mentioned above appears again since it is attributed to the mismatch of the spin structures. 
\end{itemize}

Finally, we discuss tunnel conductance in a normal metal/(QAH/SC) junction as a probe to distinguish the crystalline symmetry-protected helical Majorana edge modes from the chiral ones; the former exhibits a sharp zero bias conductance peak being mediated by the zero energy states protected by crystalline symmetry, whereas the latter  does not due to the interference of the zero energy states~\cite{yamakage_2014}.

This paper is organized as follows. We formulate QAH/unconventional SC hybrid systems in Sec.~\ref{sec:formulation}. The symmetries and topological numbers are summarized in Secs.~\ref{sec:symmetry} and \ref{sec:topology}. The numerical method is introduced in Sec.~\ref{sec:num_method}. In Sec.~\ref{sec:overview}, we overview our findings for three characteristic parameter regimes in the QAH/unconventional SC hybrid systems. In particular, we discuss topological properties and surface states for a crystalline-symmetry-protected helical Majorana edge modes, a crystalline symmetry-protected Bogoliubov Fermi surface, multiple chiral Majorana edge modes  in Secs.~\ref{ssec:p_x_d_y}, \ref{ssec:p_x_d_z}, and \ref{ssec:chiral_wave}, respectively. Summary and relations to tunnel conductance are discussed in Sec.~\ref{sec:conclusion}.

\section{QAH/SC Model}

\subsection{Formulation}
\label{sec:formulation}
We start from a minimum two-dimensional Hamiltonian describing QAH states, 
$\mathcal{H}_{\mathrm{QAH}} = \sum_{\bm{k}}\psi_{\bm{k}}^{\dagger}H_{\mathrm{QAH}}(\bm{k})\psi_{\bm{k}}$ with $\psi_{\bm{k}} = (c_{\bm{k},\uparrow},c_{\bm{k},\downarrow})^T$ and 
\begin{align}
  H_{\rm QAH}(\bm{k}) =
m(\bm{k})\sigma_z + A(k_x\sigma_x + k_y\sigma_y) ,
\label{eq:QAH_Hamiltonian}
\end{align}
where $m(\bm{k})=m+B(k_x^2 +k_y^2)$, $A$, and $\sigma_i$ are a mass term, a spin-orbital coupling term, and the Pauli matrices in the spin space, respectively. The mass term breaks TRS and open a gap at $\bm{k}=0$; the energy spectrum is given by $E_{\rm QAH}(\bm{k}) \equiv \sqrt{A^2 (k_x^2 +k_y^2) + (m +B (k_x^2 +k_y^2))^2}$, leading to the insulating gap $E_{\rm g} = 2 |m|$. For $m/B<0$, the system belongs to the QAH phase with a single chiral edge mode on the boundary~\cite{qi_2010}.

\begin{figure}[tb]
  \centering
  \includegraphics[width=7cm]
  {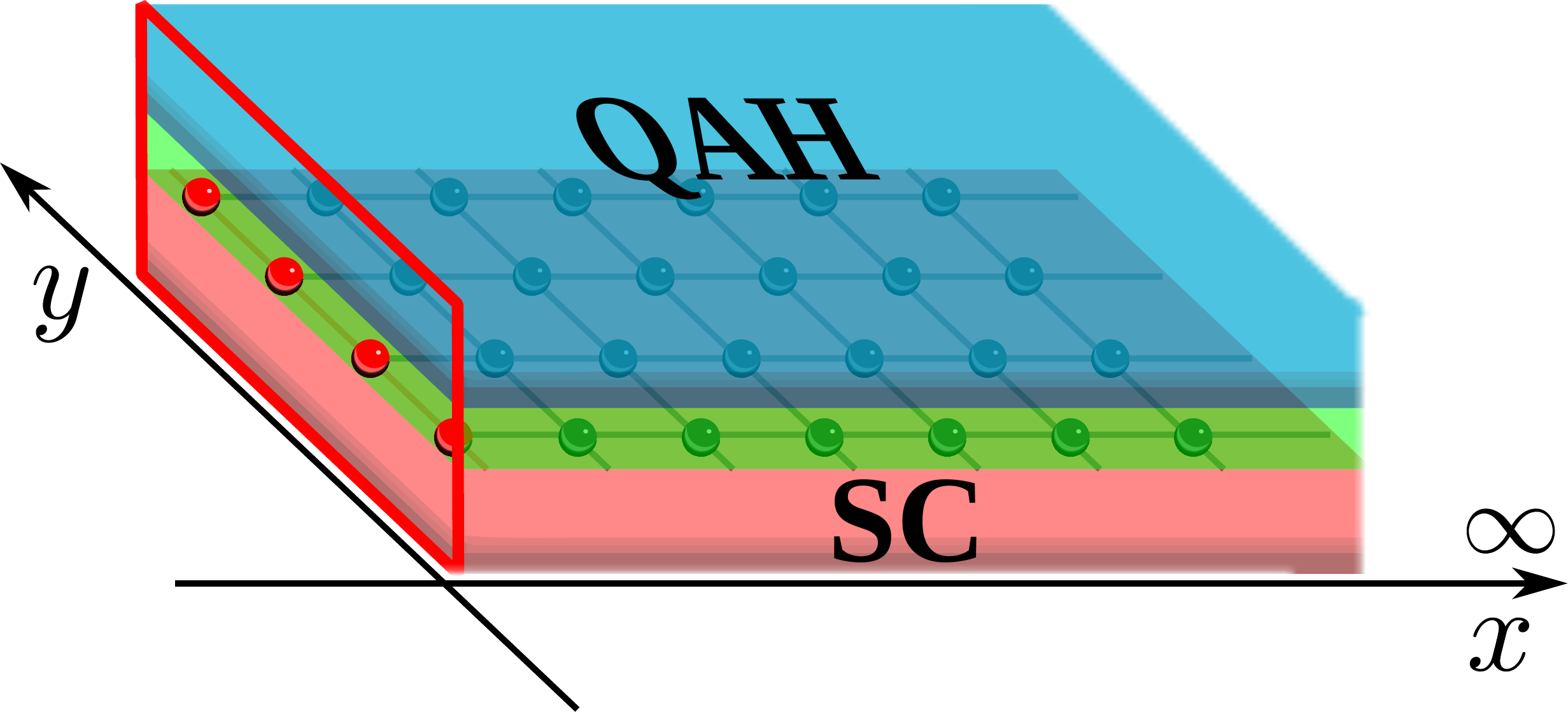}
  \caption{Schematic illustration of QAH/SC hybrid systems. }
  \label{fg:model}
\end{figure}

Using Eq.~(\ref{eq:QAH_Hamiltonian}), we model the QAH/SC hybrid system, where the QAH is supposed to be stacked on an SC; see Fig.~\ref{fg:model}. In the hybrid system, Cooper pairs in the SC are induced to the QAH due to the proximity effect, and thus the QAH/SC system can be described by the Bogoliubov-de Gennes (BdG) Hamiltonian,
$\mathcal{H}_{\mathrm{BdG}} = \tfrac{1}{2}\sum_{\bm{k}}\Psi_{\bm{k}}^{\dagger}H_{\mathrm{BdG}}(\bm{k})\Psi_{\bm{k}}$ with $\Psi_{\bm{k}} = (c_{\bm{k},\uparrow},c_{\bm{k},\downarrow},c_{-\bm{k},\uparrow}^{\dagger},c_{-\bm{k},\downarrow}^{\dagger})^T$ and 
\begin{align}
H_\mathrm{BdG}(\bm{k})=
\begin{pmatrix}
  H_\mathrm{QAH}(\bm{k}) - \mu & \Delta(\bm{k}) \\
  \Delta^{\dagger}(\bm{k}) & -H^*_\mathrm{QAH}(-\bm{k}) + \mu
\end{pmatrix}\label{eq:BdG_Hamiltonian},
\end{align}
where $\mu$ is the chemical potential and $\Delta(\bm{k})$ the induced gap function.
In the following, we consider unconventional SCs, and thus $\Delta(\bm{k})$ describes non-$s$-wave Cooper pairings such as $p$-wave, $d$-wave, chiral $p$-wave, and chiral $d$-wave pairs. 
For spin-singlet Cooper pairings, we define the gap function as
\begin{align}
 \Delta(\bm{k}) = \Delta_0 f(\bm{k}) (i\sigma_y),
\end{align}
where the form factor $f(\bm{k})$ depends on Cooper pair symmetry such that
\begin{align}
 f(\bm{k}) = 
 \begin{cases} 
 1 & \text{($s$-wave)}, \\
 k_x k_y & \text{($d_{xy}$-wave)}, \\
 k_x^2-k_y^2 & \text{($d_{x^2-y^2}$-wave)}, \\
 (k_x \pm i k_y)^2 & \text{(chiral $d\pm id$-wave)}.
 \end{cases} \label{eq:even_gap}
\end{align}
We note that the $d_{xy}$ ($d_{x^2-y^2}$)-wave pairing hosts point nodes at $k_x=0$ and $k_y=0$ ($k_x=k_y$ and $k_x = -k_y$) on the Fermi surface, whereas the chiral $d\pm id$-wave pairing have a full gap and breaks TRS spontaneously.

On the other hand, the gap function in spin-triple Cooper pairings is formally described as
\begin{align}
  \Delta(\bm{k})  = \Delta_0 \bm{d}(\bm{k}) \cdot \bm{\sigma} (i\sigma_{y}), 
\end{align}
with $\bm{d}(\bm{k}) =(d_x(\bm{k}),d_y(\bm{k}),d_z(\bm{k}))$. The $\bm{d}$ vector represents the spin degrees of freedom of Cooper pairs. The $p_{\nu}$-wave and chiral $p$-wave pairings are represented by the $\bm{d}$ vector with
\begin{align}
 d_i (\bm{k}) = 
 \begin{cases} 
  k_{\nu} & \text{($p_{\nu}$-wave)}, \\
  k_x \pm i k_y & \text{(chiral $p\pm i p$-wave)}, \\
 \end{cases}
\end{align}  
and $d_{j} =0$ when $j \neq i$, where $\nu=x,y$ and $i=x,y,z$. For the chiral $p\pm i p$-wave pair, we fix the $\bm{d}$ vector along the $z$ direction.  
Note that the $p_x$- and $p_y$-wave pairings have a pair of point nodes at $k_x=0$ and $k_y=0$ on the Fermi surface, respectively. 

\subsection{Symmetries}
\label{sec:symmetry}

In the following, we summarize symmetries of the QAH/SC hybrid systems, which are useful to define topological numbers in Sec~\ref{sec:topology}. First, let us consider symmetries of the normal Hamiltonian (\ref{eq:QAH_Hamiltonian}). In the QAH, TRS ($T=i\sigma_y \mathcal{K}$) is absent, but there exist crystalline symmetries that keep the Hamiltonian (\ref{eq:QAH_Hamiltonian}) invariant.
  Equation~(\ref{eq:QAH_Hamiltonian}) satisfies two-fold rotation symmetry in the $z$ axis,
\begin{align}
 C_{2z} H_{\rm QAH}(\bm{k}) C_{2z}^{-1} = H_{\rm QAH}(-\bm{k}), \ \ C_{2z} = i \sigma_z, \label{eq:normal_C2z}
\end{align} 
In addition, though two-fold rotation symmetries in terms of the $x$ and $y$ axis ($C_{2x}$ and $C_{2y}$) are not symmetry of the Hamiltonian, the combination with TRS keeps the Hamiltonian invariant. That is, the combined operations, defined by $A_x = C_{2x} T$ and  $A_y = C_{2y} T$, satisfy  
\begin{align}
 &A_x H_{\rm QAH}(k_x,k_y) A_x^{-1} = H_{\rm QAH}(-k_x,k_y), \ \ A_x = \sigma_z \mathcal{K}, \label{eq:normal_Ax} \\
 &A_y H_{\rm QAH}(k_x,k_y) A_y^{-1} = H_{\rm QAH}(k_x,-k_y), \ \ A_y = \mathcal{K}, \label{eq:normal_Ay}
\end{align} 
which we call magnetic-reflection symmetry in this paper. Here $\mathcal{K}$ is the complex-conjugation operator.

Next, we consider symmetries of the QAH/SC hybrid systems (\ref{eq:BdG_Hamiltonian}) that breaks TRS but preserves particle-hole symmetry (PHS):
\begin{align}
\mathcal{C}H_\mathrm{BdG}(\bm{k})\mathcal{C}^{-1} = -H_\mathrm{BdG}(-\bm{k}), \ \ \mathcal{C}=\tau_x\mathcal{K}, \label{eq:PHS}
\end{align} 
where $\tau_i$ are the Pauli matrices in the Nambu space. Hence, our systems belong to class D of the Altland-Zirnbauer symmetry classes. 

\begin{table}[tb]
\caption{Representations of crystalline symmetries in QAH/SC hybrid systems.}
 \label{tb:CS_operators}
 \begin{tabular}{c|ccccc}
 \hline \hline
 Pairing symmetry& $\mathcal{C}_{2z}$ & $\mathcal{A}_x$ & $\mathcal{A}_y$ & $\Gamma_{A_x}$ & $\Gamma_{A_y}$\\ \hline
  $s$-wave & $i\sigma_z \tau_z $ & $\sigma_z \tau_z \mathcal{K}$ & $\mathcal{K}$ & $\sigma_z \tau_y$ &  $\tau_x$ \\
  $d_{xy}$-wave & $i\sigma_z \tau_z $ & $\sigma_z \mathcal{K}$ & $\tau_z \mathcal{K}$ &$\sigma_z \tau_x$&  $\tau_y$ \\
  $d_{x^2-y^2}$-wave & $i\sigma_z \tau_z $ & $\sigma_z \tau_z \mathcal{K}$ & $\mathcal{K}$ &$\sigma_z \tau_y$&  $\tau_x$ \\
  chiral $d\pm i d$-wave & $i\sigma_z \tau_z $ & $\sigma_z \tau_z \mathcal{K}$ & $ \mathcal{K}$ &$\sigma_z \tau_y$&  $\tau_x$ \\
  $p_x$-wave ($\bm{d}\parallel\hat{\bm{x}}$) & $i\sigma_z \tau_z $ & $\sigma_z \tau_z \mathcal{K}$ & $\mathcal{K}$ &$\sigma_z \tau_y$&  $\tau_x$ \\
  $p_x$-wave ($\bm{d}\parallel\hat{\bm{y}}$) & $i\sigma_z \tau_z $ & $\sigma_z \mathcal{K}$ & $\tau_z \mathcal{K}$ &$\sigma_z \tau_x$&  $\tau_y$ \\
  $p_x$-wave ($\bm{d}\parallel\hat{\bm{z}}$) & $i\sigma_z $ & $\sigma_z \mathcal{K}$ & $\mathcal{K}$ &$\sigma_z \tau_x$& $\tau_x$ \\
  $p_y$-wave ($\bm{d}\parallel\hat{\bm{x}}$) & $i\sigma_z \tau_z $ & $\sigma_z \mathcal{K}$ & $ \tau_z \mathcal{K}$ &$\sigma_z \tau_x$&  $\tau_y$ \\
  $p_y$-wave ($\bm{d}\parallel\hat{\bm{y}}$) & $i\sigma_z \tau_z $ & $\sigma_z \tau_z \mathcal{K}$ & $\mathcal{K}$ &$\sigma_z \tau_y$&  $\tau_x$ \\
  $p_y$-wave ($\bm{d}\parallel\hat{\bm{z}}$) & $i\sigma_z $ & $\sigma_z \tau_z \mathcal{K}$ & $\tau_z \mathcal{K}$ &$\sigma_z \tau_y$&  $\tau_y$ \\
   chiral $p\pm i p$-wave & $i\sigma_z  $ & $\sigma_z \mathcal{K}$ & $\mathcal{K}$ &$\sigma_z \tau_x$&  $\tau_x$ \\ \hline \hline
 \end{tabular}
\end{table}

 In addition, the crystalline symmetries (\ref{eq:normal_C2z}), (\ref{eq:normal_Ax}), and (\ref{eq:normal_Ay}) are again symmetries of the BdG Hamiltonian, whose representations are modified due to pairing symmetry. The gap functions transform, under the action of $g = C_{2z}$, $A_x$, and $A_y$, as
\begin{align}
 g \Delta (\bm{k}) g^{T} = \pm \Delta (D_g \bm{k}), \label{eq:gap_trans}
\end{align}
where $D_g$ obeys the change of momentum in Eqs.~(\ref{eq:normal_C2z}), (\ref{eq:normal_Ax}), and (\ref{eq:normal_Ay}) and the plus (minus) sign represents the parity of the gap functions in terms of $g$. Thus, to be consistent with Eq.~(\ref{eq:gap_trans}), representations of $g$ in the Nambu space are given by   
\begin{align}
 \mathcal{G} = \begin{pmatrix} g & 0 \\ 0 & \pm g^{\ast} \end{pmatrix},
\end{align}
where $+(-)$ comes from the parity of gap functions in Eq.~(\ref{eq:gap_trans}). In the matrix representation, the BdG Hamiltonian and the particle-hole operator satisfy
\begin{align}
 &\mathcal{G} H_{\rm BdG} (\bm{k}) \mathcal{G}^{-1} = H_{\rm BdG} (D_g\bm{k}), \ \ \mathcal{C} \mathcal{G} = \pm \mathcal{G} \mathcal{C}, \label{eq:BdG_sym}
\end{align} 
where the sign reflects the parity of the gap function in Eq.~(\ref{eq:gap_trans}).
The explicit representations of $\mathcal{G} = \mathcal{C}_{2z}$, $\mathcal{A}_x$, and $\mathcal{A}_y$ are summarized in Table~\ref{tb:CS_operators}. Here we have used calligraphic fonts for the symmetries in the BdG Hamiltonian.

Furthermore, the combination of $\mathcal{G}$ and $\mathcal{C}$ constitutes chiral operators, $\Gamma_{\mathcal{A}_x} \equiv e^{i \phi} \mathcal{A}_x \mathcal{C}$ and $\Gamma_{\mathcal{A}_y} \equiv e^{i \phi'} \mathcal{A}_y \mathcal{C}$, which satisfy
\begin{align}
 \Gamma_{\mathcal{A}_x} H_{\rm BdG}(k_x,k_y) \Gamma_{\mathcal{A}_x}^{-1} = -H_{\rm BdG}(k_x,-k_y), \label{eq:BdG_gammax} \\
 \Gamma_{\mathcal{A}_y} H_{\rm BdG}(k_x,k_y) \Gamma_{\mathcal{A}_y}^{-1} = -H_{\rm BdG}(-k_x,k_y), \label{eq:BdG_gammay}
\end{align}
where we choose the phase factors $\phi$ and $\phi'$ as $\Gamma_{\mathcal{A}_x}^2=\Gamma_{\mathcal{A}_y}^2=\bm{1}_{4 \times 4}$, with $\bm{1}_{N \times N}$ being the $N \times N$ identity matrix. These chiral operators are defined for all gap functions; see the rightmost columns in Table~\ref{tb:CS_operators}. 

\subsection{Topological numbers}
\label{sec:topology}
We here discuss possible topological numbers inherent to the symmetries of the QAH/SC hybrid system.
\subsubsection{Chern number}
Since the QAH/SC hybrid systems break TRS, we can define the Chern number $\mathcal{N}$ as 
\begin{align}
  &\mathcal{N}
  =\sum_{n\in {\rm occ}}\int_{\bz}\frac{d^2\bm{k}}{2\pi}\left(\frac{\partial a_{n,y}}{\partial k_x}-\frac{\partial a_{n,x}}{\partial k_y}\right),\label{eq:Chern_number} \\
  &a_{n,i} 
  = -i\left<u_{n,\bm{k}}\right|\frac{\partial}{\partial k_i}\left|u_{n,\bm{k}}\right>,\label{eq:Berry_connection}
\end{align}
when the energy spectrum is fully gapped over the two-dimensional (2D) Brillouin zone (BZ). Here, $\ket{u_{n,\bm{k}}}$ is an eigenstate of $H_{\rm BdG}(\bm{k})$ with a band label $n$ and momentum $\bm{k}$ and the summation is taken over the occupied states. 
The Chern number  corresponds to the number of chiral Majorana edge modes through the bulk-boundary correspondence and its sign is related to the direction of edge currents; for instance, chiral Majorana edge modes associated with the Chern number $\mathcal{N}=1$ and $-1$ propagate in the opposite direction. 
In the QAH/$s$-wave SC hybrid systems, the Chern number takes $\mathcal{N}=0$, $1$, and $2$. See Fig.\ref{fg:s-wave_phase}. 

\begin{table*}[tb]
  \caption{
    Summary of surface states (SSs) and bulk gap structures (GSs) in the QAH/ unconventional SC hybrid systems, where we assume that the QAH is topologically nontrivial ($m<0$) and the semi-infinite (periodic) boundary condition is imposed in the $x$ ($y$) direction. Here, $\mathcal{N}$ and $\mathcal{W}$ represent the Chern number (\ref{eq:Chern_number}) and the winding number (\ref{eq:Winding_number}) defined at $k_y=0$. The GSs are classified as a full gap, a point node, and a line node in the 2D BZ, where the line node is a crystalline symmetry-protected Bogoliubov Fermi surfaces as discussed in Sec.~\ref{ssec:p_x_d_z}. For SSs, chiral Majorana edge modes (chiral) are stabilized by the Chern number, whereas the helical Majorana edge modes (helical) are protected by the winding number. In the metallic state regime, $p_x$-wave, $d_{xy}$-wave, and $d_{x^2-y^2}$-wave pairing host point nodes, which lead to a flat band or a sine curve state as shown in Fig.~\ref{fg:table_SDOS_px}. Note that ``$-$'' indicates no surface state and/or the Chern number being ill-defined.
    }
  \label{tb:GS_SS}
  \begin{tabular}{l|ccc|ccc|ccc} \hline \hline
    Pairing symmetry & \multicolumn{3}{c|}{$E_{\rm g}> 2|\Delta_0|$} & \multicolumn{3}{c|}{$E_{\rm g}< 2|\Delta_0|$} & \multicolumn{3}{c}{metallic state} \\
    ~ & GS & SS &($\mathcal{N},\mathcal{W}$)  & GS & SS &($\mathcal{N},\mathcal{W}$) & GS & SS &($\mathcal{N},\mathcal{W}$) \\ \hline
    $s$-wave                                       & Full gap&  chiral & $(2,0)$ &Full gap&  chiral & $(1,1)$ & Full gap&  chiral & $(1,1)$ \\
    $p_x$-wave ($\bm{d}\parallel\hat{\bm{x}}$) & Full gap&  chiral & $(2,0)$ & Full gap&  chiral & $(2,0)$ &  Point node&  flat band & $(-,1)$ \\
    $p_x$-wave ($\bm{d}\parallel\hat{\bm{y}}$) & Full gap&  chiral & $(2,0)$ & Full gap&  helical & $(0,2)$ &   Point node&  sine curve &  $(-,1)$ \\
    $p_x$-wave ($\bm{d}\parallel\hat{\bm{z}}$) & Full gap&  chiral & $(2,0)$ & Line node& $-$ & $(-,0)$ & Line node& $-$ & $(-,0)$\\
 $p_y$-wave ($\bm{d}\parallel\hat{\bm{x}}$) & Full gap&  chiral & $(2,0)$ & Full gap& $-$ & $(0,0)$ &  Point node& $-$ &$(-,0)$\\
  $p_y$-wave ($\bm{d}\parallel\hat{\bm{y}}$) & Full gap&  chiral & $(2,0)$ & Full gap& chiral & $(2,0)$ &  Point node& $-$ &$(-,0)$\\
    $p_y$-wave ($\bm{d}\parallel\hat{\bm{z}}$) & Full gap&  chiral & $(2,0)$ & Line node& $-$ & $(-,0)$ &  Line node&$-$& $(-,0)$\\
    chiral $p \pm i p$-wave                         & Full gap&  chiral & $(2,0)$ & Line node& $-$ & $(-,0)$ & Line node& $-$ &$(-,0)$ \\
    $d_{xy}$-wave                                  & Full gap&  chiral  & $(2,0)$ & Full gap&  chiral & $(2,0)$ &  Point node&  sine curve &$(-,0)$\\
    $d_{x^2-y^2}$-wave                          & Full gap&  chiral & $(2,0)$ & Full gap&  chiral & $(2,0)$ &  Point node &  sine curve &$(-,1)$\\
     chiral $d+id$-wave                           & Full gap&  chiral & $(2,0)$ & Full gap&  chiral & $(2,0)$ & Full gap&  chiral & $(-1,1)$\\
    chiral $d-id$-wave                           & Full gap& chiral & $(2,0)$ & Full gap& chiral & $(2,0)$ & Full gap& chiral & $(3,1)$\\ \hline \hline
  \end{tabular}
\end{table*}

\subsubsection{One-dimensional winding number classified by $\mathbb{Z}$}
The QAH/SC systems host additional crystalline symmetries~(\ref{eq:BdG_sym}), (\ref{eq:BdG_gammax}), and (\ref{eq:BdG_gammay}), which allow us to define crystalline symmetry-protected topological numbers. Using the chiral symmetries [Eqs.~(\ref{eq:BdG_gammax}) and (\ref{eq:BdG_gammay})], we can define a crystalline symmetry-protected one-dimensional (1D) winding number~\cite{ii_2012,shiozaki_2014}  in a 1D subspace of the BZ,
\begin{align}
 \mathcal{W}[\Gamma_{\mathcal{A}_{i}}]
  =\frac{i}{4\pi }\int_{-\pi}^{\pi}dk_i \trace\left[\Gamma_{\mathcal{A}_{i}} H_\mathrm{BdG}^{-1}(\bm{k})\frac{\partial}{\partial k_i}H_\mathrm{BdG} (\bm{k})\right], 
  \label{eq:Winding_number}
\end{align}
where we fix $k_x=0$ for $\Gamma_{\mathcal{A}_y}$ and $k_y=0$ for $\Gamma_{\mathcal{A}_x}$.  
The nonzero 1D winding number ensures the existence of zero energy modes at edges of the 1D subspace through the bulk-boundary correspondence. The 1D winding number intrinsically relates to the Chern number so that $|\mathcal{W}[\Gamma_{\mathcal{A}_{i}}]|$ counts the number of chiral Majorana edge modes that cut across $k_x=0$ or $k_y=0$ in the surface BZ. For instance, when we consider the QAH/$s$-wave SC hybrid system and impose the open (periodic) boundary condition in the $x$ ($y$) direction, $\mathcal{W}[\Gamma_{\mathcal{A}_{x}}]$ describes zero modes at $k_y=0$ in the surface BZ; the 1D winding number takes $\mathcal{W}[\Gamma_{\mathcal{A}_{x}}]=1$ when $\mathcal{N}=1$, while $\mathcal{W}[\Gamma_{\mathcal{A}_{x}}]=0$ when $\mathcal{N}=0,2$~\cite{ii_2012}, because only a single chiral Majorana edge mode cuts across $k_y=0$. 
Hereafter, we abbreviate $\mathcal{W}[\Gamma_{\mathcal{A}_{i}}]$ as $\mathcal{W}$ unless otherwise specified.

\subsubsection{Zero-dimensional topological number classified by $\mathbb{Z}_2$}

In addition, the QAH/SC hybrid systems satisfy two-fold rotation symmetry,
\begin{align}
\mathcal{C}_{2z}H_{\rm BdG}(\bm{k})\mathcal{C}_{2z}^{-1} = H_{\rm BdG}(-\bm{k}), \label{eq:C2z_sym}
\end{align}
which leads to a crystalline symmetry-protected zero-dimensional (0D) $\mathbb{Z}_2$ invariant when the parity of the gap function is odd under $C_{2z}$, i.e., $\{\mathcal{C}_{2z},\mathcal{C}\}=0$. Combining Eq.~(\ref{eq:C2z_sym}) with $\mathcal{C} = \tau_x \mathcal{K}$ in Eq.~(\ref{eq:PHS}), the BdG Hamiltonian satisfies
\begin{align}
 (-i \tau_x \mathcal{C}_{2z}) H_{\rm BdG}(\bm{k}) = -[(-i \tau_x \mathcal{C}_{2z}) H_{\rm BdG}(\bm{k})]^T, \label{eq:pfaffian_cond}
\end{align}
where we add $-i$ to make the operator real. Equation~(\ref{eq:pfaffian_cond}) indicates that the matrix is screw-symmetric, and thus we can define a Pfaffian invariant,
\begin{align}
  (-1)^{\nu}
  =\sgn \left\{\frac{\Pf\left[ (-i \tau_x \mathcal{C}_{2z}) H_{\rm BdG} (\bm{k}')\right]}{\Pf\left[  (-i \tau_x \mathcal{C}_{2z}) H_{\rm BdG} (\bm{k}'')\right]}\right\},
  \label{eq:Z2_2D}
\end{align}
which takes $\nu=0,1 \in \mathbb{Z}_2$ since $\Pf \left[  (-i \tau_x \mathcal{C}_{2z}) H_{\rm BdG} (\bm{k})\right] \in \mathbb{R}$. Here, $\bm{k}'$ and $\bm{k}''$ are arbitrary wave numbers in the 2D BZ. If $\nu = 1 \mod 2$, there exists a gapless point on a line connecting $\bm{k}'$ and $\bm{k}''$; namely, one of the wave number is enclosed by a line node.  In view of topology, the line node is analogous to the Bogoliubov Fermi surface~\cite{kobayashi_2014,zhao_2016,agterberg_2017,bzdusek_2017,timm_2017,brydon_2018}. Thus, we call it a crystalline symmetry-protected  Bogoliubov Fermi surface.

\subsection{Numerical method}
\label{sec:num_method}
We numerically calculate the surface density of states (SDOSs) and those topological numbers to explore topological surface states through the bulk-boundary correspondence. We perform the numerical calculation on a square lattice by replacing $k_i\to\sin k_i$ and $k_i^2\to 2(1-\cos k_i)$ in Eq.~(\ref{eq:BdG_Hamiltonian}). 

For SDOSs, we impose the periodic boundary condition for the $y$ axis and the semi-infinite boundary condition for the $x$ axis; see Fig.\ref{fg:model}. Using the surface Green's function $G_s(k_y,E)$, the angle-resolved SDOS is defined by  
\begin{align}
  \rho(k_y,E) = -\frac{1}{\pi}\mathrm{Im}\left[P_\mathrm{e}G^\mathrm{R}(k_y,E)\right],
\end{align}
with the retarded Green's function,
\begin{align}
 G^\mathrm{R}(k_y,E) = G_s(k_y,E+i \eta) 
\end{align}
where $E$ is the energy of the semi-infinite system, $\eta$ is an infinitesimal positive number, and $P_\mathrm{e}=(\tau_0+\tau_z)/2$ is the projection onto the electron states. The surface Green's function in the semi-infinite system is obtained using the Umerski's method~\cite{umerski_1997}.

For topological numbers, the Chern number (\ref{eq:Chern_number}) is calculated by the gauge invariant method~\cite{fukui_2005a}, where 
the Chern number is defined on the discretized Brillouin zone with the mesh size $N_x \times N_y$, which approaches the exact Chern number when the mesh is sufficiently large. We choose $N_x=N_y=512$ in this paper. The obtained Chern number is consistent with the results of SDOSs. Similarly, the winding number (\ref{eq:Winding_number}) is calculated on the discretized Brillouin zone, where we directly replace the integral to the summation since Eq.~(\ref{eq:Winding_number}) does not depend on the choice of the gauge. In the numerical calculation, we fix $k_y=0$ and take a replacement $\int dk_x \to \frac{2\pi}{N_x}\sum$ with $N_x = 10000$.

In the following, we fix $A=0.1$ and $B=1$ and choose other parameters as ($m, \Delta_0, \mu$) = ($\pm0.05,0.025,0$), ($\pm0.05,0.025,0.01$), ($\pm0.05,0.175,0$), ($\pm0.05,0.175, 0.01$) and ($\pm0.05, 0.025,0.06$). The first and second parameters belong to the regime $E_{\rm g} > 2|\Delta_0|$, the third and fourth to $E_{\rm g} < 2|\Delta_0|$, and the fifth to the metallic states, where $E_{\rm g} = 2 |m|$ is the insulating gap and $m <0$ $(m>0)$ describes the topologically nontrivial (trivial) phases of the QAH.  For $E_{\rm g} > 2|\Delta_0|$ and $E_{\rm g} < 2|\Delta_0|$, we also check the dependence of the chemical potential because an accidental degeneracy of surface zero energy states occurs when $\mu=0$, while it disappears when $\mu \neq 0$~\cite{yamakage_2014}.

\section{Result}
\label{sec:result}

\subsection{Overview}
\label{sec:overview}

\begin{figure*}
  \centering
  \includegraphics[width=0.7\linewidth]
  {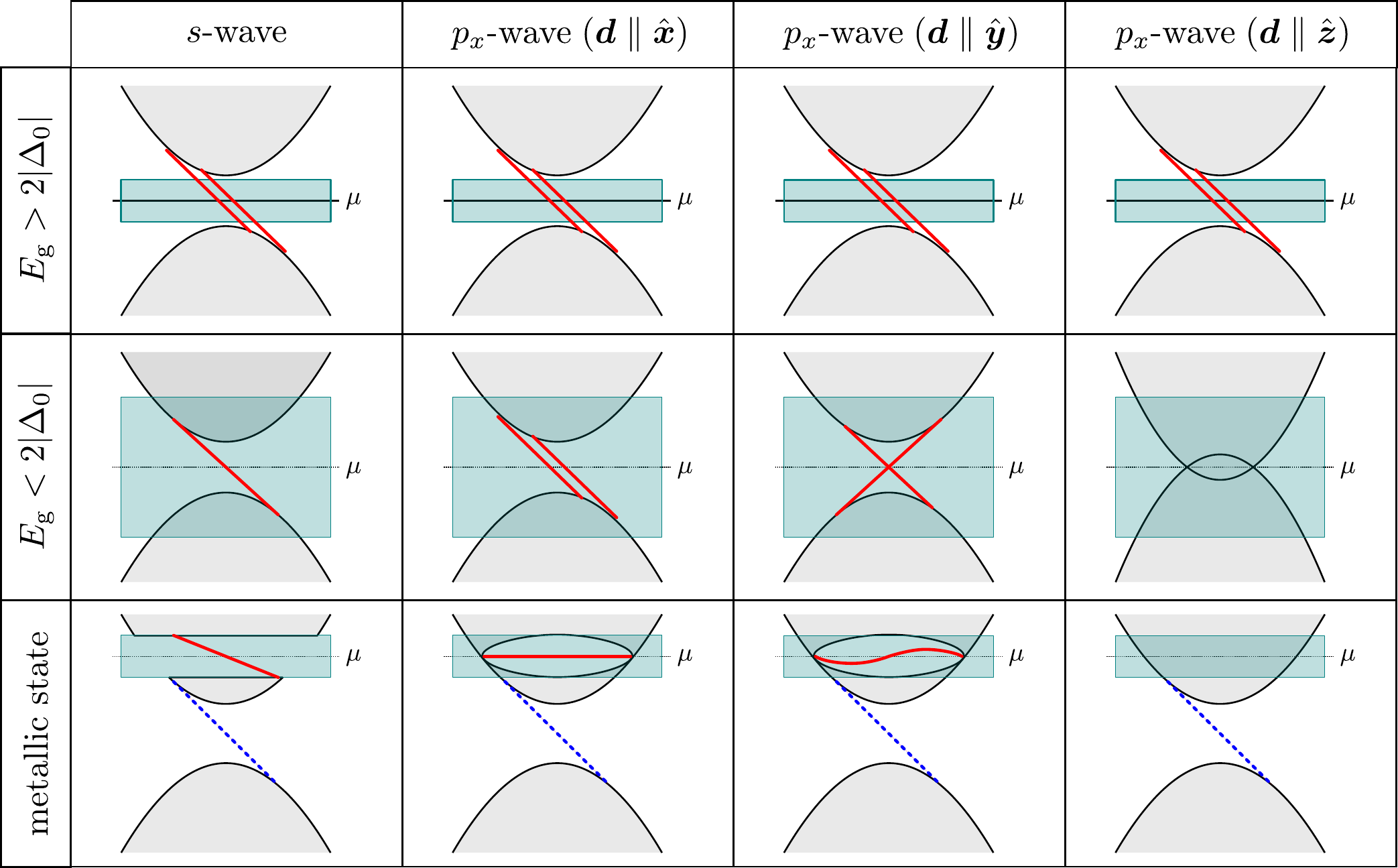}
   \caption{Schematic illustration of SDOSs in the QAH/$s$-wave and $p_x$-wave SC hybrid systems. For the $p_x$-wave pairing, we choose the direction of the $\bm{d}$ vector as $\bm{d} \parallel \hat{\bm{x}}$, $\bm{d} \parallel \hat{\bm{y}}$, and $\bm{d} \parallel \hat{\bm{z}}$.  We illustrate surface states for three parameter regimes: $E_{\rm g} > 2|\Delta_0|$, $E_{\rm g} < 2|\Delta_0|$, and the metallic state, where $E_{\rm g}$ is the insulating gap and $\Delta_0$ is the amplitude of the gap function. The red solid (blue dotted) lines represent edge states for superconducting QAHs (QAHs).  The blue transparent areas describe the energy range in which the gap function $\Delta_0$ affects. We note that the direction of the Majorana edge modes is opposite to that predicted by the Chern numbers in Table~\ref{tb:GS_SS} since the QAH/SC hybrid system is located at $x > 0$. }
  \label{fg:table_SDOS_px}
\end{figure*}

We apply the numerical methods to the QAH/unconventional SC hybrid systems. The results are summarized in Table~\ref{tb:GS_SS}, where gap structures, surface states, and topological numbers are shown in the three parameter regimes. The SDOSs on the (10) surface in the QAH/$s$-wave and $p_x$-wave SC hybrid systems are schematically shown in Fig.~\ref{fg:table_SDOS_px} and~\ref{fg:table_SDOS_py_d}. In what follow, we overview emergent topological phases for each regime.

\subsubsection{$E_{\rm g} > 2|\Delta_0|$}
 In this regime, topology of the QAH is mainly contributed to the surface states. We observe two chiral Majorana edge modes for all the hybrid systems when $m <0$. Their stability is ensured by the Chern number $\mathcal{N}=2$.

 \subsubsection{$E_{\rm g} < 2|\Delta_0|$}
  Once the magnitude of the pair potential exceeds the insulating gap, we find topological phases unique to unconventional pairings. For the $p_x$-wave ($\bm{d}\parallel\hat{\bm{x}}$), $p_y$-wave ($\bm{d}\parallel \hat{\bm{y}}$), and $d$-wave pairings, the two  chiral  Majorana edge modes appear again. On the other hand, a helical Majorana edge mode emerges for the $p_x$-wave pairing ($\bm{d}\parallel\hat{\bm{y}}$).  In this phase, the Chern number is zero, while the 1D crystalline-symmetry-protected winding number is nonzero. Thus, only the magnetic-reflection symmetry protects the helical Majorana edge mode. Note that the crystalline-symmetry-protected helical Majorana edge mode is distinguished from one protected by TRS since the system breaks TRS; see Sec.~\ref{ssec:p_x_d_y} for more discussion. 
  
  Moreover, we find a line nodal phase, i.e., a completely gapless phase, for the $p_x$-wave ($\bm{d}\parallel \hat{\bm{z}}$), $p_y$-wave ($\bm{d}\parallel \hat{\bm{z}}$), and chiral $p$-wave pairings even when $\Delta_0$ is finite. The line node originates from the mismatch of the spin structures between the QAH and the $\bm{d}$ vector. We will show that the line node is protected by the 0D topological number in Sec.~\ref{ssec:p_x_d_z}.

\begin{figure*}
  \includegraphics[width=0.7\linewidth]
  {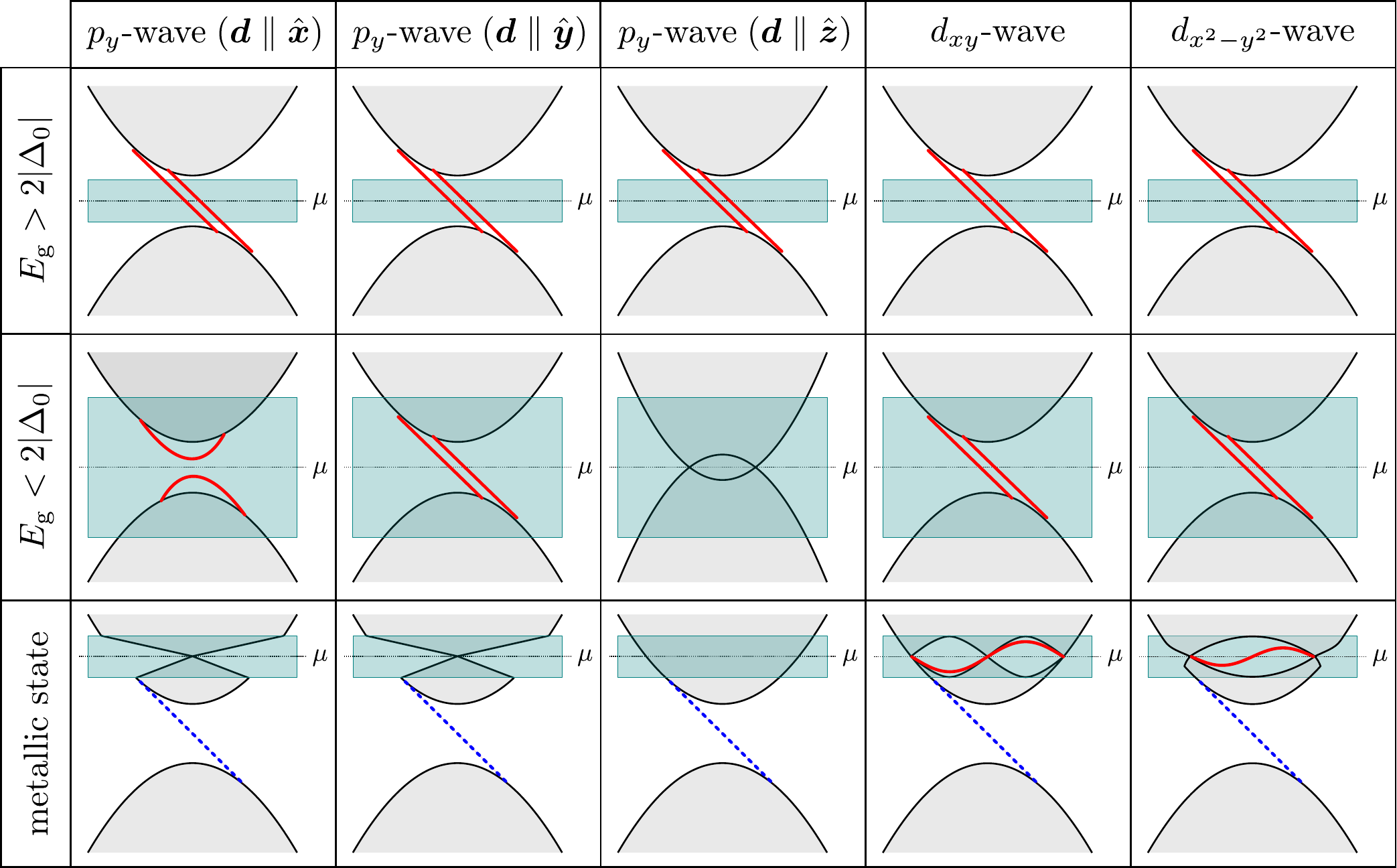}
   \caption{Schematic illustration of SDOSs in the QAH/unconventional SC hybrid systems with $p_y$-wave, $d_{xy}$-wave, and $d_{x^2-y^2}$-wave pairings.}
  \label{fg:table_SDOS_py_d}
\end{figure*}

\subsubsection{Metallic state}
 We tune the chemical potential to the metallic regime, i.e., the Fermi level lies above or below the insulating gap. In this regime, the hybrid systems realize superconducting states inherent to unconventional pairings; the gap structures reflect the pairing symmetries except for the gapless phase. For gap functions with point nodes, surface states associated with point nodes emerge~\cite{sato_2011}, where the spin-orbit coupling in the QAH gives a moderate change for some pairing symmetries. For instance, the $p_x$-wave pair ($\bm{d}\parallel \hat{\bm{y}}$) has a surface sine curve state. The surface sine curve state is a dispersed surface state terminating at the point nodes (see Fig.~\ref{fg:table_SDOS_px}) and is attributed to the spin-orbit coupling term in the QAH, which breaks the chiral symmetry (\ref{eq:BdG_gammax}) and can open a gap at $k_y \neq 0$ in the surface. The same modifications also occur for the $d_{xy}$-wave and $d_{x^2-y^2}$-wave pairings, whose surface density of states exhibit a sine curve state~\cite{sato_2009,sato_2010} (see Fig.~\ref{fg:table_SDOS_py_d}). Note that the sine curve states appear in the QAH/$d_{x^2-y^2}$-wave SC hybrid systems even when the $d_{x^2-y^2}$-wave SC does not show any surface state on the (10) surface~\cite{CRHu1994,kashiwaya_2000}. 
 
  On the other hand, a fully gapped SC with multiple chiral Majorana edge modes is realized when the SC side has chiral $d\pm id$-wave pairings, which host the nontrivial Chern number $\mathcal{N} = \mp 2$. The numerical calculation shows a sum rule of the Chern numbers: $1-2 = -1$ for the chiral $d + id$-wave pairs and $1+2 = 3$ for the chiral $d - id$-wave pairs, where the first (second) number represents the Chern number of QAH (chiral SCs). Note that the sum rule also holds true for the $s$-wave case as $1+0=1$. To understand the sum rule of the QAH/chiral SC hybrid systems, we discuss an effective Hamiltonian projected onto the Fermi level in Sec.~\ref{ssec:chiral_wave}; the effective Hamiltonian explains the sum rule as the phase winding of an effective gap function.

\begin{figure}[tb]
  \centering
  \includegraphics[width=0.9\linewidth]
  {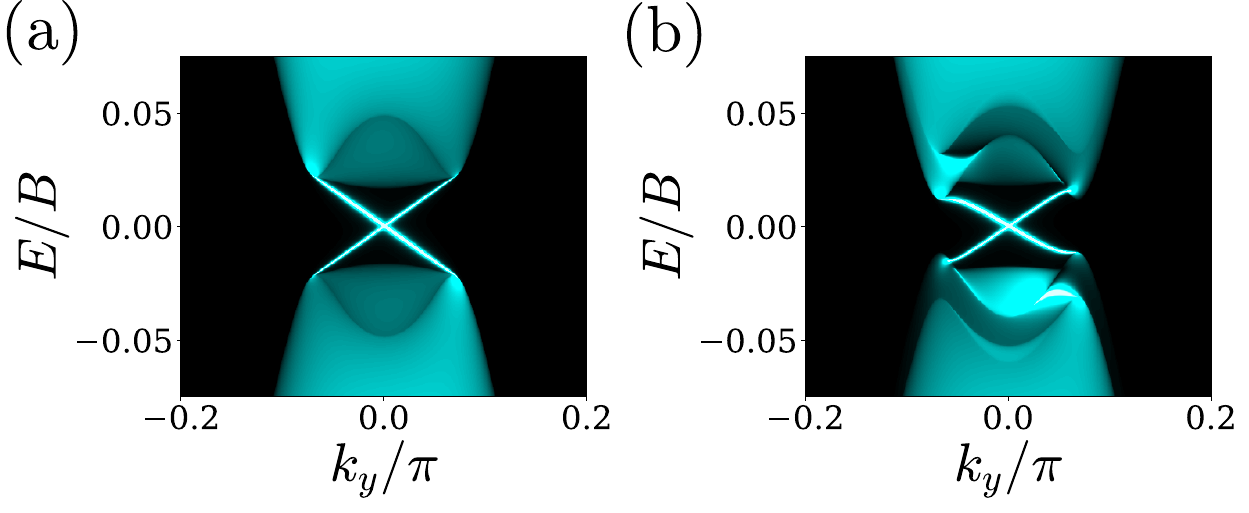}
  \caption{SDOSs on the $(10)$ surface for the QAH/$p_x$-wave SC ($\bm{d} \parallel \hat{\bm{y}}$) hybrid systems, where the parameters are chosen as $(m, \Delta_0, \mu)=(-0.05,0.175,0)$ in (a) and ($-0.05,0.175, 0.01$) in (b), respectively.}
  \label{fg: SDOS_px_y}
\end{figure}

\subsection{Crystalline-symmetry-protected helical Majorana edge modes}
\label{ssec:p_x_d_y}

It is known from the K-theoretical classification that 2D class-D SCs with a magnetic-reflection symmetry are classified by double integer numbers $\mathbb{Z} \times \mathbb{Z}$~\cite{shiozaki_2014}, where the first and second numbers represent the Chern number $\mathcal{N}$ and the 1D winding number $\mathcal{W}$ defined in Sec.~\ref{sec:topology}. Those topological numbers are not independent each other, and a minimal set of topological numbers is given by $e_1=(1,1)$ and $e_2=(1,-1) \in \mathbb{Z} \times \mathbb{Z}$~\cite{shiozaki_2014}. Other topological phases can be constructed from their combinations: $l e_1+m e_2$ ($l,m \in \mathbb{Z}$). Thus, we find a topological phase without the Chern number when the topological number is proportional to $e_1 - e_2$. In this phase, the Chern numbers of $e_1$ and $e_2$ are canceled out, whereas the 1D winding number remains nonzero and takes an even integer. As a result, two chiral Majorana edge modes flow in the opposite direction, but a gap opening perturbation between them is prohibited by the 1D winding number protected by the magnetic reflection symmetry; the resultant surface state turns out to be a crystalline-symmetry-protected helical Majorana edge mode even when TRS is absent. 

From the calculation of topological numbers, we find the topological phase with $(\mathcal{N},\mathcal{W})=(0,2)$ when the pairing symmetry in the SC side is the $p_x$ wave pairing ($\bm{d} \parallel \hat{\bm{y}}$); see Table~\ref{tb:GS_SS}. As expected, the obtained SDOS shows the crystalline-symmetry-protected helical Majorana edge mode with the band crossing at $k_y=0$ in Fig.~\ref{fg: SDOS_px_y} (a). Furthermore, when $\mu \neq 0$, each counter-propagating edge mode has different velocities since the surface states do not need to respect TRS; see Fig.~\ref{fg: SDOS_px_y} (b). This property makes a sharp contrast to the conventional helical Majorana edge mode protected by TRS.   

\begin{figure}[tb]
  \centering
  \includegraphics[width=0.8\linewidth]
  {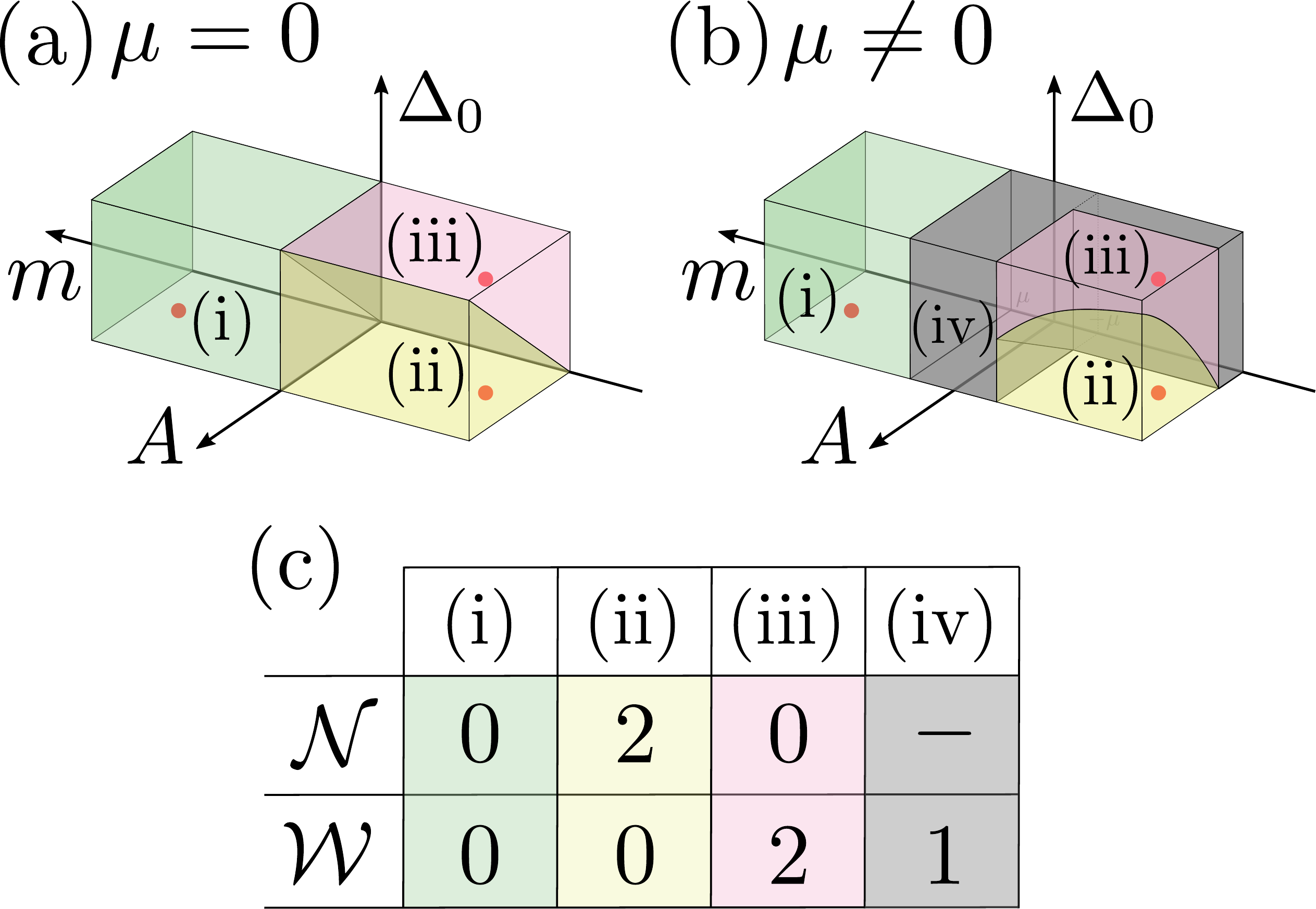}
  \caption{Topological phase diagrams of the QAH/$p_x$-wave SC ($\bm{d} \parallel \hat{\bm{y}}$) hybrid systems for (a) $\mu =0$ and (b) $\mu \neq 0$ as a function of the mass term $m$, the spin-orbit coupling $A$, and the gap function $\Delta_0$. (c) represents the topological numbers in each phase: (i) a fully-gap topologically trivial phases, (ii) a fully-gap topological phase with the two chiral Majorana edge modes, (iii) one with the crystalline-symmetry-protected helical Majorana edge modes, and (iv) a point node phase with a sine curve state.}
  \label{fg:phase_px_y}
\end{figure}

We show a topological phase diagram of the QAH/$p_x$ wave ($\bm{d} \parallel \hat{\bm{y}}$) SC hybrid systems in Fig.~\ref{fg:phase_px_y}, where the phase boundary is determined from the gap closing of energy spectrum as it signals topological phase transitions. When $\mu =0$, topological phases are divided into three regimes: (i) a fully-gap topologically trivial phase classified by $(\mathcal{N},\mathcal{W})=(0,0)$ for $m > 0$, (ii) a fully-gap topological phase with the two chiral Majorana edge modes by $(2,0)$ for $A>\Delta_0$ and $m<0$, and (iii) one with the crystalline-symmetry-protected helical Majorana edge mode by $(0,2)$ for$A<\Delta_0$ and $m<0$. See Fig.~\ref{fg:phase_px_y} (a). On the other hand, when $\mu \neq 0$, (iv) a point node phase with $\mathcal{W}=1$ appears in between the topologically trivial and nontrivial phases. In this phase, a surface sine curve state associated with the point nodes appears. The fully-gapped topological phases remain stable in a parameter regime; see Fig.~\ref{fg:phase_px_y} (b).

\begin{figure}[tb]
  \includegraphics[width=0.9\linewidth]{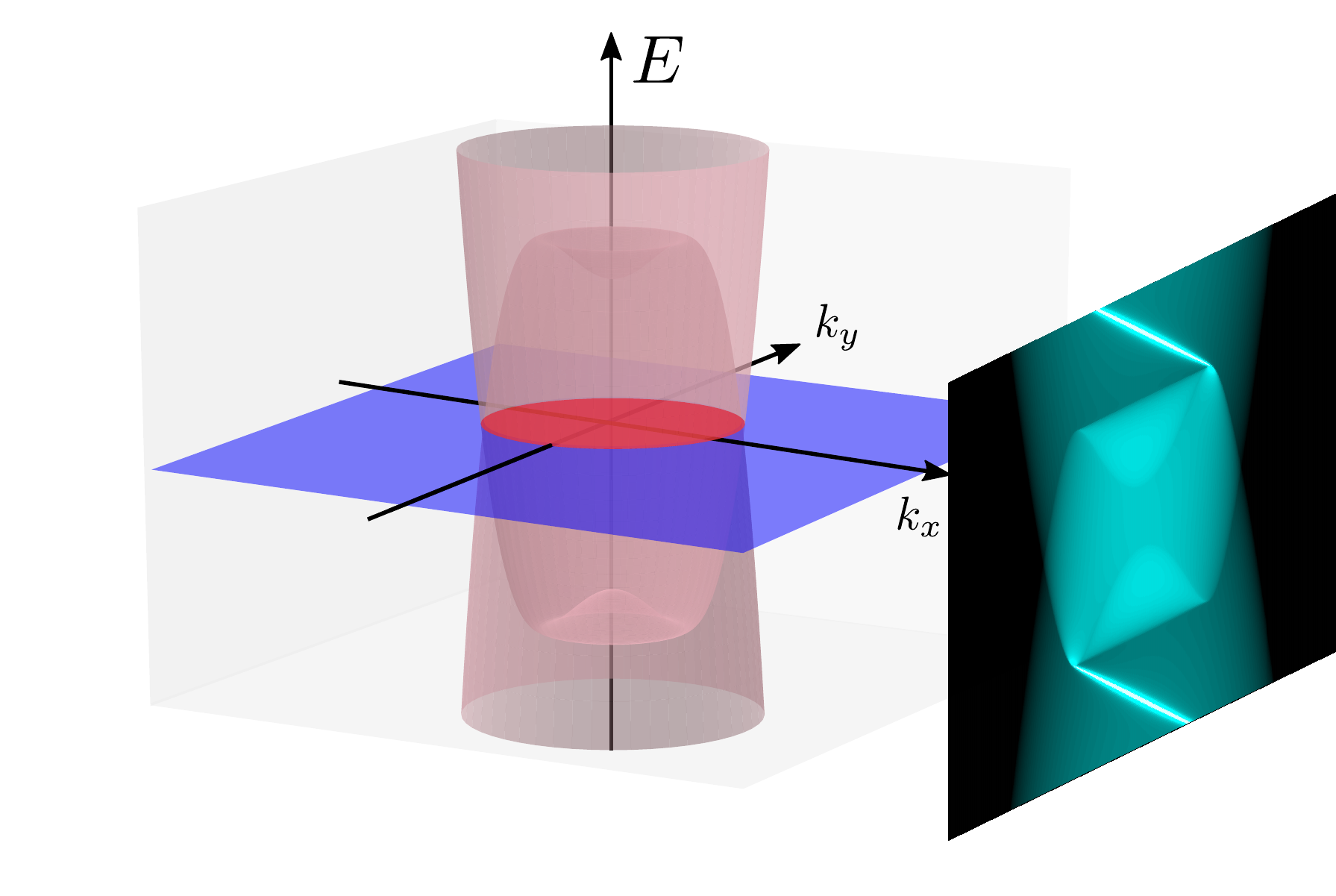}
  \caption{
    The bulk energy spectrum (left) and the SDOS on the (10) surface (right) in the QAH/$p_y$-wave ($\bm{d}\parallel \hat{\bm{z}}$) hybrid systems with the parameter $(m, \Delta_0, \mu)= (-0.05, 0.025,0.06)$. The numerical calculation shows a line node (a gapless phase) due to the mismatch of the spin structure between the QAH state and $\bm{d}$ vector.
    }
  \label{fg:Z2_correspond}
 \end{figure}
 
\subsection{Crystalline-symmetry-protected Bogoliubov Fermi surface}
\label{ssec:p_x_d_z}
Our numerical calculations show a line node phase, i.e., a completely gapless phase, when the $\bm{d}$ vector is parallel to the $z$ axis. See Fig.~\ref{fg:Z2_correspond}. The gapless phase originates from the mismatch of spin structures between the QAH and the gap function. We here discuss topological stability of the line node. The QAH/$p$-wave ($\bm{d} \parallel \hat{\bm{z}}$) SC hybrid system possesses neither TRS nor spatial-inversion symmetry. Instead, it has the two-fold rotation symmetry (\ref{eq:normal_C2z}), which plays the same role of the spatial-inversion symmetry in 2D space.  When the SC side has the $\bm{d}$ vector parallel to the $z$ axis, it yields the commutation relation $\{\mathcal{C},\mathcal{C}_{2z}\}=0$, which enables us to define a crystalline-symmetry-protected 0D $\mathbb{Z}_2$ invariant~(\ref{eq:Z2_2D}) that protects a line node even for odd-parity SCs. We note that the $\mathbb{Z}_2$ invariant is a generalization of that defined in three-dimensional even-parity SCs without TRS~\cite{kobayashi_2014,zhao_2016,agterberg_2017,bzdusek_2017,timm_2017,brydon_2018}.

\begin{figure}[tb]
  \includegraphics[width=0.9\linewidth]{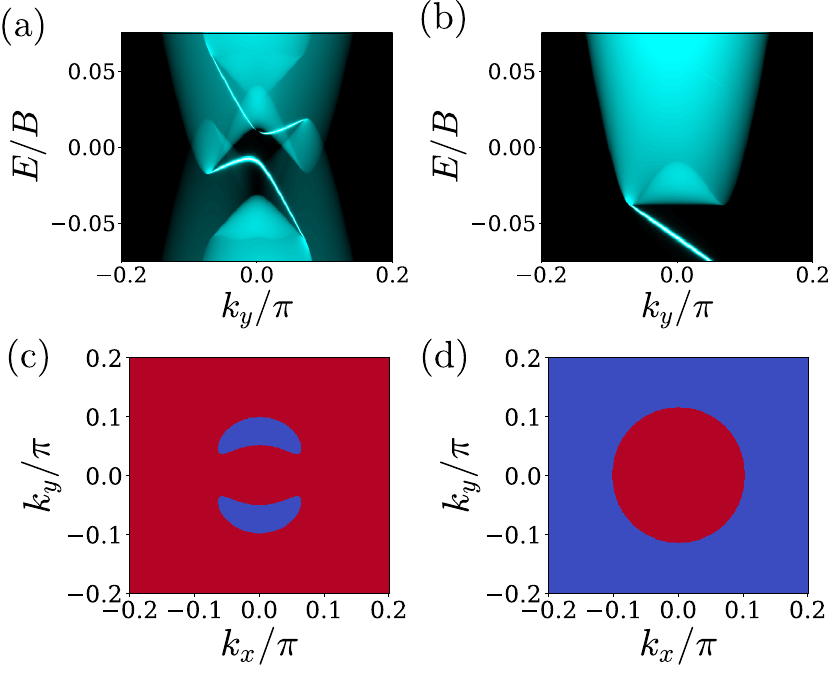}
  \caption{
    SDOSs on the (10) surface and sign of the Pfaffian in the QAH/$p_y$-wave ($\bm{d}\parallel \hat{\bm{z}}$) hybrid systems, where the parameters are chosen as $(m, \Delta_0, \mu)=(-0.05,0.175,0)$ in (a,c) and ($-0.05, 0.025,0.06$) in (c,d); these parameters belong to the cases of $E_{\rm g} < 2|\Delta_0|$ and the metallic state, respectively. In (a,b), the SDOSs exhibit a line node on the intersection of electron and hole bands. In (c,d), we show the sign of the Pfaffian in Eq.~(\ref{eq:Z2_2D}) at $E=0$, where $\bm{k}''$ is fixed to $(0,0)$. The red (blue) color represents the region with $\nu=0$ ($\nu=1$). The sign change, which is the boundary between the red and blue regions, corresponds to the line node
    }
  \label{fg:Z2}
\end{figure}

To verify that the line node is protected by the crystalline-symmetry-protected $\mathbb{Z}_2$ invariant, we numerically calculate Eq.~(\ref{eq:Z2_2D}) in the gapless phases in the QAH/$p_y$-wave ($\bm{d} \parallel \hat{\bm{z}}$) hybrid system. Figures~\ref{fg:Z2} (a) and (b) show the SDOSs for $E_{\rm g} < 2|\Delta_0|$ and the metallic state, respectively. Both cases exhibit line nodes on the intersection of electron and hole bands. Figures~\ref{fg:Z2} (c) and (d) illustrate the sign of the Pfaffian in Eq.~(\ref{eq:Z2_2D}) for $E_{\rm g} < 2|\Delta_0|$ and the metallic state, where we fix $\bm{k}'' = (0,0)$ and vary $\bm{k}'$. The line nodes appear on the boundary of the regions with $\nu=0$ (red) and $\nu=1$ (blue); namely, they are protected by the crystalline-symmetry-protected $\mathbb{Z}_2$ invariant.

\subsection{Multiple chiral Majorana edge modes}
\label{ssec:chiral_wave}

When the chemical potential lies above or below the insulating gap, the normal Hamiltonian forms a Fermi surface and a superconducting gap arises on it. Then, there arise superconducting states inherent to pairing symmetries in conjunction with a modification by the spin-orbit coupling in the QAH. In this regime, we find multiple chiral Majorana edge modes in the QAH/chiral $d \pm id$-wave SC hybrid systems since the chiral $d \pm id$-wave SCs have nonzero Chern numbers, which manifest itself chiral Majorana edge modes. However, the number of chiral Majorana edge modes in the hybrid system is different from that in the chiral $d \pm id$-wave SC. This is because the Chern number in the QAH also affect the bulk topology, and thus the number of chiral Majorana edge modes obeys a sum rule of the Chern numbers in the QAH and the chiral $d \pm id$-wave SC. 

\begin{figure}[tb]
    \centering
    \includegraphics[width=0.9\linewidth]
    {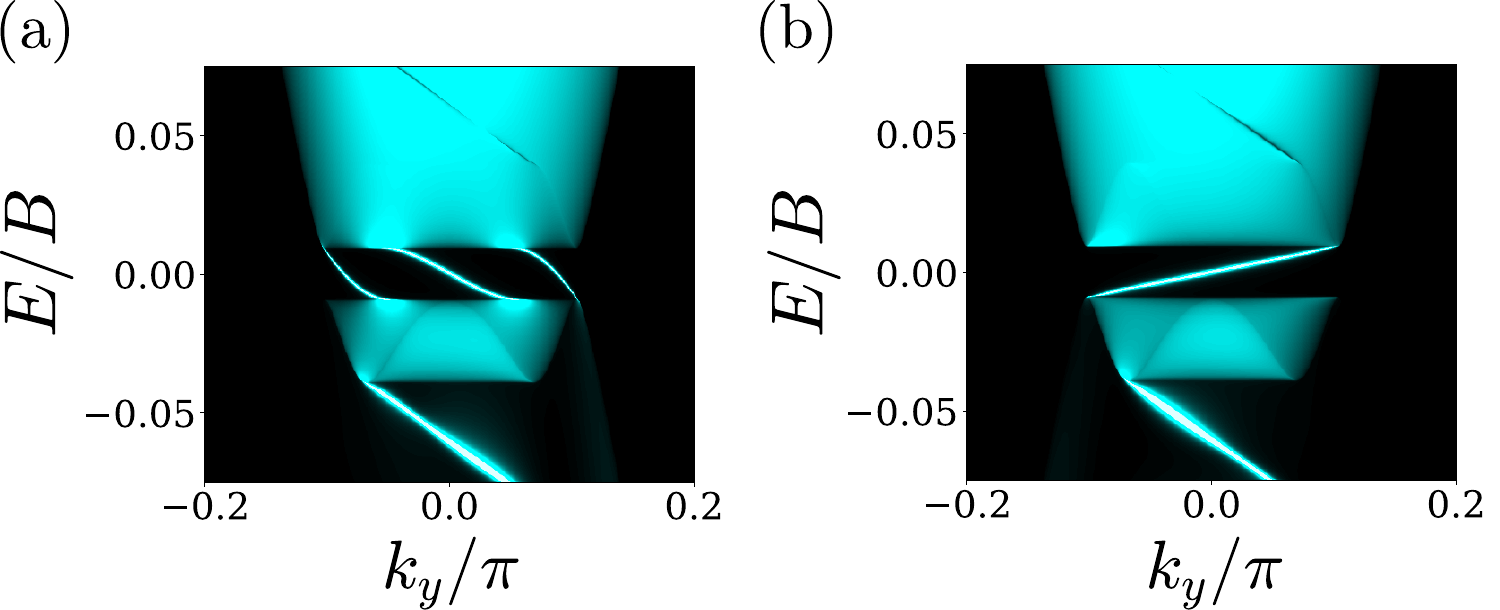}
  \caption{
   SDOSs on the $(10)$ surface in the QAH/$d \pm i d$-wave SC hybrid systems with $(m, \Delta_0, \mu)=(-0.05, 0.175,0.06)$. The SDOSs exhibit (a) three chiral Majorana edge modes for the $d - i d$-wave pairing and (b) a single chiral Majorana edge mode for the $d + i d$-wave pairing. Here, we enlarge $\Delta_0$ to check the chiral Majorana edge modes. The superconducting gap on the Fermi surface is given by Eq.~(\ref{eq:proj_gap}).
    }
  \label{fg:chiral}
\end{figure}

Figure~\ref{fg:chiral} shows the SDOSs of QAH/chiral $d \pm id$-wave SC hybrid systems in the metallic regime. We find three chiral Majorana edge modes $(dE/dk_y <0)$ for the $d - id$-wave pairing and a single chiral Majorana edge mode $(dE/dk_y >0)$ for the $d + id$-wave pairing. Since the chiral $d \pm id$-wave SCs have the Chern number $\mathcal{N} = \mp 2$, the Chern number in the hybrid systems is shifted by $1$.

To understand the sum rule of the Chern numbers, we employ the analogy between superconducting Dirac Hamiltonians and spinless chiral $p$-wave superconductors~\cite{fu_2008,Tanaka2009}. The analogy is made by projecting the BdG Hamiltonian (\ref{eq:BdG_Hamiltonian}) onto the Fermi surface. Provided that $\mu \sim v |\bm{k}|$ and $\mu \gg E_{\rm g} =2 |m|$, we can ignore $m(\bm{k})$, and the normal Hamiltonian is diagonalized by the unitary transformation,
\begin{align}
 \left(\begin{array}{@{\,} c @{\,}} \tilde{c}_{\bm{k},\uparrow} \\ \tilde{c}_{\bm{k},\downarrow} \end{array} \right) \equiv  \frac{1}{\sqrt{2}} \begin{pmatrix} 1 & e^{-i \theta_{\bm{k}}} \\ e^{i \theta_{\bm{k}}} & -1\end{pmatrix} \left(\begin{array}{@{\,} c @{\,}} c_{\bm{k},\uparrow} \\ c_{\bm{k},\downarrow} \end{array} \right),
\end{align}
where $ \theta_{\bm{k}} = \arctan(k_y/k_x)$. The projected Hamiltonian with the basis $\tilde{c}_{\bm{k},\uparrow}$ is then described as
\begin{align}
\tilde{H}_{\rm BdG} &= (v |\bm{k}| - \mu)\tilde{c}^{\dagger}_{\bm{k},\uparrow}\tilde{c}_{\bm{k},\uparrow} \notag \\
&+ \frac{\Delta_0}{2} \left\{f(\bm{k}) e^{-i \theta_{\bm{k}}} \tilde{c}_{\bm{k},\uparrow}^{\dagger}\tilde{c}_{\bm{k},\uparrow}^{\dagger} + {\rm h.c.}\right\}. \label{eq:proj_Hami}
\end{align}
Combining with Eq.~(\ref{eq:even_gap}), Eq.~(\ref{eq:proj_Hami}) has the effective gap function as
\begin{align}
  \tilde{\Delta}(\bm{k}) \equiv 
 \begin{cases} 
 \Delta_0 e^{-i \theta_{\bm{k}}} & \text{($s$-wave)}, \\ 
 \Delta_0 |\bm{k}|^2 e^{i \theta_{\bm{k}}} & \text{(chiral $d+ id$-wave)}, \\ 
 \Delta_0 |\bm{k}|^2 e^{-3i \theta_{\bm{k}}} & \text{(chiral $d- id$-wave)}.
 \end{cases}
\end{align}
The results are consistent with the numerical calculation and show that the sum rule of the Chern number is interpreted as the change of the phase winding in the effective gap function. In a similar way, in the hole-doping regime ($\mu \ll -E_{\rm g}$), we also have the sum rule of the Chern number by reversing  the sign of the phase winding in the QAH. Note that the similar results hold true even when $m(\bm{k}) \neq 0$. In this case, the effective gap function is obtained as 
\begin{align}
\tilde{\Delta}(\bm{k}) = \frac{A|\bm{k}|}{E_{\rm QAH}} \Delta_0 f(\bm{k}) e^{-i \theta_{\bm{k}}}. \label{eq:proj_gap}
\end{align}

\section{Discussion and conclusion}
\label{sec:conclusion}
\begin{figure*}[htb]
  \centering
  \includegraphics[width=0.9\hsize]{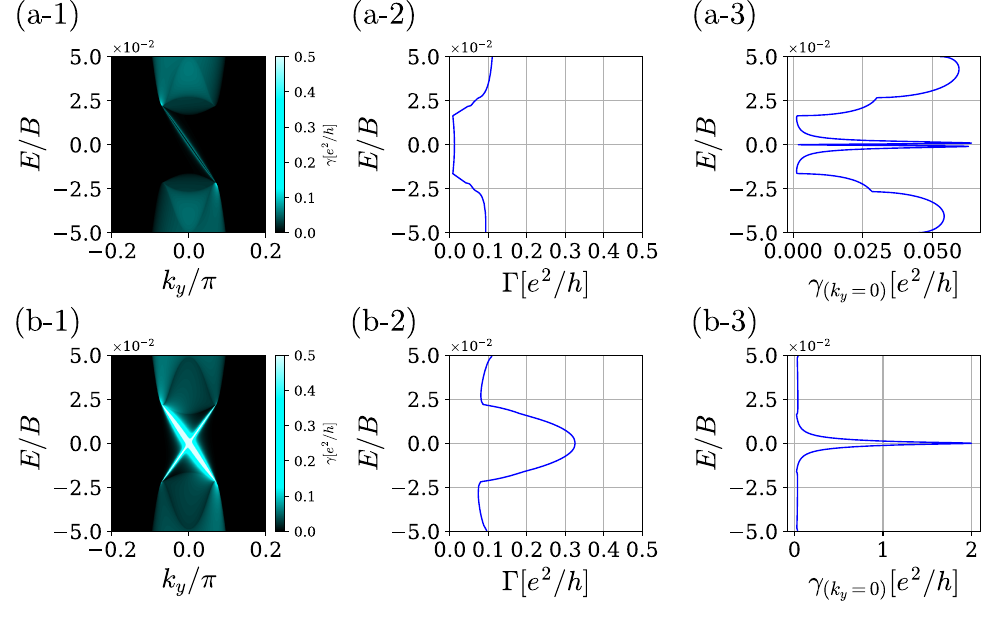} 
  \caption{
   Tunneling conductance in the normal metal/(QAH/$p_x$-wave SC ($\bm{d}\parallel\hat{\bm{y}}$)) junction. (a-1), (a-2), and (a-3) [(b-1), (b-2), and (b-3)] show the angle-resolved conductance, the total conductance, and the angle-resolved conductance at $k_y =0$ in the case of $(\mathcal{N},\mathcal{W})=(2,0)$ [$(0,2)$]. Here, we define the normal metal as $H_{\rm N} = 2t_{\rm N}(2-\cos k_x - \cos k_y) - \mu_{\rm N}$, whose parameters are chosen as $(t_{\rm N}, \mu_{\rm N})=(10,10)$. In the QAH/$p_x$-wave SC ($\bm{d}\parallel\hat{\bm{y}}$) hybrid system, we choose $(A,B,m,\Delta_0,\mu) =(0.1,1,-0.05,0.025,0)$ for (a-1), (a-2), and (a-3) and $(0.1,1,-0.05,0.175,0)$ for (b-1), (b-2), and (b-3). 
   }  
  \label{fg:Cond_px_y}
\end{figure*}

In this paper, we have studied the QAH/unconventional SC hybrid systems with $p_x$-wave, $p_y$-wave, $d_{xy}$-wave, $d_{x^2-y^2}$-wave, chiral $p\pm ip$-wave, and chiral $d\pm id$-wave pairings. By focusing on the three characteristic parameter regimes and numerically calculating the SDOSs and the topological numbers, we have found a variety of topological phases, with exotic topological phases enabled by the interplay between the QAH and unconventional SCs. The gap structure, surface states and topological numbers are summarized in Table~\ref{tb:GS_SS} and the characteristic surface states are illustrated in Fig.~\ref{fg:table_SDOS_px}. We have unveiled the distinctive topological phases with the crystalline-symmetry-protected helical Majorana edge modes [Sec.~\ref{ssec:p_x_d_y}], the crystalline-symmetry-protected Bogoliubov Fermi surfaces (line nodes) [Sec.~\ref{ssec:p_x_d_z}], and the multiple chiral Majorana edge modes [Sec.~\ref{ssec:chiral_wave}]. 
 In particular, unlike the (pseudo-)TRS protected helical Majorana edge modes~\cite{qi_2009,huang_2018}, the crystalline-symmetry-protected helical Majorana edge modes do not respect TRS and allow the counter-propagating edge currents with different velocities. 
A large variety of topological phases offer a platform for systematic study of topological phases and enable us to design exotic topological SCs experimentally.

Finally, we discuss tunnel conductance in a junction between a normal metal ($x>0$) and the QAH/unconventional SC hybrid systems ($x <0$) as a probe of the topological phases,
 which can be determined from the SDOSs and the topological numbers~\cite{ii_2012,ohashi_2020,matsumoto_1995,tanaka_1995,kashiwaya_2000,tanuma_2001,yada_2011,sato_2011,kobayashi_2015,tamura_2017,Lapp2020}. The behavior of tunnel conductance is briefly summarized as follow. Hereafter, we assume a normal metal without spin-orbit coupling having a quadratic dispersion and an insulating barrier between a normal metal and the QAH/unconventional hybrid system in the metallic regime. The periodic boundary condition is imposed for $y$ axis, and the angle-resolved conductance is defined in terms of $k_y$.  
\begin{itemize}
\item Tunnel conductance depends on the 1D winding number when the QAH/SC hybrid systems have a full gap and show topological surface states~\cite{ii_2012,ohashi_2020}; the angle-resolved conductance enhances in the gap and shows a zero-bias conductance peak at $k_y=0$ when $\mathcal{W}$ is nonzero, while the vanishing of conductance occurs when $\mathcal{W}$ is zero due to the interference of degenerate Majorana fermions~\cite{yamakage_2014}. This scenario can be applied to the topological phases with $(\mathcal{N},\mathcal{W}) = (-1,1)$, $(1,1)$, $(2,0)$, $(0,2)$, and $(3,1)$, including the QAH/$p_x$-wave ($\bm{d} \parallel \hat{\bm{y}}$) SC hybrid systems showing the crystalline-symmetry-protected helical Majorana edge modes. 

\item Tunnel conductance directly reflects the bulk density of states when the superconducting state has nodes but a surface state is absent~\cite{kashiwaya_2000}. In particular, the crystalline-symmetry-protected Bogoliubov Fermi surface behaves like a normal Fermi surface, and thus can be observed as a finite zero-bias conductance in the absence of the insulating barrier~\cite{Lapp2020}.

\item Tunnel conductance exhibits a sharp zero-bias conductance peak (a broadened zero-bias conductance peak) when the superconducting state has point nodes involving a surface flat band (a surface sine-curve state). Such behaviors have been studied in the context of unconventional SCs~\cite{matsumoto_1995,tanaka_1995,kashiwaya_2000,tanuma_2001,yada_2011,sato_2011,kobayashi_2015,tamura_2017}.
\end{itemize}

  We numerically calculate tunnel conductance in a junction between a normal metal  ($x >0$) and the QAH/$p_x$-wave SC ($\bm{d}\parallel\hat{\bm{y}}$) hybrid system ($x <0$) using the Lee-Fisher formula~\cite{lee_1981,ii_2012} to understand the difference between the chiral and helical Majorana edge modes classified by $(\mathcal{N},\mathcal{W}) = (2,0)$ and $(0,2)$, where the angle-resolved conductance and the total conductance are defined by $\gamma(k_y,E)$ and $\Gamma(E) = \sum_{k_y} \gamma(k_y,E)$; see Ref.~\onlinecite{ii_2012} for the explicit expressions.  
 
 Figure~\ref{fg:Cond_px_y} shows the tunnel conductance in the normal metal/(QAH/$p_x$-wave SC ($\bm{d}\parallel\hat{\bm{y}}$)) junction. We here consider the two parameter regimes:  $E_{\rm g} > 2|\Delta_0|$ and  $E_{\rm g} < 2|\Delta_0|$ for the purposes of comparison; their topological numbers are given by $(\mathcal{N},\mathcal{W})=(2,0)$ and $(\mathcal{N},\mathcal{W})=(0,2)$, respectively. For $(\mathcal{N},\mathcal{W})=(2,0)$, we find two chiral Majorana edge modes on the (10) surface (Fig.~\ref{fg:Cond_px_y} (a-1)). The total conductance decreases and is close to zero in the gap (see Fig.~\ref{fg:Cond_px_y} (a-3)), whereas the angle-resolved conductance at $k_y =0$ drastically vanishes at $E=0$ (Fig.~\ref{fg:Cond_px_y} (a-3)). The vanishing of the conductance results from that the degeneracy of zero modes at $k_y=0$ is accidental and not protected by any symmetry~\cite{ii_2012,yamakage_2014}. On the other hand, for $(\mathcal{N},\mathcal{W})=(0,2)$, the crystalline-symmetry-protected helical Majorana edge modes appears on the (10) surface (Fig.~\ref{fg:Cond_px_y} (b-1)) and significantly affect the conductance; the total conductance increase in the gap and peaks at $E=0$ (Fig.~\ref{fg:Cond_px_y} (b-2)), and  the angle-resolved conductance at $k_y=0$ exhibits a sharp zero bias peak at $E=0$ owing to the nonzero 1D winding number (Fig.~\ref{fg:Cond_px_y} (b-3)). The results are consistent with the previous study~\cite{ii_2012,ohashi_2020}, and thus  the chiral and helical Majorana edge modes are distinguishable though the tunnel conductance.

\section*{Acknowledgements}
R.O. was supported by JSPS KAKENHI Grants No. JP21J15526. 
S.K. was supported by JSPS KAKENHI Grants Nos. JP19K14612, JP19H01824 and the CREST project (JPMJCR16F2, JPMJCR19T2) from Japan Science and Technology Agency (JST).
Y.T. was supported by Scientific Research (A) (KAKENHI Grant No.JP20H00131), Scientific Research (B) (KAKENHI Grants No. JP18H01176 and No. JP20H01857), Japan-RFBR Bilateral Joint Research Projects/Seminars No. 19-52-50026, and the JSPS Core-to-Core program “Oxide Superspin” international network.

\bibliographystyle{apsrev4-1}

\bibliography{qah-sc}

\begin{thebibliography}{124}%
\makeatletter
\providecommand \@ifxundefined [1]{%
 \@ifx{#1\undefined}
}%
\providecommand \@ifnum [1]{%
 \ifnum #1\expandafter \@firstoftwo
 \else \expandafter \@secondoftwo
 \fi
}%
\providecommand \@ifx [1]{%
 \ifx #1\expandafter \@firstoftwo
 \else \expandafter \@secondoftwo
 \fi
}%
\providecommand \natexlab [1]{#1}%
\providecommand \enquote  [1]{``#1''}%
\providecommand \bibnamefont  [1]{#1}%
\providecommand \bibfnamefont [1]{#1}%
\providecommand \citenamefont [1]{#1}%
\providecommand \href@noop [0]{\@secondoftwo}%
\providecommand \href [0]{\begingroup \@sanitize@url \@href}%
\providecommand \@href[1]{\@@startlink{#1}\@@href}%
\providecommand \@@href[1]{\endgroup#1\@@endlink}%
\providecommand \@sanitize@url [0]{\catcode `\\12\catcode `\$12\catcode
  `\&12\catcode `\#12\catcode `\^12\catcode `\_12\catcode `\%12\relax}%
\providecommand \@@startlink[1]{}%
\providecommand \@@endlink[0]{}%
\providecommand \url  [0]{\begingroup\@sanitize@url \@url }%
\providecommand \@url [1]{\endgroup\@href {#1}{\urlprefix }}%
\providecommand \urlprefix  [0]{URL }%
\providecommand \Eprint [0]{\href }%
\providecommand \doibase [0]{http://dx.doi.org/}%
\providecommand \selectlanguage [0]{\@gobble}%
\providecommand \bibinfo  [0]{\@secondoftwo}%
\providecommand \bibfield  [0]{\@secondoftwo}%
\providecommand \translation [1]{[#1]}%
\providecommand \BibitemOpen [0]{}%
\providecommand \bibitemStop [0]{}%
\providecommand \bibitemNoStop [0]{.\EOS\space}%
\providecommand \EOS [0]{\spacefactor3000\relax}%
\providecommand \BibitemShut  [1]{\csname bibitem#1\endcsname}%
\let\auto@bib@innerbib\@empty
\bibitem [{\citenamefont {Qi}\ and\ \citenamefont {Zhang}(2011)}]{qi_2011}%
  \BibitemOpen
  \bibfield  {author} {\bibinfo {author} {\bibfnamefont {X.-L.}\ \bibnamefont
  {Qi}}\ and\ \bibinfo {author} {\bibfnamefont {S.-C.}\ \bibnamefont {Zhang}},\
  }\href {\doibase 10.1103/RevModPhys.83.1057} {\bibfield  {journal} {\bibinfo
  {journal} {Rev. Mod. Phys.}\ }\textbf {\bibinfo {volume} {83}},\ \bibinfo
  {pages} {1057} (\bibinfo {year} {2011})}\BibitemShut {NoStop}%
\bibitem [{\citenamefont {Tanaka}\ \emph {et~al.}(2012)\citenamefont {Tanaka},
  \citenamefont {Sato},\ and\ \citenamefont {Nagaosa}}]{tanaka_2012}%
  \BibitemOpen
  \bibfield  {author} {\bibinfo {author} {\bibfnamefont {Y.}~\bibnamefont
  {Tanaka}}, \bibinfo {author} {\bibfnamefont {M.}~\bibnamefont {Sato}}, \ and\
  \bibinfo {author} {\bibfnamefont {N.}~\bibnamefont {Nagaosa}},\ }\href
  {\doibase 10.1143/JPSJ.81.011013} {\bibfield  {journal} {\bibinfo  {journal}
  {Journal of the Physical Society of Japan}\ }\textbf {\bibinfo {volume}
  {81}},\ \bibinfo {pages} {011013} (\bibinfo {year} {2012})}\BibitemShut
  {NoStop}%
\bibitem [{\citenamefont {Alicea}(2012)}]{alicea_2012}%
  \BibitemOpen
  \bibfield  {author} {\bibinfo {author} {\bibfnamefont {J.}~\bibnamefont
  {Alicea}},\ }\href@noop {} {\bibfield  {journal} {\bibinfo  {journal} {Rep.
  Prog. Phys.}\ }\textbf {\bibinfo {volume} {75}},\ \bibinfo {pages} {076501}
  (\bibinfo {year} {2012})}\BibitemShut {NoStop}%
\bibitem [{\citenamefont {Elliott}\ and\ \citenamefont
  {Franz}(2015)}]{elliott_2015}%
  \BibitemOpen
  \bibfield  {author} {\bibinfo {author} {\bibfnamefont {S.~R.}\ \bibnamefont
  {Elliott}}\ and\ \bibinfo {author} {\bibfnamefont {M.}~\bibnamefont
  {Franz}},\ }\href {\doibase 10.1103/RevModPhys.87.137} {\bibfield  {journal}
  {\bibinfo  {journal} {Rev. Mod. Phys.}\ }\textbf {\bibinfo {volume} {87}},\
  \bibinfo {pages} {137} (\bibinfo {year} {2015})}\BibitemShut {NoStop}%
\bibitem [{\citenamefont {Sato}\ and\ \citenamefont
  {Fujimoto}(2016)}]{sato_2016}%
  \BibitemOpen
  \bibfield  {author} {\bibinfo {author} {\bibfnamefont {M.}~\bibnamefont
  {Sato}}\ and\ \bibinfo {author} {\bibfnamefont {S.}~\bibnamefont
  {Fujimoto}},\ }\href {\doibase 10.7566/JPSJ.85.072001} {\bibfield  {journal}
  {\bibinfo  {journal} {J. Phys. Soc. Jpn.}\ }\textbf {\bibinfo {volume}
  {85}},\ \bibinfo {pages} {072001} (\bibinfo {year} {2016})}\BibitemShut
  {NoStop}%
\bibitem [{\citenamefont {Mizushima}\ \emph {et~al.}(2016)\citenamefont
  {Mizushima}, \citenamefont {Tsutsumi}, \citenamefont {Kawakami},
  \citenamefont {Sato}, \citenamefont {Ichioka},\ and\ \citenamefont
  {Machida}}]{mizushima_2016}%
  \BibitemOpen
  \bibfield  {author} {\bibinfo {author} {\bibfnamefont {T.}~\bibnamefont
  {Mizushima}}, \bibinfo {author} {\bibfnamefont {Y.}~\bibnamefont {Tsutsumi}},
  \bibinfo {author} {\bibfnamefont {T.}~\bibnamefont {Kawakami}}, \bibinfo
  {author} {\bibfnamefont {M.}~\bibnamefont {Sato}}, \bibinfo {author}
  {\bibfnamefont {M.}~\bibnamefont {Ichioka}}, \ and\ \bibinfo {author}
  {\bibfnamefont {K.}~\bibnamefont {Machida}},\ }\href {\doibase
  10.7566/JPSJ.85.022001} {\bibfield  {journal} {\bibinfo  {journal} {J. Phys.
  Soc. Jpn.}\ }\textbf {\bibinfo {volume} {85}},\ \bibinfo {pages} {022001}
  (\bibinfo {year} {2016})}\BibitemShut {NoStop}%
\bibitem [{\citenamefont {Sato}\ and\ \citenamefont {Ando}(2017)}]{sato_2017}%
  \BibitemOpen
  \bibfield  {author} {\bibinfo {author} {\bibfnamefont {M.}~\bibnamefont
  {Sato}}\ and\ \bibinfo {author} {\bibfnamefont {Y.}~\bibnamefont {Ando}},\
  }\href {http://stacks.iop.org/0034-4885/80/i=7/a=076501} {\bibfield
  {journal} {\bibinfo  {journal} {Rep. Prog. Phys.}\ }\textbf {\bibinfo
  {volume} {80}},\ \bibinfo {pages} {076501} (\bibinfo {year}
  {2017})}\BibitemShut {NoStop}%
\bibitem [{\citenamefont {Chiu}\ \emph {et~al.}(2016)\citenamefont {Chiu},
  \citenamefont {Teo}, \citenamefont {Schnyder},\ and\ \citenamefont
  {Ryu}}]{chiu_2016}%
  \BibitemOpen
  \bibfield  {author} {\bibinfo {author} {\bibfnamefont {C.-K.}\ \bibnamefont
  {Chiu}}, \bibinfo {author} {\bibfnamefont {J.~C.~Y.}\ \bibnamefont {Teo}},
  \bibinfo {author} {\bibfnamefont {A.~P.}\ \bibnamefont {Schnyder}}, \ and\
  \bibinfo {author} {\bibfnamefont {S.}~\bibnamefont {Ryu}},\ }\href {\doibase
  10.1103/RevModPhys.88.035005} {\bibfield  {journal} {\bibinfo  {journal}
  {Rev. Mod. Phys.}\ }\textbf {\bibinfo {volume} {88}},\ \bibinfo {pages}
  {035005} (\bibinfo {year} {2016})}\BibitemShut {NoStop}%
\bibitem [{\citenamefont {Moore}\ and\ \citenamefont
  {Read}(1991)}]{moore_1991}%
  \BibitemOpen
  \bibfield  {author} {\bibinfo {author} {\bibfnamefont {G.}~\bibnamefont
  {Moore}}\ and\ \bibinfo {author} {\bibfnamefont {N.}~\bibnamefont {Read}},\
  }\href@noop {} {\bibfield  {journal} {\bibinfo  {journal} {Nuclear Physics
  B}\ }\textbf {\bibinfo {volume} {360}},\ \bibinfo {pages} {362} (\bibinfo
  {year} {1991})}\BibitemShut {NoStop}%
\bibitem [{\citenamefont {Read}\ and\ \citenamefont {Green}(2000)}]{read_2000}%
  \BibitemOpen
  \bibfield  {author} {\bibinfo {author} {\bibfnamefont {N.}~\bibnamefont
  {Read}}\ and\ \bibinfo {author} {\bibfnamefont {D.}~\bibnamefont {Green}},\
  }\href {\doibase 10.1103/PhysRevB.61.10267} {\bibfield  {journal} {\bibinfo
  {journal} {Phys. Rev. B}\ }\textbf {\bibinfo {volume} {61}},\ \bibinfo
  {pages} {10267} (\bibinfo {year} {2000})}\BibitemShut {NoStop}%
\bibitem [{\citenamefont {Ivanov}(2001)}]{ivanov_2001}%
  \BibitemOpen
  \bibfield  {author} {\bibinfo {author} {\bibfnamefont {D.~A.}\ \bibnamefont
  {Ivanov}},\ }\href {\doibase 10.1103/PhysRevLett.86.268} {\bibfield
  {journal} {\bibinfo  {journal} {Phys. Rev. Lett.}\ }\textbf {\bibinfo
  {volume} {86}},\ \bibinfo {pages} {268} (\bibinfo {year} {2001})}\BibitemShut
  {NoStop}%
\bibitem [{\citenamefont {Stone}\ and\ \citenamefont
  {Chung}(2006)}]{stone_2006}%
  \BibitemOpen
  \bibfield  {author} {\bibinfo {author} {\bibfnamefont {M.}~\bibnamefont
  {Stone}}\ and\ \bibinfo {author} {\bibfnamefont {S.-B.}\ \bibnamefont
  {Chung}},\ }\href {\doibase 10.1103/PhysRevB.73.014505} {\bibfield  {journal}
  {\bibinfo  {journal} {Phys. Rev. B}\ }\textbf {\bibinfo {volume} {73}},\
  \bibinfo {pages} {014505} (\bibinfo {year} {2006})}\BibitemShut {NoStop}%
\bibitem [{\citenamefont {Fujimoto}(2008)}]{fujimoto2008}%
  \BibitemOpen
  \bibfield  {author} {\bibinfo {author} {\bibfnamefont {S.}~\bibnamefont
  {Fujimoto}},\ }\href {\doibase 10.1103/PhysRevB.77.220501} {\bibfield
  {journal} {\bibinfo  {journal} {Phys. Rev. B}\ }\textbf {\bibinfo {volume}
  {77}},\ \bibinfo {pages} {220501(R)} (\bibinfo {year} {2008})}\BibitemShut
  {NoStop}%
\bibitem [{\citenamefont {Alicea}\ \emph {et~al.}(2011)\citenamefont {Alicea},
  \citenamefont {Oreg}, \citenamefont {Refael}, \citenamefont {Von~Oppen},\
  and\ \citenamefont {Fisher}}]{alicea_2011}%
  \BibitemOpen
  \bibfield  {author} {\bibinfo {author} {\bibfnamefont {J.}~\bibnamefont
  {Alicea}}, \bibinfo {author} {\bibfnamefont {Y.}~\bibnamefont {Oreg}},
  \bibinfo {author} {\bibfnamefont {G.}~\bibnamefont {Refael}}, \bibinfo
  {author} {\bibfnamefont {F.}~\bibnamefont {Von~Oppen}}, \ and\ \bibinfo
  {author} {\bibfnamefont {M.~P.}\ \bibnamefont {Fisher}},\ }\href@noop {}
  {\bibfield  {journal} {\bibinfo  {journal} {Nature Physics}\ }\textbf
  {\bibinfo {volume} {7}},\ \bibinfo {pages} {412} (\bibinfo {year}
  {2011})}\BibitemShut {NoStop}%
\bibitem [{\citenamefont {Nayak}\ \emph {et~al.}(2008)\citenamefont {Nayak},
  \citenamefont {Simon}, \citenamefont {Stern}, \citenamefont {Freedman},\ and\
  \citenamefont {Das~Sarma}}]{nayak_2008}%
  \BibitemOpen
  \bibfield  {author} {\bibinfo {author} {\bibfnamefont {C.}~\bibnamefont
  {Nayak}}, \bibinfo {author} {\bibfnamefont {S.~H.}\ \bibnamefont {Simon}},
  \bibinfo {author} {\bibfnamefont {A.}~\bibnamefont {Stern}}, \bibinfo
  {author} {\bibfnamefont {M.}~\bibnamefont {Freedman}}, \ and\ \bibinfo
  {author} {\bibfnamefont {S.}~\bibnamefont {Das~Sarma}},\ }\href {\doibase
  10.1103/RevModPhys.80.1083} {\bibfield  {journal} {\bibinfo  {journal} {Rev.
  Mod. Phys.}\ }\textbf {\bibinfo {volume} {80}},\ \bibinfo {pages} {1083}
  (\bibinfo {year} {2008})}\BibitemShut {NoStop}%
\bibitem [{\citenamefont {Kitaev}(2001)}]{kitaev_2001}%
  \BibitemOpen
  \bibfield  {author} {\bibinfo {author} {\bibfnamefont {A.~Y.}\ \bibnamefont
  {Kitaev}},\ }\href@noop {} {\bibfield  {journal} {\bibinfo  {journal}
  {Physics-Uspekhi}\ }\textbf {\bibinfo {volume} {44}},\ \bibinfo {pages} {131}
  (\bibinfo {year} {2001})}\BibitemShut {NoStop}%
\bibitem [{\citenamefont {Sato}\ and\ \citenamefont
  {Fujimoto}(2009)}]{sato_2009}%
  \BibitemOpen
  \bibfield  {author} {\bibinfo {author} {\bibfnamefont {M.}~\bibnamefont
  {Sato}}\ and\ \bibinfo {author} {\bibfnamefont {S.}~\bibnamefont
  {Fujimoto}},\ }\href {\doibase 10.1103/PhysRevB.79.094504} {\bibfield
  {journal} {\bibinfo  {journal} {Phys. Rev. B}\ }\textbf {\bibinfo {volume}
  {79}},\ \bibinfo {pages} {094504} (\bibinfo {year} {2009})}\BibitemShut
  {NoStop}%
\bibitem [{\citenamefont {Sau}\ \emph {et~al.}(2010)\citenamefont {Sau},
  \citenamefont {Lutchyn}, \citenamefont {Tewari},\ and\ \citenamefont
  {Das~Sarma}}]{sau_2010}%
  \BibitemOpen
  \bibfield  {author} {\bibinfo {author} {\bibfnamefont {J.~D.}\ \bibnamefont
  {Sau}}, \bibinfo {author} {\bibfnamefont {R.~M.}\ \bibnamefont {Lutchyn}},
  \bibinfo {author} {\bibfnamefont {S.}~\bibnamefont {Tewari}}, \ and\ \bibinfo
  {author} {\bibfnamefont {S.}~\bibnamefont {Das~Sarma}},\ }\href {\doibase
  10.1103/PhysRevLett.104.040502} {\bibfield  {journal} {\bibinfo  {journal}
  {Phys. Rev. Lett.}\ }\textbf {\bibinfo {volume} {104}},\ \bibinfo {pages}
  {040502} (\bibinfo {year} {2010})}\BibitemShut {NoStop}%
\bibitem [{\citenamefont {Alicea}(2010)}]{alicea_2010}%
  \BibitemOpen
  \bibfield  {author} {\bibinfo {author} {\bibfnamefont {J.}~\bibnamefont
  {Alicea}},\ }\href {\doibase 10.1103/PhysRevB.81.125318} {\bibfield
  {journal} {\bibinfo  {journal} {Phys. Rev. B}\ }\textbf {\bibinfo {volume}
  {81}},\ \bibinfo {pages} {125318} (\bibinfo {year} {2010})}\BibitemShut
  {NoStop}%
\bibitem [{\citenamefont {Oreg}\ \emph {et~al.}(2010)\citenamefont {Oreg},
  \citenamefont {Refael},\ and\ \citenamefont {von Oppen}}]{oreg_2010}%
  \BibitemOpen
  \bibfield  {author} {\bibinfo {author} {\bibfnamefont {Y.}~\bibnamefont
  {Oreg}}, \bibinfo {author} {\bibfnamefont {G.}~\bibnamefont {Refael}}, \ and\
  \bibinfo {author} {\bibfnamefont {F.}~\bibnamefont {von Oppen}},\ }\href
  {\doibase 10.1103/PhysRevLett.105.177002} {\bibfield  {journal} {\bibinfo
  {journal} {Phys. Rev. Lett.}\ }\textbf {\bibinfo {volume} {105}},\ \bibinfo
  {pages} {177002} (\bibinfo {year} {2010})}\BibitemShut {NoStop}%
\bibitem [{\citenamefont {Cook}\ and\ \citenamefont {Franz}(2011)}]{cook_2011}%
  \BibitemOpen
  \bibfield  {author} {\bibinfo {author} {\bibfnamefont {A.}~\bibnamefont
  {Cook}}\ and\ \bibinfo {author} {\bibfnamefont {M.}~\bibnamefont {Franz}},\
  }\href {\doibase 10.1103/PhysRevB.84.201105} {\bibfield  {journal} {\bibinfo
  {journal} {Phys. Rev. B}\ }\textbf {\bibinfo {volume} {84}},\ \bibinfo
  {pages} {201105(R)} (\bibinfo {year} {2011})}\BibitemShut {NoStop}%
\bibitem [{\citenamefont {Mourik}\ \emph {et~al.}(2012)\citenamefont {Mourik},
  \citenamefont {Zuo}, \citenamefont {Frolov}, \citenamefont {Plissard},
  \citenamefont {Bakkers},\ and\ \citenamefont {Kouwenhoven}}]{mourik_2012}%
  \BibitemOpen
  \bibfield  {author} {\bibinfo {author} {\bibfnamefont {V.}~\bibnamefont
  {Mourik}}, \bibinfo {author} {\bibfnamefont {K.}~\bibnamefont {Zuo}},
  \bibinfo {author} {\bibfnamefont {S.~M.}\ \bibnamefont {Frolov}}, \bibinfo
  {author} {\bibfnamefont {S.~R.}\ \bibnamefont {Plissard}}, \bibinfo {author}
  {\bibfnamefont {E.~P. A.~M.}\ \bibnamefont {Bakkers}}, \ and\ \bibinfo
  {author} {\bibfnamefont {L.~P.}\ \bibnamefont {Kouwenhoven}},\ }\href
  {\doibase 10.1126/science.1222360} {\bibfield  {journal} {\bibinfo  {journal}
  {Science}\ }\textbf {\bibinfo {volume} {336}},\ \bibinfo {pages} {1003}
  (\bibinfo {year} {2012})}\BibitemShut {NoStop}%
\bibitem [{\citenamefont {Das}\ \emph {et~al.}(2012)\citenamefont {Das},
  \citenamefont {Ronen}, \citenamefont {Most}, \citenamefont {Oreg},
  \citenamefont {Heiblum},\ and\ \citenamefont {Shtrikman}}]{das_2012}%
  \BibitemOpen
  \bibfield  {author} {\bibinfo {author} {\bibfnamefont {A.}~\bibnamefont
  {Das}}, \bibinfo {author} {\bibfnamefont {Y.}~\bibnamefont {Ronen}}, \bibinfo
  {author} {\bibfnamefont {Y.}~\bibnamefont {Most}}, \bibinfo {author}
  {\bibfnamefont {Y.}~\bibnamefont {Oreg}}, \bibinfo {author} {\bibfnamefont
  {M.}~\bibnamefont {Heiblum}}, \ and\ \bibinfo {author} {\bibfnamefont
  {H.}~\bibnamefont {Shtrikman}},\ }\href@noop {} {\bibfield  {journal}
  {\bibinfo  {journal} {Nature Physics}\ }\textbf {\bibinfo {volume} {8}},\
  \bibinfo {pages} {887} (\bibinfo {year} {2012})}\BibitemShut {NoStop}%
\bibitem [{\citenamefont {Rokhinson}\ \emph {et~al.}(2012)\citenamefont
  {Rokhinson}, \citenamefont {Liu},\ and\ \citenamefont
  {Furdyna}}]{rokhinson_2012}%
  \BibitemOpen
  \bibfield  {author} {\bibinfo {author} {\bibfnamefont {L.~P.}\ \bibnamefont
  {Rokhinson}}, \bibinfo {author} {\bibfnamefont {X.}~\bibnamefont {Liu}}, \
  and\ \bibinfo {author} {\bibfnamefont {J.~K.}\ \bibnamefont {Furdyna}},\
  }\href@noop {} {\bibfield  {journal} {\bibinfo  {journal} {Nature Physics}\
  }\textbf {\bibinfo {volume} {8}},\ \bibinfo {pages} {795} (\bibinfo {year}
  {2012})}\BibitemShut {NoStop}%
\bibitem [{\citenamefont {Deng}\ \emph {et~al.}(2012)\citenamefont {Deng},
  \citenamefont {Yu}, \citenamefont {Huang}, \citenamefont {Larsson},
  \citenamefont {Caroff},\ and\ \citenamefont {Xu}}]{deng_2012}%
  \BibitemOpen
  \bibfield  {author} {\bibinfo {author} {\bibfnamefont {M.}~\bibnamefont
  {Deng}}, \bibinfo {author} {\bibfnamefont {C.}~\bibnamefont {Yu}}, \bibinfo
  {author} {\bibfnamefont {G.}~\bibnamefont {Huang}}, \bibinfo {author}
  {\bibfnamefont {M.}~\bibnamefont {Larsson}}, \bibinfo {author} {\bibfnamefont
  {P.}~\bibnamefont {Caroff}}, \ and\ \bibinfo {author} {\bibfnamefont
  {H.}~\bibnamefont {Xu}},\ }\href@noop {} {\bibfield  {journal} {\bibinfo
  {journal} {Nano letters}\ }\textbf {\bibinfo {volume} {12}},\ \bibinfo
  {pages} {6414} (\bibinfo {year} {2012})}\BibitemShut {NoStop}%
\bibitem [{\citenamefont {Finck}\ \emph {et~al.}(2013)\citenamefont {Finck},
  \citenamefont {Van~Harlingen}, \citenamefont {Mohseni}, \citenamefont
  {Jung},\ and\ \citenamefont {Li}}]{finck_2013}%
  \BibitemOpen
  \bibfield  {author} {\bibinfo {author} {\bibfnamefont {A.~D.~K.}\
  \bibnamefont {Finck}}, \bibinfo {author} {\bibfnamefont {D.~J.}\ \bibnamefont
  {Van~Harlingen}}, \bibinfo {author} {\bibfnamefont {P.~K.}\ \bibnamefont
  {Mohseni}}, \bibinfo {author} {\bibfnamefont {K.}~\bibnamefont {Jung}}, \
  and\ \bibinfo {author} {\bibfnamefont {X.}~\bibnamefont {Li}},\ }\href
  {\doibase 10.1103/PhysRevLett.110.126406} {\bibfield  {journal} {\bibinfo
  {journal} {Phys. Rev. Lett.}\ }\textbf {\bibinfo {volume} {110}},\ \bibinfo
  {pages} {126406} (\bibinfo {year} {2013})}\BibitemShut {NoStop}%
\bibitem [{\citenamefont {Deng}\ \emph {et~al.}(2016)\citenamefont {Deng},
  \citenamefont {Vaitiekenas}, \citenamefont {Hansen}, \citenamefont {Danon},
  \citenamefont {Leijnse}, \citenamefont {Flensberg}, \citenamefont {Nyg{\r
  a}rd}, \citenamefont {Krogstrup},\ and\ \citenamefont {Marcus}}]{Deng_2016}%
  \BibitemOpen
  \bibfield  {author} {\bibinfo {author} {\bibfnamefont {M.~T.}\ \bibnamefont
  {Deng}}, \bibinfo {author} {\bibfnamefont {S.}~\bibnamefont {Vaitiekenas}},
  \bibinfo {author} {\bibfnamefont {E.~B.}\ \bibnamefont {Hansen}}, \bibinfo
  {author} {\bibfnamefont {J.}~\bibnamefont {Danon}}, \bibinfo {author}
  {\bibfnamefont {M.}~\bibnamefont {Leijnse}}, \bibinfo {author} {\bibfnamefont
  {K.}~\bibnamefont {Flensberg}}, \bibinfo {author} {\bibfnamefont
  {J.}~\bibnamefont {Nyg{\r a}rd}}, \bibinfo {author} {\bibfnamefont
  {P.}~\bibnamefont {Krogstrup}}, \ and\ \bibinfo {author} {\bibfnamefont
  {C.~M.}\ \bibnamefont {Marcus}},\ }\href {\doibase 10.1126/science.aaf3961}
  {\bibfield  {journal} {\bibinfo  {journal} {Science}\ }\textbf {\bibinfo
  {volume} {354}},\ \bibinfo {pages} {1557} (\bibinfo {year}
  {2016})}\BibitemShut {NoStop}%
\bibitem [{\citenamefont {Choy}\ \emph {et~al.}(2011)\citenamefont {Choy},
  \citenamefont {Edge}, \citenamefont {Akhmerov},\ and\ \citenamefont
  {Beenakker}}]{choy_2011}%
  \BibitemOpen
  \bibfield  {author} {\bibinfo {author} {\bibfnamefont {T.-P.}\ \bibnamefont
  {Choy}}, \bibinfo {author} {\bibfnamefont {J.~M.}\ \bibnamefont {Edge}},
  \bibinfo {author} {\bibfnamefont {A.~R.}\ \bibnamefont {Akhmerov}}, \ and\
  \bibinfo {author} {\bibfnamefont {C.~W.~J.}\ \bibnamefont {Beenakker}},\
  }\href {\doibase 10.1103/PhysRevB.84.195442} {\bibfield  {journal} {\bibinfo
  {journal} {Phys. Rev. B}\ }\textbf {\bibinfo {volume} {84}},\ \bibinfo
  {pages} {195442} (\bibinfo {year} {2011})}\BibitemShut {NoStop}%
\bibitem [{\citenamefont {Braunecker}\ and\ \citenamefont
  {Simon}(2013)}]{braunecker_2013}%
  \BibitemOpen
  \bibfield  {author} {\bibinfo {author} {\bibfnamefont {B.}~\bibnamefont
  {Braunecker}}\ and\ \bibinfo {author} {\bibfnamefont {P.}~\bibnamefont
  {Simon}},\ }\href {\doibase 10.1103/PhysRevLett.111.147202} {\bibfield
  {journal} {\bibinfo  {journal} {Phys. Rev. Lett.}\ }\textbf {\bibinfo
  {volume} {111}},\ \bibinfo {pages} {147202} (\bibinfo {year}
  {2013})}\BibitemShut {NoStop}%
\bibitem [{\citenamefont {Klinovaja}\ \emph {et~al.}(2013)\citenamefont
  {Klinovaja}, \citenamefont {Stano}, \citenamefont {Yazdani},\ and\
  \citenamefont {Loss}}]{klinovaja_2013}%
  \BibitemOpen
  \bibfield  {author} {\bibinfo {author} {\bibfnamefont {J.}~\bibnamefont
  {Klinovaja}}, \bibinfo {author} {\bibfnamefont {P.}~\bibnamefont {Stano}},
  \bibinfo {author} {\bibfnamefont {A.}~\bibnamefont {Yazdani}}, \ and\
  \bibinfo {author} {\bibfnamefont {D.}~\bibnamefont {Loss}},\ }\href {\doibase
  10.1103/PhysRevLett.111.186805} {\bibfield  {journal} {\bibinfo  {journal}
  {Phys. Rev. Lett.}\ }\textbf {\bibinfo {volume} {111}},\ \bibinfo {pages}
  {186805} (\bibinfo {year} {2013})}\BibitemShut {NoStop}%
\bibitem [{\citenamefont {Nadj-Perge}\ \emph {et~al.}(2013)\citenamefont
  {Nadj-Perge}, \citenamefont {Drozdov}, \citenamefont {Bernevig},\ and\
  \citenamefont {Yazdani}}]{nadj_2013}%
  \BibitemOpen
  \bibfield  {author} {\bibinfo {author} {\bibfnamefont {S.}~\bibnamefont
  {Nadj-Perge}}, \bibinfo {author} {\bibfnamefont {I.~K.}\ \bibnamefont
  {Drozdov}}, \bibinfo {author} {\bibfnamefont {B.~A.}\ \bibnamefont
  {Bernevig}}, \ and\ \bibinfo {author} {\bibfnamefont {A.}~\bibnamefont
  {Yazdani}},\ }\href {\doibase 10.1103/PhysRevB.88.020407} {\bibfield
  {journal} {\bibinfo  {journal} {Phys. Rev. B}\ }\textbf {\bibinfo {volume}
  {88}},\ \bibinfo {pages} {020407(R)} (\bibinfo {year} {2013})}\BibitemShut
  {NoStop}%
\bibitem [{\citenamefont {Vazifeh}\ and\ \citenamefont
  {Franz}(2013)}]{vazifeh_2013}%
  \BibitemOpen
  \bibfield  {author} {\bibinfo {author} {\bibfnamefont {M.~M.}\ \bibnamefont
  {Vazifeh}}\ and\ \bibinfo {author} {\bibfnamefont {M.}~\bibnamefont
  {Franz}},\ }\href {\doibase 10.1103/PhysRevLett.111.206802} {\bibfield
  {journal} {\bibinfo  {journal} {Phys. Rev. Lett.}\ }\textbf {\bibinfo
  {volume} {111}},\ \bibinfo {pages} {206802} (\bibinfo {year}
  {2013})}\BibitemShut {NoStop}%
\bibitem [{\citenamefont {Nadj-Perge}\ \emph {et~al.}(2014)\citenamefont
  {Nadj-Perge}, \citenamefont {Drozdov}, \citenamefont {Li}, \citenamefont
  {Chen}, \citenamefont {Jeon}, \citenamefont {Seo}, \citenamefont {MacDonald},
  \citenamefont {Bernevig},\ and\ \citenamefont {Yazdani}}]{nadj_2014}%
  \BibitemOpen
  \bibfield  {author} {\bibinfo {author} {\bibfnamefont {S.}~\bibnamefont
  {Nadj-Perge}}, \bibinfo {author} {\bibfnamefont {I.~K.}\ \bibnamefont
  {Drozdov}}, \bibinfo {author} {\bibfnamefont {J.}~\bibnamefont {Li}},
  \bibinfo {author} {\bibfnamefont {H.}~\bibnamefont {Chen}}, \bibinfo {author}
  {\bibfnamefont {S.}~\bibnamefont {Jeon}}, \bibinfo {author} {\bibfnamefont
  {J.}~\bibnamefont {Seo}}, \bibinfo {author} {\bibfnamefont {A.~H.}\
  \bibnamefont {MacDonald}}, \bibinfo {author} {\bibfnamefont {B.~A.}\
  \bibnamefont {Bernevig}}, \ and\ \bibinfo {author} {\bibfnamefont
  {A.}~\bibnamefont {Yazdani}},\ }\href {\doibase 10.1126/science.1259327}
  {\bibfield  {journal} {\bibinfo  {journal} {Science}\ }\textbf {\bibinfo
  {volume} {346}},\ \bibinfo {pages} {602} (\bibinfo {year}
  {2014})}\BibitemShut {NoStop}%
\bibitem [{\citenamefont {Ruby}\ \emph {et~al.}(2015)\citenamefont {Ruby},
  \citenamefont {Pientka}, \citenamefont {Peng}, \citenamefont {von Oppen},
  \citenamefont {Heinrich},\ and\ \citenamefont {Franke}}]{ruby_2015}%
  \BibitemOpen
  \bibfield  {author} {\bibinfo {author} {\bibfnamefont {M.}~\bibnamefont
  {Ruby}}, \bibinfo {author} {\bibfnamefont {F.}~\bibnamefont {Pientka}},
  \bibinfo {author} {\bibfnamefont {Y.}~\bibnamefont {Peng}}, \bibinfo {author}
  {\bibfnamefont {F.}~\bibnamefont {von Oppen}}, \bibinfo {author}
  {\bibfnamefont {B.~W.}\ \bibnamefont {Heinrich}}, \ and\ \bibinfo {author}
  {\bibfnamefont {K.~J.}\ \bibnamefont {Franke}},\ }\href {\doibase
  10.1103/PhysRevLett.115.197204} {\bibfield  {journal} {\bibinfo  {journal}
  {Phys. Rev. Lett.}\ }\textbf {\bibinfo {volume} {115}},\ \bibinfo {pages}
  {197204} (\bibinfo {year} {2015})}\BibitemShut {NoStop}%
\bibitem [{\citenamefont {Pawlak}\ \emph {et~al.}(2016)\citenamefont {Pawlak},
  \citenamefont {Kisiel}, \citenamefont {Klinovaja}, \citenamefont {Meier},
  \citenamefont {Kawai}, \citenamefont {Glatzel}, \citenamefont {Loss},\ and\
  \citenamefont {Meyer}}]{pawlak_2016}%
  \BibitemOpen
  \bibfield  {author} {\bibinfo {author} {\bibfnamefont {R.}~\bibnamefont
  {Pawlak}}, \bibinfo {author} {\bibfnamefont {M.}~\bibnamefont {Kisiel}},
  \bibinfo {author} {\bibfnamefont {J.}~\bibnamefont {Klinovaja}}, \bibinfo
  {author} {\bibfnamefont {T.}~\bibnamefont {Meier}}, \bibinfo {author}
  {\bibfnamefont {S.}~\bibnamefont {Kawai}}, \bibinfo {author} {\bibfnamefont
  {T.}~\bibnamefont {Glatzel}}, \bibinfo {author} {\bibfnamefont
  {D.}~\bibnamefont {Loss}}, \ and\ \bibinfo {author} {\bibfnamefont
  {E.}~\bibnamefont {Meyer}},\ }\href@noop {} {\bibfield  {journal} {\bibinfo
  {journal} {npj Quantum Information}\ }\textbf {\bibinfo {volume} {2}},\
  \bibinfo {pages} {1} (\bibinfo {year} {2016})}\BibitemShut {NoStop}%
\bibitem [{\citenamefont {Jeon}\ \emph {et~al.}(2017)\citenamefont {Jeon},
  \citenamefont {Xie}, \citenamefont {Li}, \citenamefont {Wang}, \citenamefont
  {Bernevig},\ and\ \citenamefont {Yazdani}}]{jeon_2017}%
  \BibitemOpen
  \bibfield  {author} {\bibinfo {author} {\bibfnamefont {S.}~\bibnamefont
  {Jeon}}, \bibinfo {author} {\bibfnamefont {Y.}~\bibnamefont {Xie}}, \bibinfo
  {author} {\bibfnamefont {J.}~\bibnamefont {Li}}, \bibinfo {author}
  {\bibfnamefont {Z.}~\bibnamefont {Wang}}, \bibinfo {author} {\bibfnamefont
  {B.~A.}\ \bibnamefont {Bernevig}}, \ and\ \bibinfo {author} {\bibfnamefont
  {A.}~\bibnamefont {Yazdani}},\ }\href {\doibase 10.1126/science.aan3670}
  {\bibfield  {journal} {\bibinfo  {journal} {Science}\ }\textbf {\bibinfo
  {volume} {358}},\ \bibinfo {pages} {772} (\bibinfo {year}
  {2017})}\BibitemShut {NoStop}%
\bibitem [{\citenamefont {Kim}\ \emph {et~al.}(2018)\citenamefont {Kim},
  \citenamefont {Palacio-Morales}, \citenamefont {Posske}, \citenamefont
  {R{\'o}zsa}, \citenamefont {Palot{\'a}s}, \citenamefont {Szunyogh},
  \citenamefont {Thorwart},\ and\ \citenamefont {Wiesendanger}}]{kim_2018}%
  \BibitemOpen
  \bibfield  {author} {\bibinfo {author} {\bibfnamefont {H.}~\bibnamefont
  {Kim}}, \bibinfo {author} {\bibfnamefont {A.}~\bibnamefont
  {Palacio-Morales}}, \bibinfo {author} {\bibfnamefont {T.}~\bibnamefont
  {Posske}}, \bibinfo {author} {\bibfnamefont {L.}~\bibnamefont {R{\'o}zsa}},
  \bibinfo {author} {\bibfnamefont {K.}~\bibnamefont {Palot{\'a}s}}, \bibinfo
  {author} {\bibfnamefont {L.}~\bibnamefont {Szunyogh}}, \bibinfo {author}
  {\bibfnamefont {M.}~\bibnamefont {Thorwart}}, \ and\ \bibinfo {author}
  {\bibfnamefont {R.}~\bibnamefont {Wiesendanger}},\ }\href {\doibase
  10.1126/sciadv.aar5251} {\bibfield  {journal} {\bibinfo  {journal} {Science
  Advances}\ }\textbf {\bibinfo {volume} {4}} (\bibinfo {year} {2018}),\
  10.1126/sciadv.aar5251}\BibitemShut {NoStop}%
\bibitem [{\citenamefont {Haldane}(1988)}]{haldane_1988}%
  \BibitemOpen
  \bibfield  {author} {\bibinfo {author} {\bibfnamefont {F.~D.~M.}\
  \bibnamefont {Haldane}},\ }\href {\doibase 10.1103/PhysRevLett.61.2015}
  {\bibfield  {journal} {\bibinfo  {journal} {Phys. Rev. Lett.}\ }\textbf
  {\bibinfo {volume} {61}},\ \bibinfo {pages} {2015} (\bibinfo {year}
  {1988})}\BibitemShut {NoStop}%
\bibitem [{\citenamefont {Yu}\ \emph {et~al.}(2010)\citenamefont {Yu},
  \citenamefont {Zhang}, \citenamefont {Zhang}, \citenamefont {Zhang},
  \citenamefont {Dai},\ and\ \citenamefont {Fang}}]{yu_2010}%
  \BibitemOpen
  \bibfield  {author} {\bibinfo {author} {\bibfnamefont {R.}~\bibnamefont
  {Yu}}, \bibinfo {author} {\bibfnamefont {W.}~\bibnamefont {Zhang}}, \bibinfo
  {author} {\bibfnamefont {H.-J.}\ \bibnamefont {Zhang}}, \bibinfo {author}
  {\bibfnamefont {S.-C.}\ \bibnamefont {Zhang}}, \bibinfo {author}
  {\bibfnamefont {X.}~\bibnamefont {Dai}}, \ and\ \bibinfo {author}
  {\bibfnamefont {Z.}~\bibnamefont {Fang}},\ }\href {\doibase
  10.1126/science.1187485} {\bibfield  {journal} {\bibinfo  {journal}
  {Science}\ }\textbf {\bibinfo {volume} {329}},\ \bibinfo {pages} {61}
  (\bibinfo {year} {2010})}\BibitemShut {NoStop}%
\bibitem [{\citenamefont {Chang}\ \emph {et~al.}(2013)\citenamefont {Chang},
  \citenamefont {Zhang}, \citenamefont {Feng}, \citenamefont {Shen},
  \citenamefont {Zhang}, \citenamefont {Guo}, \citenamefont {Li}, \citenamefont
  {Ou}, \citenamefont {Wei}, \citenamefont {Wang}, \citenamefont {Ji},
  \citenamefont {Feng}, \citenamefont {Ji}, \citenamefont {Chen}, \citenamefont
  {Jia}, \citenamefont {Dai}, \citenamefont {Fang}, \citenamefont {Zhang},
  \citenamefont {He}, \citenamefont {Wang}, \citenamefont {Lu}, \citenamefont
  {Ma},\ and\ \citenamefont {Xue}}]{chang_2013}%
  \BibitemOpen
  \bibfield  {author} {\bibinfo {author} {\bibfnamefont {C.-Z.}\ \bibnamefont
  {Chang}}, \bibinfo {author} {\bibfnamefont {J.}~\bibnamefont {Zhang}},
  \bibinfo {author} {\bibfnamefont {X.}~\bibnamefont {Feng}}, \bibinfo {author}
  {\bibfnamefont {J.}~\bibnamefont {Shen}}, \bibinfo {author} {\bibfnamefont
  {Z.}~\bibnamefont {Zhang}}, \bibinfo {author} {\bibfnamefont
  {M.}~\bibnamefont {Guo}}, \bibinfo {author} {\bibfnamefont {K.}~\bibnamefont
  {Li}}, \bibinfo {author} {\bibfnamefont {Y.}~\bibnamefont {Ou}}, \bibinfo
  {author} {\bibfnamefont {P.}~\bibnamefont {Wei}}, \bibinfo {author}
  {\bibfnamefont {L.-L.}\ \bibnamefont {Wang}}, \bibinfo {author}
  {\bibfnamefont {Z.-Q.}\ \bibnamefont {Ji}}, \bibinfo {author} {\bibfnamefont
  {Y.}~\bibnamefont {Feng}}, \bibinfo {author} {\bibfnamefont {S.}~\bibnamefont
  {Ji}}, \bibinfo {author} {\bibfnamefont {X.}~\bibnamefont {Chen}}, \bibinfo
  {author} {\bibfnamefont {J.}~\bibnamefont {Jia}}, \bibinfo {author}
  {\bibfnamefont {X.}~\bibnamefont {Dai}}, \bibinfo {author} {\bibfnamefont
  {Z.}~\bibnamefont {Fang}}, \bibinfo {author} {\bibfnamefont {S.-C.}\
  \bibnamefont {Zhang}}, \bibinfo {author} {\bibfnamefont {K.}~\bibnamefont
  {He}}, \bibinfo {author} {\bibfnamefont {Y.}~\bibnamefont {Wang}}, \bibinfo
  {author} {\bibfnamefont {L.}~\bibnamefont {Lu}}, \bibinfo {author}
  {\bibfnamefont {X.-C.}\ \bibnamefont {Ma}}, \ and\ \bibinfo {author}
  {\bibfnamefont {Q.-K.}\ \bibnamefont {Xue}},\ }\href {\doibase
  10.1126/science.1234414} {\bibfield  {journal} {\bibinfo  {journal}
  {Science}\ }\textbf {\bibinfo {volume} {340}},\ \bibinfo {pages} {167}
  (\bibinfo {year} {2013})}\BibitemShut {NoStop}%
\bibitem [{\citenamefont {Qi}\ \emph {et~al.}(2010)\citenamefont {Qi},
  \citenamefont {Hughes},\ and\ \citenamefont {Zhang}}]{qi_2010}%
  \BibitemOpen
  \bibfield  {author} {\bibinfo {author} {\bibfnamefont {X.-L.}\ \bibnamefont
  {Qi}}, \bibinfo {author} {\bibfnamefont {T.~L.}\ \bibnamefont {Hughes}}, \
  and\ \bibinfo {author} {\bibfnamefont {S.-C.}\ \bibnamefont {Zhang}},\ }\href
  {\doibase 10.1103/PhysRevB.82.184516} {\bibfield  {journal} {\bibinfo
  {journal} {Phys. Rev. B}\ }\textbf {\bibinfo {volume} {82}},\ \bibinfo
  {pages} {184516} (\bibinfo {year} {2010})}\BibitemShut {NoStop}%
\bibitem [{\citenamefont {Ii}\ \emph {et~al.}(2011)\citenamefont {Ii},
  \citenamefont {Yada}, \citenamefont {Sato},\ and\ \citenamefont
  {Tanaka}}]{ii_2011}%
  \BibitemOpen
  \bibfield  {author} {\bibinfo {author} {\bibfnamefont {A.}~\bibnamefont
  {Ii}}, \bibinfo {author} {\bibfnamefont {K.}~\bibnamefont {Yada}}, \bibinfo
  {author} {\bibfnamefont {M.}~\bibnamefont {Sato}}, \ and\ \bibinfo {author}
  {\bibfnamefont {Y.}~\bibnamefont {Tanaka}},\ }\href {\doibase
  10.1103/PhysRevB.83.224524} {\bibfield  {journal} {\bibinfo  {journal} {Phys.
  Rev. B}\ }\textbf {\bibinfo {volume} {83}},\ \bibinfo {pages} {224524}
  (\bibinfo {year} {2011})}\BibitemShut {NoStop}%
\bibitem [{\citenamefont {Ii}\ \emph {et~al.}(2012)\citenamefont {Ii},
  \citenamefont {Yamakage}, \citenamefont {Yada}, \citenamefont {Sato},\ and\
  \citenamefont {Tanaka}}]{ii_2012}%
  \BibitemOpen
  \bibfield  {author} {\bibinfo {author} {\bibfnamefont {A.}~\bibnamefont
  {Ii}}, \bibinfo {author} {\bibfnamefont {A.}~\bibnamefont {Yamakage}},
  \bibinfo {author} {\bibfnamefont {K.}~\bibnamefont {Yada}}, \bibinfo {author}
  {\bibfnamefont {M.}~\bibnamefont {Sato}}, \ and\ \bibinfo {author}
  {\bibfnamefont {Y.}~\bibnamefont {Tanaka}},\ }\href {\doibase
  10.1103/PhysRevB.86.174512} {\bibfield  {journal} {\bibinfo  {journal} {Phys.
  Rev. B}\ }\textbf {\bibinfo {volume} {86}},\ \bibinfo {pages} {174512}
  (\bibinfo {year} {2012})}\BibitemShut {NoStop}%
\bibitem [{\citenamefont {He}\ \emph {et~al.}(2014)\citenamefont {He},
  \citenamefont {Wu}, \citenamefont {Choy}, \citenamefont {Liu}, \citenamefont
  {Tanaka},\ and\ \citenamefont {Law}}]{he_2014}%
  \BibitemOpen
  \bibfield  {author} {\bibinfo {author} {\bibfnamefont {J.~J.}\ \bibnamefont
  {He}}, \bibinfo {author} {\bibfnamefont {J.}~\bibnamefont {Wu}}, \bibinfo
  {author} {\bibfnamefont {T.-P.}\ \bibnamefont {Choy}}, \bibinfo {author}
  {\bibfnamefont {X.-J.}\ \bibnamefont {Liu}}, \bibinfo {author} {\bibfnamefont
  {Y.}~\bibnamefont {Tanaka}}, \ and\ \bibinfo {author} {\bibfnamefont {K.~T.}\
  \bibnamefont {Law}},\ }\href {\doibase 10.1038/ncomms4232} {\bibfield
  {journal} {\bibinfo  {journal} {Nature Communications}\ }\textbf {\bibinfo
  {volume} {5}},\ \bibinfo {pages} {3232} (\bibinfo {year} {2014})}\BibitemShut
  {NoStop}%
\bibitem [{\citenamefont {Altland}\ and\ \citenamefont
  {Zirnbauer}(1997)}]{altland_1997}%
  \BibitemOpen
  \bibfield  {author} {\bibinfo {author} {\bibfnamefont {A.}~\bibnamefont
  {Altland}}\ and\ \bibinfo {author} {\bibfnamefont {M.~R.}\ \bibnamefont
  {Zirnbauer}},\ }\href {\doibase 10.1103/PhysRevB.55.1142} {\bibfield
  {journal} {\bibinfo  {journal} {Phys. Rev. B}\ }\textbf {\bibinfo {volume}
  {55}},\ \bibinfo {pages} {1142} (\bibinfo {year} {1997})}\BibitemShut
  {NoStop}%
\bibitem [{\citenamefont {Zirnbauer}(1996)}]{zirnbauer_1996}%
  \BibitemOpen
  \bibfield  {author} {\bibinfo {author} {\bibfnamefont {M.~R.}\ \bibnamefont
  {Zirnbauer}},\ }\href {\doibase 10.1063/1.531675} {\bibfield  {journal}
  {\bibinfo  {journal} {Journal of Mathematical Physics}\ }\textbf {\bibinfo
  {volume} {37}},\ \bibinfo {pages} {4986} (\bibinfo {year}
  {1996})}\BibitemShut {NoStop}%
\bibitem [{\citenamefont {Schnyder}\ \emph {et~al.}(2008)\citenamefont
  {Schnyder}, \citenamefont {Ryu}, \citenamefont {Furusaki},\ and\
  \citenamefont {Ludwig}}]{schnyder_2008}%
  \BibitemOpen
  \bibfield  {author} {\bibinfo {author} {\bibfnamefont {A.~P.}\ \bibnamefont
  {Schnyder}}, \bibinfo {author} {\bibfnamefont {S.}~\bibnamefont {Ryu}},
  \bibinfo {author} {\bibfnamefont {A.}~\bibnamefont {Furusaki}}, \ and\
  \bibinfo {author} {\bibfnamefont {A.~W.~W.}\ \bibnamefont {Ludwig}},\ }\href
  {\doibase 10.1103/PhysRevB.78.195125} {\bibfield  {journal} {\bibinfo
  {journal} {Phys. Rev. B}\ }\textbf {\bibinfo {volume} {78}},\ \bibinfo
  {pages} {195125} (\bibinfo {year} {2008})}\BibitemShut {NoStop}%
\bibitem [{\citenamefont {Kitaev}(2009)}]{kitaev_2009}%
  \BibitemOpen
  \bibfield  {author} {\bibinfo {author} {\bibfnamefont {A.}~\bibnamefont
  {Kitaev}},\ }\href {\doibase 10.1063/1.3149495} {\bibfield  {journal}
  {\bibinfo  {journal} {AIP Conference Proceedings}\ }\textbf {\bibinfo
  {volume} {1134}},\ \bibinfo {pages} {22} (\bibinfo {year}
  {2009})}\BibitemShut {NoStop}%
\bibitem [{\citenamefont {Schnyder}\ \emph {et~al.}(2009)\citenamefont
  {Schnyder}, \citenamefont {Ryu}, \citenamefont {Furusaki},\ and\
  \citenamefont {Ludwig}}]{schnyder_2009}%
  \BibitemOpen
  \bibfield  {author} {\bibinfo {author} {\bibfnamefont {A.~P.}\ \bibnamefont
  {Schnyder}}, \bibinfo {author} {\bibfnamefont {S.}~\bibnamefont {Ryu}},
  \bibinfo {author} {\bibfnamefont {A.}~\bibnamefont {Furusaki}}, \ and\
  \bibinfo {author} {\bibfnamefont {A.~W.~W.}\ \bibnamefont {Ludwig}},\ }\href
  {\doibase 10.1063/1.3149481} {\bibfield  {journal} {\bibinfo  {journal} {AIP
  Conference Proceedings}\ }\textbf {\bibinfo {volume} {1134}},\ \bibinfo
  {pages} {10} (\bibinfo {year} {2009})}\BibitemShut {NoStop}%
\bibitem [{\citenamefont {Ryu}\ \emph {et~al.}(2010)\citenamefont {Ryu},
  \citenamefont {Schnyder}, \citenamefont {Furusaki},\ and\ \citenamefont
  {Ludwig}}]{ryu_2010}%
  \BibitemOpen
  \bibfield  {author} {\bibinfo {author} {\bibfnamefont {S.}~\bibnamefont
  {Ryu}}, \bibinfo {author} {\bibfnamefont {A.~P.}\ \bibnamefont {Schnyder}},
  \bibinfo {author} {\bibfnamefont {A.}~\bibnamefont {Furusaki}}, \ and\
  \bibinfo {author} {\bibfnamefont {A.~W.~W.}\ \bibnamefont {Ludwig}},\ }\href
  {http://stacks.iop.org/1367-2630/12/i=6/a=065010} {\bibfield  {journal}
  {\bibinfo  {journal} {New Journal of Physics}\ }\textbf {\bibinfo {volume}
  {12}},\ \bibinfo {pages} {065010} (\bibinfo {year} {2010})}\BibitemShut
  {NoStop}%
\bibitem [{\citenamefont {He}\ \emph {et~al.}(2017)\citenamefont {He},
  \citenamefont {Pan}, \citenamefont {Stern}, \citenamefont {Burks},
  \citenamefont {Che}, \citenamefont {Yin}, \citenamefont {Wang}, \citenamefont
  {Lian}, \citenamefont {Zhou}, \citenamefont {Choi}, \citenamefont {Murata},
  \citenamefont {Kou}, \citenamefont {Chen}, \citenamefont {Nie}, \citenamefont
  {Shao}, \citenamefont {Fan}, \citenamefont {Zhang}, \citenamefont {Liu},
  \citenamefont {Xia},\ and\ \citenamefont {Wang}}]{he_2017}%
  \BibitemOpen
  \bibfield  {author} {\bibinfo {author} {\bibfnamefont {Q.~L.}\ \bibnamefont
  {He}}, \bibinfo {author} {\bibfnamefont {L.}~\bibnamefont {Pan}}, \bibinfo
  {author} {\bibfnamefont {A.~L.}\ \bibnamefont {Stern}}, \bibinfo {author}
  {\bibfnamefont {E.~C.}\ \bibnamefont {Burks}}, \bibinfo {author}
  {\bibfnamefont {X.}~\bibnamefont {Che}}, \bibinfo {author} {\bibfnamefont
  {G.}~\bibnamefont {Yin}}, \bibinfo {author} {\bibfnamefont {J.}~\bibnamefont
  {Wang}}, \bibinfo {author} {\bibfnamefont {B.}~\bibnamefont {Lian}}, \bibinfo
  {author} {\bibfnamefont {Q.}~\bibnamefont {Zhou}}, \bibinfo {author}
  {\bibfnamefont {E.~S.}\ \bibnamefont {Choi}}, \bibinfo {author}
  {\bibfnamefont {K.}~\bibnamefont {Murata}}, \bibinfo {author} {\bibfnamefont
  {X.}~\bibnamefont {Kou}}, \bibinfo {author} {\bibfnamefont {Z.}~\bibnamefont
  {Chen}}, \bibinfo {author} {\bibfnamefont {T.}~\bibnamefont {Nie}}, \bibinfo
  {author} {\bibfnamefont {Q.}~\bibnamefont {Shao}}, \bibinfo {author}
  {\bibfnamefont {Y.}~\bibnamefont {Fan}}, \bibinfo {author} {\bibfnamefont
  {S.-C.}\ \bibnamefont {Zhang}}, \bibinfo {author} {\bibfnamefont
  {K.}~\bibnamefont {Liu}}, \bibinfo {author} {\bibfnamefont {J.}~\bibnamefont
  {Xia}}, \ and\ \bibinfo {author} {\bibfnamefont {K.~L.}\ \bibnamefont
  {Wang}},\ }\href {\doibase 10.1126/science.aag2792} {\bibfield  {journal}
  {\bibinfo  {journal} {Science}\ }\textbf {\bibinfo {volume} {357}},\ \bibinfo
  {pages} {294} (\bibinfo {year} {2017})}\BibitemShut {NoStop}%
\bibitem [{\citenamefont {Chung}\ \emph {et~al.}(2011)\citenamefont {Chung},
  \citenamefont {Qi}, \citenamefont {Maciejko},\ and\ \citenamefont
  {Zhang}}]{chung_2011}%
  \BibitemOpen
  \bibfield  {author} {\bibinfo {author} {\bibfnamefont {S.~B.}\ \bibnamefont
  {Chung}}, \bibinfo {author} {\bibfnamefont {X.-L.}\ \bibnamefont {Qi}},
  \bibinfo {author} {\bibfnamefont {J.}~\bibnamefont {Maciejko}}, \ and\
  \bibinfo {author} {\bibfnamefont {S.-C.}\ \bibnamefont {Zhang}},\ }\href
  {\doibase 10.1103/PhysRevB.83.100512} {\bibfield  {journal} {\bibinfo
  {journal} {Phys. Rev. B}\ }\textbf {\bibinfo {volume} {83}},\ \bibinfo
  {pages} {100512(R)} (\bibinfo {year} {2011})}\BibitemShut {NoStop}%
\bibitem [{\citenamefont {Wang}\ \emph {et~al.}(2015)\citenamefont {Wang},
  \citenamefont {Zhou}, \citenamefont {Lian},\ and\ \citenamefont
  {Zhang}}]{wang_2015}%
  \BibitemOpen
  \bibfield  {author} {\bibinfo {author} {\bibfnamefont {J.}~\bibnamefont
  {Wang}}, \bibinfo {author} {\bibfnamefont {Q.}~\bibnamefont {Zhou}}, \bibinfo
  {author} {\bibfnamefont {B.}~\bibnamefont {Lian}}, \ and\ \bibinfo {author}
  {\bibfnamefont {S.-C.}\ \bibnamefont {Zhang}},\ }\href {\doibase
  10.1103/PhysRevB.92.064520} {\bibfield  {journal} {\bibinfo  {journal} {Phys.
  Rev. B}\ }\textbf {\bibinfo {volume} {92}},\ \bibinfo {pages} {064520}
  (\bibinfo {year} {2015})}\BibitemShut {NoStop}%
\bibitem [{\citenamefont {Lian}\ \emph {et~al.}(2018)\citenamefont {Lian},
  \citenamefont {Sun}, \citenamefont {Vaezi}, \citenamefont {Qi},\ and\
  \citenamefont {Zhang}}]{lian_2018}%
  \BibitemOpen
  \bibfield  {author} {\bibinfo {author} {\bibfnamefont {B.}~\bibnamefont
  {Lian}}, \bibinfo {author} {\bibfnamefont {X.-Q.}\ \bibnamefont {Sun}},
  \bibinfo {author} {\bibfnamefont {A.}~\bibnamefont {Vaezi}}, \bibinfo
  {author} {\bibfnamefont {X.-L.}\ \bibnamefont {Qi}}, \ and\ \bibinfo {author}
  {\bibfnamefont {S.-C.}\ \bibnamefont {Zhang}},\ }\href {\doibase
  10.1073/pnas.1810003115} {\bibfield  {journal} {\bibinfo  {journal}
  {Proceedings of the National Academy of Sciences}\ }\textbf {\bibinfo
  {volume} {115}},\ \bibinfo {pages} {10938} (\bibinfo {year}
  {2018})}\BibitemShut {NoStop}%
\bibitem [{\citenamefont {Yasuda}\ \emph {et~al.}(2017)\citenamefont {Yasuda},
  \citenamefont {Mogi}, \citenamefont {Yoshimi}, \citenamefont {Tsukazaki},
  \citenamefont {Takahashi}, \citenamefont {Kawasaki}, \citenamefont {Kagawa},\
  and\ \citenamefont {Tokura}}]{yasuda_2017}%
  \BibitemOpen
  \bibfield  {author} {\bibinfo {author} {\bibfnamefont {K.}~\bibnamefont
  {Yasuda}}, \bibinfo {author} {\bibfnamefont {M.}~\bibnamefont {Mogi}},
  \bibinfo {author} {\bibfnamefont {R.}~\bibnamefont {Yoshimi}}, \bibinfo
  {author} {\bibfnamefont {A.}~\bibnamefont {Tsukazaki}}, \bibinfo {author}
  {\bibfnamefont {K.~S.}\ \bibnamefont {Takahashi}}, \bibinfo {author}
  {\bibfnamefont {M.}~\bibnamefont {Kawasaki}}, \bibinfo {author}
  {\bibfnamefont {F.}~\bibnamefont {Kagawa}}, \ and\ \bibinfo {author}
  {\bibfnamefont {Y.}~\bibnamefont {Tokura}},\ }\href {\doibase
  10.1126/science.aan5991} {\bibfield  {journal} {\bibinfo  {journal}
  {Science}\ }\textbf {\bibinfo {volume} {358}},\ \bibinfo {pages} {1311}
  (\bibinfo {year} {2017})}\BibitemShut {NoStop}%
\bibitem [{\citenamefont {Ji}\ and\ \citenamefont {Wen}(2018)}]{ji_2018}%
  \BibitemOpen
  \bibfield  {author} {\bibinfo {author} {\bibfnamefont {W.}~\bibnamefont
  {Ji}}\ and\ \bibinfo {author} {\bibfnamefont {X.-G.}\ \bibnamefont {Wen}},\
  }\href {\doibase 10.1103/PhysRevLett.120.107002} {\bibfield  {journal}
  {\bibinfo  {journal} {Phys. Rev. Lett.}\ }\textbf {\bibinfo {volume} {120}},\
  \bibinfo {pages} {107002} (\bibinfo {year} {2018})}\BibitemShut {NoStop}%
\bibitem [{\citenamefont {Huang}\ \emph {et~al.}(2018)\citenamefont {Huang},
  \citenamefont {Setiawan},\ and\ \citenamefont {Sau}}]{huang_2018}%
  \BibitemOpen
  \bibfield  {author} {\bibinfo {author} {\bibfnamefont {Y.}~\bibnamefont
  {Huang}}, \bibinfo {author} {\bibfnamefont {F.}~\bibnamefont {Setiawan}}, \
  and\ \bibinfo {author} {\bibfnamefont {J.~D.}\ \bibnamefont {Sau}},\ }\href
  {\doibase 10.1103/PhysRevB.97.100501} {\bibfield  {journal} {\bibinfo
  {journal} {Phys. Rev. B}\ }\textbf {\bibinfo {volume} {97}},\ \bibinfo
  {pages} {100501(R)} (\bibinfo {year} {2018})}\BibitemShut {NoStop}%
\bibitem [{\citenamefont {Kayyalha}\ \emph {et~al.}(2020)\citenamefont
  {Kayyalha}, \citenamefont {Xiao}, \citenamefont {Zhang}, \citenamefont
  {Shin}, \citenamefont {Jiang}, \citenamefont {Wang}, \citenamefont {Zhao},
  \citenamefont {Xiao}, \citenamefont {Zhang}, \citenamefont {Fijalkowski},
  \citenamefont {Mandal}, \citenamefont {Winnerlein}, \citenamefont {Gould},
  \citenamefont {Li}, \citenamefont {Molenkamp}, \citenamefont {Chan},
  \citenamefont {Samarth},\ and\ \citenamefont {Chang}}]{kayyalha_2020}%
  \BibitemOpen
  \bibfield  {author} {\bibinfo {author} {\bibfnamefont {M.}~\bibnamefont
  {Kayyalha}}, \bibinfo {author} {\bibfnamefont {D.}~\bibnamefont {Xiao}},
  \bibinfo {author} {\bibfnamefont {R.}~\bibnamefont {Zhang}}, \bibinfo
  {author} {\bibfnamefont {J.}~\bibnamefont {Shin}}, \bibinfo {author}
  {\bibfnamefont {J.}~\bibnamefont {Jiang}}, \bibinfo {author} {\bibfnamefont
  {F.}~\bibnamefont {Wang}}, \bibinfo {author} {\bibfnamefont {Y.-F.}\
  \bibnamefont {Zhao}}, \bibinfo {author} {\bibfnamefont {R.}~\bibnamefont
  {Xiao}}, \bibinfo {author} {\bibfnamefont {L.}~\bibnamefont {Zhang}},
  \bibinfo {author} {\bibfnamefont {K.~M.}\ \bibnamefont {Fijalkowski}},
  \bibinfo {author} {\bibfnamefont {P.}~\bibnamefont {Mandal}}, \bibinfo
  {author} {\bibfnamefont {M.}~\bibnamefont {Winnerlein}}, \bibinfo {author}
  {\bibfnamefont {C.}~\bibnamefont {Gould}}, \bibinfo {author} {\bibfnamefont
  {Q.}~\bibnamefont {Li}}, \bibinfo {author} {\bibfnamefont {L.~W.}\
  \bibnamefont {Molenkamp}}, \bibinfo {author} {\bibfnamefont {M.~H.~W.}\
  \bibnamefont {Chan}}, \bibinfo {author} {\bibfnamefont {N.}~\bibnamefont
  {Samarth}}, \ and\ \bibinfo {author} {\bibfnamefont {C.-Z.}\ \bibnamefont
  {Chang}},\ }\href {\doibase 10.1126/science.aax6361} {\bibfield  {journal}
  {\bibinfo  {journal} {Science}\ }\textbf {\bibinfo {volume} {367}},\ \bibinfo
  {pages} {64} (\bibinfo {year} {2020})}\BibitemShut {NoStop}%
\bibitem [{\citenamefont {Shen}\ \emph {et~al.}(2020)\citenamefont {Shen},
  \citenamefont {Lyu}, \citenamefont {Gao}, \citenamefont {Xie}, \citenamefont
  {Chen}, \citenamefont {Cho}, \citenamefont {Atanov}, \citenamefont {Chen},
  \citenamefont {Liu}, \citenamefont {Hu}, \citenamefont {Yip}, \citenamefont
  {Goh}, \citenamefont {He}, \citenamefont {Pan}, \citenamefont {Wang},
  \citenamefont {Law},\ and\ \citenamefont {Lortz}}]{shen_2020}%
  \BibitemOpen
  \bibfield  {author} {\bibinfo {author} {\bibfnamefont {J.}~\bibnamefont
  {Shen}}, \bibinfo {author} {\bibfnamefont {J.}~\bibnamefont {Lyu}}, \bibinfo
  {author} {\bibfnamefont {J.~Z.}\ \bibnamefont {Gao}}, \bibinfo {author}
  {\bibfnamefont {Y.-M.}\ \bibnamefont {Xie}}, \bibinfo {author} {\bibfnamefont
  {C.-Z.}\ \bibnamefont {Chen}}, \bibinfo {author} {\bibfnamefont {C.-w.}\
  \bibnamefont {Cho}}, \bibinfo {author} {\bibfnamefont {O.}~\bibnamefont
  {Atanov}}, \bibinfo {author} {\bibfnamefont {Z.}~\bibnamefont {Chen}},
  \bibinfo {author} {\bibfnamefont {K.}~\bibnamefont {Liu}}, \bibinfo {author}
  {\bibfnamefont {Y.~J.}\ \bibnamefont {Hu}}, \bibinfo {author} {\bibfnamefont
  {K.~Y.}\ \bibnamefont {Yip}}, \bibinfo {author} {\bibfnamefont {S.~K.}\
  \bibnamefont {Goh}}, \bibinfo {author} {\bibfnamefont {Q.~L.}\ \bibnamefont
  {He}}, \bibinfo {author} {\bibfnamefont {L.}~\bibnamefont {Pan}}, \bibinfo
  {author} {\bibfnamefont {K.~L.}\ \bibnamefont {Wang}}, \bibinfo {author}
  {\bibfnamefont {K.~T.}\ \bibnamefont {Law}}, \ and\ \bibinfo {author}
  {\bibfnamefont {R.}~\bibnamefont {Lortz}},\ }\href {\doibase
  10.1073/pnas.1910967117} {\bibfield  {journal} {\bibinfo  {journal}
  {Proceedings of the National Academy of Sciences}\ }\textbf {\bibinfo
  {volume} {117}},\ \bibinfo {pages} {238} (\bibinfo {year}
  {2020})}\BibitemShut {NoStop}%
\bibitem [{\citenamefont {Lado}\ and\ \citenamefont
  {Sigrist}(2018)}]{lado_2018}%
  \BibitemOpen
  \bibfield  {author} {\bibinfo {author} {\bibfnamefont {J.~L.}\ \bibnamefont
  {Lado}}\ and\ \bibinfo {author} {\bibfnamefont {M.}~\bibnamefont {Sigrist}},\
  }\href {\doibase 10.1103/PhysRevLett.121.037002} {\bibfield  {journal}
  {\bibinfo  {journal} {Phys. Rev. Lett.}\ }\textbf {\bibinfo {volume} {121}},\
  \bibinfo {pages} {037002} (\bibinfo {year} {2018})}\BibitemShut {NoStop}%
\bibitem [{\citenamefont {Wang}\ and\ \citenamefont {Lian}(2018)}]{wang_2018}%
  \BibitemOpen
  \bibfield  {author} {\bibinfo {author} {\bibfnamefont {J.}~\bibnamefont
  {Wang}}\ and\ \bibinfo {author} {\bibfnamefont {B.}~\bibnamefont {Lian}},\
  }\href {\doibase 10.1103/PhysRevLett.121.256801} {\bibfield  {journal}
  {\bibinfo  {journal} {Phys. Rev. Lett.}\ }\textbf {\bibinfo {volume} {121}},\
  \bibinfo {pages} {256801} (\bibinfo {year} {2018})}\BibitemShut {NoStop}%
\bibitem [{\citenamefont {He}\ \emph {et~al.}(2019)\citenamefont {He},
  \citenamefont {Liang}, \citenamefont {Tanaka},\ and\ \citenamefont
  {Nagaosa}}]{he_2019}%
  \BibitemOpen
  \bibfield  {author} {\bibinfo {author} {\bibfnamefont {J.~J.}\ \bibnamefont
  {He}}, \bibinfo {author} {\bibfnamefont {T.}~\bibnamefont {Liang}}, \bibinfo
  {author} {\bibfnamefont {Y.}~\bibnamefont {Tanaka}}, \ and\ \bibinfo {author}
  {\bibfnamefont {N.}~\bibnamefont {Nagaosa}},\ }\href@noop {} {\bibfield
  {journal} {\bibinfo  {journal} {Communications Physics}\ }\textbf {\bibinfo
  {volume} {2}},\ \bibinfo {pages} {1} (\bibinfo {year} {2019})}\BibitemShut
  {NoStop}%
\bibitem [{\citenamefont {Fu}\ and\ \citenamefont {Berg}(2010)}]{Fu2010}%
  \BibitemOpen
  \bibfield  {author} {\bibinfo {author} {\bibfnamefont {L.}~\bibnamefont
  {Fu}}\ and\ \bibinfo {author} {\bibfnamefont {E.}~\bibnamefont {Berg}},\
  }\href {\doibase 10.1103/PhysRevLett.105.097001} {\bibfield  {journal}
  {\bibinfo  {journal} {Phys. Rev. Lett.}\ }\textbf {\bibinfo {volume} {105}},\
  \bibinfo {pages} {097001} (\bibinfo {year} {2010})}\BibitemShut {NoStop}%
\bibitem [{\citenamefont {Sato}(2010)}]{Sato2010}%
  \BibitemOpen
  \bibfield  {author} {\bibinfo {author} {\bibfnamefont {M.}~\bibnamefont
  {Sato}},\ }\href {\doibase 10.1103/PhysRevB.81.220504} {\bibfield  {journal}
  {\bibinfo  {journal} {Phys. Rev. B}\ }\textbf {\bibinfo {volume} {81}},\
  \bibinfo {pages} {220504(R)} (\bibinfo {year} {2010})}\BibitemShut {NoStop}%
\bibitem [{\citenamefont {Yamakage}\ \emph {et~al.}(2012)\citenamefont
  {Yamakage}, \citenamefont {Yada}, \citenamefont {Sato},\ and\ \citenamefont
  {Tanaka}}]{Yamakage_2012}%
  \BibitemOpen
  \bibfield  {author} {\bibinfo {author} {\bibfnamefont {A.}~\bibnamefont
  {Yamakage}}, \bibinfo {author} {\bibfnamefont {K.}~\bibnamefont {Yada}},
  \bibinfo {author} {\bibfnamefont {M.}~\bibnamefont {Sato}}, \ and\ \bibinfo
  {author} {\bibfnamefont {Y.}~\bibnamefont {Tanaka}},\ }\href {\doibase
  10.1103/PhysRevB.85.180509} {\bibfield  {journal} {\bibinfo  {journal} {Phys.
  Rev. B}\ }\textbf {\bibinfo {volume} {85}},\ \bibinfo {pages} {180509(R)}
  (\bibinfo {year} {2012})}\BibitemShut {NoStop}%
\bibitem [{\citenamefont {Hashimoto}\ \emph {et~al.}(2015)\citenamefont
  {Hashimoto}, \citenamefont {Yada}, \citenamefont {Sato},\ and\ \citenamefont
  {Tanaka}}]{Hashimoto_2015}%
  \BibitemOpen
  \bibfield  {author} {\bibinfo {author} {\bibfnamefont {T.}~\bibnamefont
  {Hashimoto}}, \bibinfo {author} {\bibfnamefont {K.}~\bibnamefont {Yada}},
  \bibinfo {author} {\bibfnamefont {M.}~\bibnamefont {Sato}}, \ and\ \bibinfo
  {author} {\bibfnamefont {Y.}~\bibnamefont {Tanaka}},\ }\href {\doibase
  10.1103/PhysRevB.92.174527} {\bibfield  {journal} {\bibinfo  {journal} {Phys.
  Rev. B}\ }\textbf {\bibinfo {volume} {92}},\ \bibinfo {pages} {174527}
  (\bibinfo {year} {2015})}\BibitemShut {NoStop}%
\bibitem [{\citenamefont {Lu}\ \emph {et~al.}(2015)\citenamefont {Lu},
  \citenamefont {Yada}, \citenamefont {Sato},\ and\ \citenamefont
  {Tanaka}}]{Lu_2015}%
  \BibitemOpen
  \bibfield  {author} {\bibinfo {author} {\bibfnamefont {B.}~\bibnamefont
  {Lu}}, \bibinfo {author} {\bibfnamefont {K.}~\bibnamefont {Yada}}, \bibinfo
  {author} {\bibfnamefont {M.}~\bibnamefont {Sato}}, \ and\ \bibinfo {author}
  {\bibfnamefont {Y.}~\bibnamefont {Tanaka}},\ }\href {\doibase
  10.1103/PhysRevLett.114.096804} {\bibfield  {journal} {\bibinfo  {journal}
  {Phys. Rev. Lett.}\ }\textbf {\bibinfo {volume} {114}},\ \bibinfo {pages}
  {096804} (\bibinfo {year} {2015})}\BibitemShut {NoStop}%
\bibitem [{\citenamefont {Kobayashi}\ and\ \citenamefont
  {Sato}(2015)}]{Kobayashi2015}%
  \BibitemOpen
  \bibfield  {author} {\bibinfo {author} {\bibfnamefont {S.}~\bibnamefont
  {Kobayashi}}\ and\ \bibinfo {author} {\bibfnamefont {M.}~\bibnamefont
  {Sato}},\ }\href {\doibase 10.1103/PhysRevLett.115.187001} {\bibfield
  {journal} {\bibinfo  {journal} {Phys. Rev. Lett.}\ }\textbf {\bibinfo
  {volume} {115}},\ \bibinfo {pages} {187001} (\bibinfo {year}
  {2015})}\BibitemShut {NoStop}%
\bibitem [{\citenamefont {Hashimoto}\ \emph {et~al.}(2016)\citenamefont
  {Hashimoto}, \citenamefont {Kobayashi}, \citenamefont {Tanaka},\ and\
  \citenamefont {Sato}}]{Hashimoto_2016}%
  \BibitemOpen
  \bibfield  {author} {\bibinfo {author} {\bibfnamefont {T.}~\bibnamefont
  {Hashimoto}}, \bibinfo {author} {\bibfnamefont {S.}~\bibnamefont
  {Kobayashi}}, \bibinfo {author} {\bibfnamefont {Y.}~\bibnamefont {Tanaka}}, \
  and\ \bibinfo {author} {\bibfnamefont {M.}~\bibnamefont {Sato}},\ }\href
  {\doibase 10.1103/PhysRevB.94.014510} {\bibfield  {journal} {\bibinfo
  {journal} {Phys. Rev. B}\ }\textbf {\bibinfo {volume} {94}},\ \bibinfo
  {pages} {014510} (\bibinfo {year} {2016})}\BibitemShut {NoStop}%
\bibitem [{\citenamefont {Oudah}\ \emph {et~al.}(2016)\citenamefont {Oudah},
  \citenamefont {Ikeda}, \citenamefont {Hausmann}, \citenamefont {Yonezawa},
  \citenamefont {Fukumoto}, \citenamefont {Kobayashi}, \citenamefont {Sato},\
  and\ \citenamefont {Maeno}}]{Oudah2016}%
  \BibitemOpen
  \bibfield  {author} {\bibinfo {author} {\bibfnamefont {M.}~\bibnamefont
  {Oudah}}, \bibinfo {author} {\bibfnamefont {A.}~\bibnamefont {Ikeda}},
  \bibinfo {author} {\bibfnamefont {J.~N.}\ \bibnamefont {Hausmann}}, \bibinfo
  {author} {\bibfnamefont {S.}~\bibnamefont {Yonezawa}}, \bibinfo {author}
  {\bibfnamefont {T.}~\bibnamefont {Fukumoto}}, \bibinfo {author}
  {\bibfnamefont {S.}~\bibnamefont {Kobayashi}}, \bibinfo {author}
  {\bibfnamefont {M.}~\bibnamefont {Sato}}, \ and\ \bibinfo {author}
  {\bibfnamefont {Y.}~\bibnamefont {Maeno}},\ }\href@noop {} {\bibfield
  {journal} {\bibinfo  {journal} {Nature communications}\ }\textbf {\bibinfo
  {volume} {7}},\ \bibinfo {pages} {1} (\bibinfo {year} {2016})}\BibitemShut
  {NoStop}%
\bibitem [{\citenamefont {Kawakami}\ \emph {et~al.}(2018)\citenamefont
  {Kawakami}, \citenamefont {Okamura}, \citenamefont {Kobayashi},\ and\
  \citenamefont {Sato}}]{Kawakami2018}%
  \BibitemOpen
  \bibfield  {author} {\bibinfo {author} {\bibfnamefont {T.}~\bibnamefont
  {Kawakami}}, \bibinfo {author} {\bibfnamefont {T.}~\bibnamefont {Okamura}},
  \bibinfo {author} {\bibfnamefont {S.}~\bibnamefont {Kobayashi}}, \ and\
  \bibinfo {author} {\bibfnamefont {M.}~\bibnamefont {Sato}},\ }\href {\doibase
  10.1103/PhysRevX.8.041026} {\bibfield  {journal} {\bibinfo  {journal} {Phys.
  Rev. X}\ }\textbf {\bibinfo {volume} {8}},\ \bibinfo {pages} {041026}
  (\bibinfo {year} {2018})}\BibitemShut {NoStop}%
\bibitem [{\citenamefont {Kawakami}\ and\ \citenamefont
  {Sato}(2019)}]{Kawakami2019}%
  \BibitemOpen
  \bibfield  {author} {\bibinfo {author} {\bibfnamefont {T.}~\bibnamefont
  {Kawakami}}\ and\ \bibinfo {author} {\bibfnamefont {M.}~\bibnamefont
  {Sato}},\ }\href {\doibase 10.1103/PhysRevB.100.094520} {\bibfield  {journal}
  {\bibinfo  {journal} {Phys. Rev. B}\ }\textbf {\bibinfo {volume} {100}},\
  \bibinfo {pages} {094520} (\bibinfo {year} {2019})}\BibitemShut {NoStop}%
\bibitem [{\citenamefont {Bednorz}\ and\ \citenamefont
  {M{\"u}ller}(1986)}]{bednorz_1986}%
  \BibitemOpen
  \bibfield  {author} {\bibinfo {author} {\bibfnamefont {J.~G.}\ \bibnamefont
  {Bednorz}}\ and\ \bibinfo {author} {\bibfnamefont {K.~A.}\ \bibnamefont
  {M{\"u}ller}},\ }\href@noop {} {\bibfield  {journal} {\bibinfo  {journal} {Z.
  Phys. B}\ }\textbf {\bibinfo {volume} {64}},\ \bibinfo {pages} {189}
  (\bibinfo {year} {1986})}\BibitemShut {NoStop}%
\bibitem [{\citenamefont {Scalapino}(1995)}]{scalapino_1995}%
  \BibitemOpen
  \bibfield  {author} {\bibinfo {author} {\bibfnamefont {D.}~\bibnamefont
  {Scalapino}},\ }\href {\doibase https://doi.org/10.1016/0370-1573(94)00086-I}
  {\bibfield  {journal} {\bibinfo  {journal} {Physics Reports}\ }\textbf
  {\bibinfo {volume} {250}},\ \bibinfo {pages} {329 } (\bibinfo {year}
  {1995})}\BibitemShut {NoStop}%
\bibitem [{\citenamefont {Tsuei}\ and\ \citenamefont
  {Kirtley}(2000)}]{tsuei_2000}%
  \BibitemOpen
  \bibfield  {author} {\bibinfo {author} {\bibfnamefont {C.~C.}\ \bibnamefont
  {Tsuei}}\ and\ \bibinfo {author} {\bibfnamefont {J.~R.}\ \bibnamefont
  {Kirtley}},\ }\href {\doibase 10.1103/RevModPhys.72.969} {\bibfield
  {journal} {\bibinfo  {journal} {Rev. Mod. Phys.}\ }\textbf {\bibinfo {volume}
  {72}},\ \bibinfo {pages} {969} (\bibinfo {year} {2000})}\BibitemShut
  {NoStop}%
\bibitem [{\citenamefont {Maeno}\ \emph {et~al.}(1994)\citenamefont {Maeno},
  \citenamefont {Hashimoto}, \citenamefont {Yoshida}, \citenamefont
  {Nishizaki}, \citenamefont {Fujita}, \citenamefont {Bednorz},\ and\
  \citenamefont {Lichtenberg}}]{maeno_1994}%
  \BibitemOpen
  \bibfield  {author} {\bibinfo {author} {\bibfnamefont {Y.}~\bibnamefont
  {Maeno}}, \bibinfo {author} {\bibfnamefont {H.}~\bibnamefont {Hashimoto}},
  \bibinfo {author} {\bibfnamefont {K.}~\bibnamefont {Yoshida}}, \bibinfo
  {author} {\bibfnamefont {S.}~\bibnamefont {Nishizaki}}, \bibinfo {author}
  {\bibfnamefont {T.}~\bibnamefont {Fujita}}, \bibinfo {author} {\bibfnamefont
  {J.}~\bibnamefont {Bednorz}}, \ and\ \bibinfo {author} {\bibfnamefont
  {F.}~\bibnamefont {Lichtenberg}},\ }\href@noop {} {\bibfield  {journal}
  {\bibinfo  {journal} {Nature}\ }\textbf {\bibinfo {volume} {372}},\ \bibinfo
  {pages} {532} (\bibinfo {year} {1994})}\BibitemShut {NoStop}%
\bibitem [{\citenamefont {Rice}\ and\ \citenamefont
  {Sigrist}(1995)}]{rice_1995}%
  \BibitemOpen
  \bibfield  {author} {\bibinfo {author} {\bibfnamefont {T.~M.}\ \bibnamefont
  {Rice}}\ and\ \bibinfo {author} {\bibfnamefont {M.}~\bibnamefont {Sigrist}},\
  }\href {\doibase 10.1088/0953-8984/7/47/002} {\bibfield  {journal} {\bibinfo
  {journal} {Journal of Physics: Condensed Matter}\ }\textbf {\bibinfo {volume}
  {7}},\ \bibinfo {pages} {L643} (\bibinfo {year} {1995})}\BibitemShut
  {NoStop}%
\bibitem [{\citenamefont {Mackenzie}\ and\ \citenamefont
  {Maeno}(2003)}]{mackenzie_2003}%
  \BibitemOpen
  \bibfield  {author} {\bibinfo {author} {\bibfnamefont {A.~P.}\ \bibnamefont
  {Mackenzie}}\ and\ \bibinfo {author} {\bibfnamefont {Y.}~\bibnamefont
  {Maeno}},\ }\href {\doibase 10.1103/RevModPhys.75.657} {\bibfield  {journal}
  {\bibinfo  {journal} {Rev. Mod. Phys.}\ }\textbf {\bibinfo {volume} {75}},\
  \bibinfo {pages} {657} (\bibinfo {year} {2003})}\BibitemShut {NoStop}%
\bibitem [{\citenamefont {Maeno}\ \emph {et~al.}(2012)\citenamefont {Maeno},
  \citenamefont {Kittaka}, \citenamefont {Nomura}, \citenamefont {Yonezawa},\
  and\ \citenamefont {Ishida}}]{maeno_2012}%
  \BibitemOpen
  \bibfield  {author} {\bibinfo {author} {\bibfnamefont {Y.}~\bibnamefont
  {Maeno}}, \bibinfo {author} {\bibfnamefont {S.}~\bibnamefont {Kittaka}},
  \bibinfo {author} {\bibfnamefont {T.}~\bibnamefont {Nomura}}, \bibinfo
  {author} {\bibfnamefont {S.}~\bibnamefont {Yonezawa}}, \ and\ \bibinfo
  {author} {\bibfnamefont {K.}~\bibnamefont {Ishida}},\ }\href {\doibase
  10.1143/JPSJ.81.011009} {\bibfield  {journal} {\bibinfo  {journal} {Journal
  of the Physical Society of Japan}\ }\textbf {\bibinfo {volume} {81}},\
  \bibinfo {pages} {011009} (\bibinfo {year} {2012})}\BibitemShut {NoStop}%
\bibitem [{\citenamefont {Pustogow}\ \emph {et~al.}(2019)\citenamefont
  {Pustogow}, \citenamefont {Luo}, \citenamefont {Chronister}, \citenamefont
  {Su}, \citenamefont {Sokolov}, \citenamefont {Jerzembeck}, \citenamefont
  {Mackenzie}, \citenamefont {Hicks}, \citenamefont {Kikugawa}, \citenamefont
  {Raghu}, \citenamefont {Bauer},\ and\ \citenamefont {Brown}}]{pustogow_2019}%
  \BibitemOpen
  \bibfield  {author} {\bibinfo {author} {\bibfnamefont {A.}~\bibnamefont
  {Pustogow}}, \bibinfo {author} {\bibfnamefont {Y.}~\bibnamefont {Luo}},
  \bibinfo {author} {\bibfnamefont {A.}~\bibnamefont {Chronister}}, \bibinfo
  {author} {\bibfnamefont {Y.-S.}\ \bibnamefont {Su}}, \bibinfo {author}
  {\bibfnamefont {D.}~\bibnamefont {Sokolov}}, \bibinfo {author} {\bibfnamefont
  {F.}~\bibnamefont {Jerzembeck}}, \bibinfo {author} {\bibfnamefont {A.~P.}\
  \bibnamefont {Mackenzie}}, \bibinfo {author} {\bibfnamefont {C.~W.}\
  \bibnamefont {Hicks}}, \bibinfo {author} {\bibfnamefont {N.}~\bibnamefont
  {Kikugawa}}, \bibinfo {author} {\bibfnamefont {S.}~\bibnamefont {Raghu}},
  \bibinfo {author} {\bibfnamefont {E.~D.}\ \bibnamefont {Bauer}}, \ and\
  \bibinfo {author} {\bibfnamefont {S.~E.}\ \bibnamefont {Brown}},\ }\href@noop
  {} {\bibfield  {journal} {\bibinfo  {journal} {Nature}\ }\textbf {\bibinfo
  {volume} {574}},\ \bibinfo {pages} {72} (\bibinfo {year} {2019})}\BibitemShut
  {NoStop}%
\bibitem [{\citenamefont {Suh}\ \emph {et~al.}(2020)\citenamefont {Suh},
  \citenamefont {Menke}, \citenamefont {Brydon}, \citenamefont {Timm},
  \citenamefont {Ramires},\ and\ \citenamefont {Agterberg}}]{Suh2020}%
  \BibitemOpen
  \bibfield  {author} {\bibinfo {author} {\bibfnamefont {H.~G.}\ \bibnamefont
  {Suh}}, \bibinfo {author} {\bibfnamefont {H.}~\bibnamefont {Menke}}, \bibinfo
  {author} {\bibfnamefont {P.~M.~R.}\ \bibnamefont {Brydon}}, \bibinfo {author}
  {\bibfnamefont {C.}~\bibnamefont {Timm}}, \bibinfo {author} {\bibfnamefont
  {A.}~\bibnamefont {Ramires}}, \ and\ \bibinfo {author} {\bibfnamefont
  {D.~F.}\ \bibnamefont {Agterberg}},\ }\href {\doibase
  10.1103/PhysRevResearch.2.032023} {\bibfield  {journal} {\bibinfo  {journal}
  {Phys. Rev. Research}\ }\textbf {\bibinfo {volume} {2}},\ \bibinfo {pages}
  {032023(R)} (\bibinfo {year} {2020})}\BibitemShut {NoStop}%
\bibitem [{\citenamefont {Gingras}\ \emph {et~al.}(2019)\citenamefont
  {Gingras}, \citenamefont {Nourafkan}, \citenamefont {Tremblay},\ and\
  \citenamefont {C\^ot\'e}}]{Gingras2019}%
  \BibitemOpen
  \bibfield  {author} {\bibinfo {author} {\bibfnamefont {O.}~\bibnamefont
  {Gingras}}, \bibinfo {author} {\bibfnamefont {R.}~\bibnamefont {Nourafkan}},
  \bibinfo {author} {\bibfnamefont {A.~M.~S.}\ \bibnamefont {Tremblay}}, \ and\
  \bibinfo {author} {\bibfnamefont {M.}~\bibnamefont {C\^ot\'e}},\ }\href
  {\doibase 10.1103/PhysRevLett.123.217005} {\bibfield  {journal} {\bibinfo
  {journal} {Phys. Rev. Lett.}\ }\textbf {\bibinfo {volume} {123}},\ \bibinfo
  {pages} {217005} (\bibinfo {year} {2019})}\BibitemShut {NoStop}%
\bibitem [{\citenamefont {R\o{}ising}\ \emph {et~al.}(2019)\citenamefont
  {R\o{}ising}, \citenamefont {Scaffidi}, \citenamefont {Flicker},
  \citenamefont {Lange},\ and\ \citenamefont {Simon}}]{Roising2019}%
  \BibitemOpen
  \bibfield  {author} {\bibinfo {author} {\bibfnamefont {H.~S.}\ \bibnamefont
  {R\o{}ising}}, \bibinfo {author} {\bibfnamefont {T.}~\bibnamefont
  {Scaffidi}}, \bibinfo {author} {\bibfnamefont {F.}~\bibnamefont {Flicker}},
  \bibinfo {author} {\bibfnamefont {G.~F.}\ \bibnamefont {Lange}}, \ and\
  \bibinfo {author} {\bibfnamefont {S.~H.}\ \bibnamefont {Simon}},\ }\href
  {\doibase 10.1103/PhysRevResearch.1.033108} {\bibfield  {journal} {\bibinfo
  {journal} {Phys. Rev. Research}\ }\textbf {\bibinfo {volume} {1}},\ \bibinfo
  {pages} {033108} (\bibinfo {year} {2019})}\BibitemShut {NoStop}%
\bibitem [{\citenamefont {Ghosh}\ \emph {et~al.}(2021)\citenamefont {Ghosh},
  \citenamefont {Shekhter}, \citenamefont {Jerzembeck}, \citenamefont
  {Kikugawa}, \citenamefont {Sokolov}, \citenamefont {Brando}, \citenamefont
  {Mackenzie}, \citenamefont {Hicks},\ and\ \citenamefont
  {Ramshaw}}]{Ghosh2021}%
  \BibitemOpen
  \bibfield  {author} {\bibinfo {author} {\bibfnamefont {S.}~\bibnamefont
  {Ghosh}}, \bibinfo {author} {\bibfnamefont {A.}~\bibnamefont {Shekhter}},
  \bibinfo {author} {\bibfnamefont {F.}~\bibnamefont {Jerzembeck}}, \bibinfo
  {author} {\bibfnamefont {N.}~\bibnamefont {Kikugawa}}, \bibinfo {author}
  {\bibfnamefont {D.~A.}\ \bibnamefont {Sokolov}}, \bibinfo {author}
  {\bibfnamefont {M.}~\bibnamefont {Brando}}, \bibinfo {author} {\bibfnamefont
  {A.}~\bibnamefont {Mackenzie}}, \bibinfo {author} {\bibfnamefont {C.~W.}\
  \bibnamefont {Hicks}}, \ and\ \bibinfo {author} {\bibfnamefont
  {B.}~\bibnamefont {Ramshaw}},\ }\href@noop {} {\bibfield  {journal} {\bibinfo
   {journal} {Nature Physics}\ }\textbf {\bibinfo {volume} {17}},\ \bibinfo
  {pages} {199} (\bibinfo {year} {2021})}\BibitemShut {NoStop}%
\bibitem [{\citenamefont {Grinenko}\ \emph {et~al.}(2021)\citenamefont
  {Grinenko}, \citenamefont {Ghosh}, \citenamefont {Sarkar}, \citenamefont
  {Orain}, \citenamefont {Nikitin}, \citenamefont {Elender}, \citenamefont
  {Das}, \citenamefont {Guguchia}, \citenamefont {Br{\"u}ckner}, \citenamefont
  {Barber}, \citenamefont {Park}, \citenamefont {Kikugawa}, \citenamefont
  {Sokolov}, \citenamefont {Bobowski}, \citenamefont {Miyoshi}, \citenamefont
  {Maeno}, \citenamefont {Mackenzie}, \citenamefont {Luetkens}, \citenamefont
  {Hicks},\ and\ \citenamefont {Klauss}}]{Grinenko2021}%
  \BibitemOpen
  \bibfield  {author} {\bibinfo {author} {\bibfnamefont {V.}~\bibnamefont
  {Grinenko}}, \bibinfo {author} {\bibfnamefont {S.}~\bibnamefont {Ghosh}},
  \bibinfo {author} {\bibfnamefont {R.}~\bibnamefont {Sarkar}}, \bibinfo
  {author} {\bibfnamefont {J.-C.}\ \bibnamefont {Orain}}, \bibinfo {author}
  {\bibfnamefont {A.}~\bibnamefont {Nikitin}}, \bibinfo {author} {\bibfnamefont
  {M.}~\bibnamefont {Elender}}, \bibinfo {author} {\bibfnamefont
  {D.}~\bibnamefont {Das}}, \bibinfo {author} {\bibfnamefont {Z.}~\bibnamefont
  {Guguchia}}, \bibinfo {author} {\bibfnamefont {F.}~\bibnamefont
  {Br{\"u}ckner}}, \bibinfo {author} {\bibfnamefont {M.~E.}\ \bibnamefont
  {Barber}}, \bibinfo {author} {\bibfnamefont {J.}~\bibnamefont {Park}},
  \bibinfo {author} {\bibfnamefont {N.}~\bibnamefont {Kikugawa}}, \bibinfo
  {author} {\bibfnamefont {D.~A.}\ \bibnamefont {Sokolov}}, \bibinfo {author}
  {\bibfnamefont {J.~S.}\ \bibnamefont {Bobowski}}, \bibinfo {author}
  {\bibfnamefont {T.}~\bibnamefont {Miyoshi}}, \bibinfo {author} {\bibfnamefont
  {Y.}~\bibnamefont {Maeno}}, \bibinfo {author} {\bibfnamefont {A.~P.}\
  \bibnamefont {Mackenzie}}, \bibinfo {author} {\bibfnamefont {H.}~\bibnamefont
  {Luetkens}}, \bibinfo {author} {\bibfnamefont {C.~W.}\ \bibnamefont {Hicks}},
  \ and\ \bibinfo {author} {\bibfnamefont {H.-H.}\ \bibnamefont {Klauss}},\
  }\href@noop {} {\bibfield  {journal} {\bibinfo  {journal} {Nature Physics}\
  }\textbf {\bibinfo {volume} {17}},\ \bibinfo {pages} {748} (\bibinfo {year}
  {2021})}\BibitemShut {NoStop}%
\bibitem [{\citenamefont {Benhabib}\ \emph {et~al.}(2021)\citenamefont
  {Benhabib}, \citenamefont {Lupien}, \citenamefont {Paul}, \citenamefont
  {Berges}, \citenamefont {Dion}, \citenamefont {Nardone}, \citenamefont
  {Zitouni}, \citenamefont {Mao}, \citenamefont {Maeno}, \citenamefont
  {Georges}, \citenamefont {Taillefer},\ and\ \citenamefont
  {Proust}}]{Benhabib2021}%
  \BibitemOpen
  \bibfield  {author} {\bibinfo {author} {\bibfnamefont {S.}~\bibnamefont
  {Benhabib}}, \bibinfo {author} {\bibfnamefont {C.}~\bibnamefont {Lupien}},
  \bibinfo {author} {\bibfnamefont {I.}~\bibnamefont {Paul}}, \bibinfo {author}
  {\bibfnamefont {L.}~\bibnamefont {Berges}}, \bibinfo {author} {\bibfnamefont
  {M.}~\bibnamefont {Dion}}, \bibinfo {author} {\bibfnamefont {M.}~\bibnamefont
  {Nardone}}, \bibinfo {author} {\bibfnamefont {A.}~\bibnamefont {Zitouni}},
  \bibinfo {author} {\bibfnamefont {Z.}~\bibnamefont {Mao}}, \bibinfo {author}
  {\bibfnamefont {Y.}~\bibnamefont {Maeno}}, \bibinfo {author} {\bibfnamefont
  {A.}~\bibnamefont {Georges}}, \bibinfo {author} {\bibfnamefont
  {L.}~\bibnamefont {Taillefer}}, \ and\ \bibinfo {author} {\bibfnamefont
  {C.}~\bibnamefont {Proust}},\ }\href@noop {} {\bibfield  {journal} {\bibinfo
  {journal} {Nature physics}\ }\textbf {\bibinfo {volume} {17}},\ \bibinfo
  {pages} {194} (\bibinfo {year} {2021})}\BibitemShut {NoStop}%
\bibitem [{\citenamefont {Nishikubo}\ \emph {et~al.}(2011)\citenamefont
  {Nishikubo}, \citenamefont {Kudo},\ and\ \citenamefont
  {Nohara}}]{nishikubo_2011}%
  \BibitemOpen
  \bibfield  {author} {\bibinfo {author} {\bibfnamefont {Y.}~\bibnamefont
  {Nishikubo}}, \bibinfo {author} {\bibfnamefont {K.}~\bibnamefont {Kudo}}, \
  and\ \bibinfo {author} {\bibfnamefont {M.}~\bibnamefont {Nohara}},\ }\href
  {\doibase 10.1143/JPSJ.80.055002} {\bibfield  {journal} {\bibinfo  {journal}
  {J. Phys. Soc. Jpn.}\ }\textbf {\bibinfo {volume} {80}},\ \bibinfo {pages}
  {055002} (\bibinfo {year} {2011})}\BibitemShut {NoStop}%
\bibitem [{\citenamefont {Biswas}\ \emph {et~al.}(2013)\citenamefont {Biswas},
  \citenamefont {Luetkens}, \citenamefont {Neupert}, \citenamefont {St\"urzer},
  \citenamefont {Baines}, \citenamefont {Pascua}, \citenamefont {Schnyder},
  \citenamefont {Fischer}, \citenamefont {Goryo}, \citenamefont {Lees},
  \citenamefont {Maeter}, \citenamefont {Br\"uckner}, \citenamefont {Klauss},
  \citenamefont {Nicklas}, \citenamefont {Baker}, \citenamefont {Hillier},
  \citenamefont {Sigrist}, \citenamefont {Amato},\ and\ \citenamefont
  {Johrendt}}]{biswas_2013}%
  \BibitemOpen
  \bibfield  {author} {\bibinfo {author} {\bibfnamefont {P.~K.}\ \bibnamefont
  {Biswas}}, \bibinfo {author} {\bibfnamefont {H.}~\bibnamefont {Luetkens}},
  \bibinfo {author} {\bibfnamefont {T.}~\bibnamefont {Neupert}}, \bibinfo
  {author} {\bibfnamefont {T.}~\bibnamefont {St\"urzer}}, \bibinfo {author}
  {\bibfnamefont {C.}~\bibnamefont {Baines}}, \bibinfo {author} {\bibfnamefont
  {G.}~\bibnamefont {Pascua}}, \bibinfo {author} {\bibfnamefont {A.~P.}\
  \bibnamefont {Schnyder}}, \bibinfo {author} {\bibfnamefont {M.~H.}\
  \bibnamefont {Fischer}}, \bibinfo {author} {\bibfnamefont {J.}~\bibnamefont
  {Goryo}}, \bibinfo {author} {\bibfnamefont {M.~R.}\ \bibnamefont {Lees}},
  \bibinfo {author} {\bibfnamefont {H.}~\bibnamefont {Maeter}}, \bibinfo
  {author} {\bibfnamefont {F.}~\bibnamefont {Br\"uckner}}, \bibinfo {author}
  {\bibfnamefont {H.-H.}\ \bibnamefont {Klauss}}, \bibinfo {author}
  {\bibfnamefont {M.}~\bibnamefont {Nicklas}}, \bibinfo {author} {\bibfnamefont
  {P.~J.}\ \bibnamefont {Baker}}, \bibinfo {author} {\bibfnamefont {A.~D.}\
  \bibnamefont {Hillier}}, \bibinfo {author} {\bibfnamefont {M.}~\bibnamefont
  {Sigrist}}, \bibinfo {author} {\bibfnamefont {A.}~\bibnamefont {Amato}}, \
  and\ \bibinfo {author} {\bibfnamefont {D.}~\bibnamefont {Johrendt}},\ }\href
  {\doibase 10.1103/PhysRevB.87.180503} {\bibfield  {journal} {\bibinfo
  {journal} {Phys. Rev. B}\ }\textbf {\bibinfo {volume} {87}},\ \bibinfo
  {pages} {180503(R)} (\bibinfo {year} {2013})}\BibitemShut {NoStop}%
\bibitem [{\citenamefont {Fischer}\ \emph {et~al.}(2014)\citenamefont
  {Fischer}, \citenamefont {Neupert}, \citenamefont {Platt}, \citenamefont
  {Schnyder}, \citenamefont {Hanke}, \citenamefont {Goryo}, \citenamefont
  {Thomale},\ and\ \citenamefont {Sigrist}}]{fischer_2014}%
  \BibitemOpen
  \bibfield  {author} {\bibinfo {author} {\bibfnamefont {M.~H.}\ \bibnamefont
  {Fischer}}, \bibinfo {author} {\bibfnamefont {T.}~\bibnamefont {Neupert}},
  \bibinfo {author} {\bibfnamefont {C.}~\bibnamefont {Platt}}, \bibinfo
  {author} {\bibfnamefont {A.~P.}\ \bibnamefont {Schnyder}}, \bibinfo {author}
  {\bibfnamefont {W.}~\bibnamefont {Hanke}}, \bibinfo {author} {\bibfnamefont
  {J.}~\bibnamefont {Goryo}}, \bibinfo {author} {\bibfnamefont
  {R.}~\bibnamefont {Thomale}}, \ and\ \bibinfo {author} {\bibfnamefont
  {M.}~\bibnamefont {Sigrist}},\ }\href {\doibase 10.1103/PhysRevB.89.020509}
  {\bibfield  {journal} {\bibinfo  {journal} {Phys. Rev. B}\ }\textbf {\bibinfo
  {volume} {89}},\ \bibinfo {pages} {020509(R)} (\bibinfo {year}
  {2014})}\BibitemShut {NoStop}%
\bibitem [{\citenamefont {Sauls}(1994)}]{sauls_1994}%
  \BibitemOpen
  \bibfield  {author} {\bibinfo {author} {\bibfnamefont {J.~A.}\ \bibnamefont
  {Sauls}},\ }\href {\doibase 10.1080/00018739400101475} {\bibfield  {journal}
  {\bibinfo  {journal} {Advances in Physics}\ }\textbf {\bibinfo {volume}
  {43}},\ \bibinfo {pages} {113} (\bibinfo {year} {1994})}\BibitemShut
  {NoStop}%
\bibitem [{\citenamefont {Joynt}\ and\ \citenamefont
  {Taillefer}(2002)}]{joynt_2002}%
  \BibitemOpen
  \bibfield  {author} {\bibinfo {author} {\bibfnamefont {R.}~\bibnamefont
  {Joynt}}\ and\ \bibinfo {author} {\bibfnamefont {L.}~\bibnamefont
  {Taillefer}},\ }\href {\doibase 10.1103/RevModPhys.74.235} {\bibfield
  {journal} {\bibinfo  {journal} {Rev. Mod. Phys.}\ }\textbf {\bibinfo {volume}
  {74}},\ \bibinfo {pages} {235} (\bibinfo {year} {2002})}\BibitemShut
  {NoStop}%
\bibitem [{\citenamefont {Strand}\ \emph {et~al.}(2010)\citenamefont {Strand},
  \citenamefont {Bahr}, \citenamefont {Van~Harlingen}, \citenamefont {Davis},
  \citenamefont {Gannon},\ and\ \citenamefont {Halperin}}]{strand_2010}%
  \BibitemOpen
  \bibfield  {author} {\bibinfo {author} {\bibfnamefont {J.~D.}\ \bibnamefont
  {Strand}}, \bibinfo {author} {\bibfnamefont {D.~J.}\ \bibnamefont {Bahr}},
  \bibinfo {author} {\bibfnamefont {D.~J.}\ \bibnamefont {Van~Harlingen}},
  \bibinfo {author} {\bibfnamefont {J.~P.}\ \bibnamefont {Davis}}, \bibinfo
  {author} {\bibfnamefont {W.~J.}\ \bibnamefont {Gannon}}, \ and\ \bibinfo
  {author} {\bibfnamefont {W.~P.}\ \bibnamefont {Halperin}},\ }\href {\doibase
  10.1126/science.1187943} {\bibfield  {journal} {\bibinfo  {journal}
  {Science}\ }\textbf {\bibinfo {volume} {328}},\ \bibinfo {pages} {1368}
  (\bibinfo {year} {2010})}\BibitemShut {NoStop}%
\bibitem [{\citenamefont {Schemm}\ \emph {et~al.}(2014)\citenamefont {Schemm},
  \citenamefont {Gannon}, \citenamefont {Wishne}, \citenamefont {Halperin},\
  and\ \citenamefont {Kapitulnik}}]{schemm_2014}%
  \BibitemOpen
  \bibfield  {author} {\bibinfo {author} {\bibfnamefont {E.~R.}\ \bibnamefont
  {Schemm}}, \bibinfo {author} {\bibfnamefont {W.~J.}\ \bibnamefont {Gannon}},
  \bibinfo {author} {\bibfnamefont {C.~M.}\ \bibnamefont {Wishne}}, \bibinfo
  {author} {\bibfnamefont {W.~P.}\ \bibnamefont {Halperin}}, \ and\ \bibinfo
  {author} {\bibfnamefont {A.}~\bibnamefont {Kapitulnik}},\ }\href {\doibase
  10.1126/science.1248552} {\bibfield  {journal} {\bibinfo  {journal}
  {Science}\ }\textbf {\bibinfo {volume} {345}},\ \bibinfo {pages} {190}
  (\bibinfo {year} {2014})},\ \Eprint
  {http://arxiv.org/abs/https://science.sciencemag.org/content/345/6193/190.full.pdf}
  {https://science.sciencemag.org/content/345/6193/190.full.pdf} \BibitemShut
  {NoStop}%
\bibitem [{\citenamefont {Kasahara}\ \emph {et~al.}(2007)\citenamefont
  {Kasahara}, \citenamefont {Iwasawa}, \citenamefont {Shishido}, \citenamefont
  {Shibauchi}, \citenamefont {Behnia}, \citenamefont {Haga}, \citenamefont
  {Matsuda}, \citenamefont {Onuki}, \citenamefont {Sigrist},\ and\
  \citenamefont {Matsuda}}]{kasahara_2007}%
  \BibitemOpen
  \bibfield  {author} {\bibinfo {author} {\bibfnamefont {Y.}~\bibnamefont
  {Kasahara}}, \bibinfo {author} {\bibfnamefont {T.}~\bibnamefont {Iwasawa}},
  \bibinfo {author} {\bibfnamefont {H.}~\bibnamefont {Shishido}}, \bibinfo
  {author} {\bibfnamefont {T.}~\bibnamefont {Shibauchi}}, \bibinfo {author}
  {\bibfnamefont {K.}~\bibnamefont {Behnia}}, \bibinfo {author} {\bibfnamefont
  {Y.}~\bibnamefont {Haga}}, \bibinfo {author} {\bibfnamefont {T.~D.}\
  \bibnamefont {Matsuda}}, \bibinfo {author} {\bibfnamefont {Y.}~\bibnamefont
  {Onuki}}, \bibinfo {author} {\bibfnamefont {M.}~\bibnamefont {Sigrist}}, \
  and\ \bibinfo {author} {\bibfnamefont {Y.}~\bibnamefont {Matsuda}},\ }\href
  {\doibase 10.1103/PhysRevLett.99.116402} {\bibfield  {journal} {\bibinfo
  {journal} {Phys. Rev. Lett.}\ }\textbf {\bibinfo {volume} {99}},\ \bibinfo
  {pages} {116402} (\bibinfo {year} {2007})}\BibitemShut {NoStop}%
\bibitem [{\citenamefont {Shibauchi}\ \emph {et~al.}(2014)\citenamefont
  {Shibauchi}, \citenamefont {Ikeda},\ and\ \citenamefont
  {Matsuda}}]{shibauchi_2014}%
  \BibitemOpen
  \bibfield  {author} {\bibinfo {author} {\bibfnamefont {T.}~\bibnamefont
  {Shibauchi}}, \bibinfo {author} {\bibfnamefont {H.}~\bibnamefont {Ikeda}}, \
  and\ \bibinfo {author} {\bibfnamefont {Y.}~\bibnamefont {Matsuda}},\ }\href
  {\doibase 10.1080/14786435.2014.887861} {\bibfield  {journal} {\bibinfo
  {journal} {Philosophical Magazine}\ }\textbf {\bibinfo {volume} {94}},\
  \bibinfo {pages} {3747} (\bibinfo {year} {2014})}\BibitemShut {NoStop}%
\bibitem [{\citenamefont {Schemm}\ \emph {et~al.}(2015)\citenamefont {Schemm},
  \citenamefont {Baumbach}, \citenamefont {Tobash}, \citenamefont {Ronning},
  \citenamefont {Bauer},\ and\ \citenamefont {Kapitulnik}}]{Schemm_2015}%
  \BibitemOpen
  \bibfield  {author} {\bibinfo {author} {\bibfnamefont {E.~R.}\ \bibnamefont
  {Schemm}}, \bibinfo {author} {\bibfnamefont {R.~E.}\ \bibnamefont
  {Baumbach}}, \bibinfo {author} {\bibfnamefont {P.~H.}\ \bibnamefont
  {Tobash}}, \bibinfo {author} {\bibfnamefont {F.}~\bibnamefont {Ronning}},
  \bibinfo {author} {\bibfnamefont {E.~D.}\ \bibnamefont {Bauer}}, \ and\
  \bibinfo {author} {\bibfnamefont {A.}~\bibnamefont {Kapitulnik}},\ }\href
  {\doibase 10.1103/PhysRevB.91.140506} {\bibfield  {journal} {\bibinfo
  {journal} {Phys. Rev. B}\ }\textbf {\bibinfo {volume} {91}},\ \bibinfo
  {pages} {140506(R)} (\bibinfo {year} {2015})}\BibitemShut {NoStop}%
\bibitem [{\citenamefont {Ran}\ \emph {et~al.}(2019)\citenamefont {Ran},
  \citenamefont {Eckberg}, \citenamefont {Ding}, \citenamefont {Furukawa},
  \citenamefont {Metz}, \citenamefont {Saha}, \citenamefont {Liu},
  \citenamefont {Zic}, \citenamefont {Kim}, \citenamefont {Paglione},\ and\
  \citenamefont {Butch}}]{ran_2019}%
  \BibitemOpen
  \bibfield  {author} {\bibinfo {author} {\bibfnamefont {S.}~\bibnamefont
  {Ran}}, \bibinfo {author} {\bibfnamefont {C.}~\bibnamefont {Eckberg}},
  \bibinfo {author} {\bibfnamefont {Q.-P.}\ \bibnamefont {Ding}}, \bibinfo
  {author} {\bibfnamefont {Y.}~\bibnamefont {Furukawa}}, \bibinfo {author}
  {\bibfnamefont {T.}~\bibnamefont {Metz}}, \bibinfo {author} {\bibfnamefont
  {S.~R.}\ \bibnamefont {Saha}}, \bibinfo {author} {\bibfnamefont {I.-L.}\
  \bibnamefont {Liu}}, \bibinfo {author} {\bibfnamefont {M.}~\bibnamefont
  {Zic}}, \bibinfo {author} {\bibfnamefont {H.}~\bibnamefont {Kim}}, \bibinfo
  {author} {\bibfnamefont {J.}~\bibnamefont {Paglione}}, \ and\ \bibinfo
  {author} {\bibfnamefont {N.~P.}\ \bibnamefont {Butch}},\ }\href {\doibase
  10.1126/science.aav8645} {\bibfield  {journal} {\bibinfo  {journal}
  {Science}\ }\textbf {\bibinfo {volume} {365}},\ \bibinfo {pages} {684}
  (\bibinfo {year} {2019})}\BibitemShut {NoStop}%
\bibitem [{\citenamefont {Jiao}\ \emph {et~al.}(2020)\citenamefont {Jiao},
  \citenamefont {Howard}, \citenamefont {Ran}, \citenamefont {Wang},
  \citenamefont {Rodriguez}, \citenamefont {Sigrist}, \citenamefont {Wang},
  \citenamefont {Butch},\ and\ \citenamefont {Madhavan}}]{jiao_2020}%
  \BibitemOpen
  \bibfield  {author} {\bibinfo {author} {\bibfnamefont {L.}~\bibnamefont
  {Jiao}}, \bibinfo {author} {\bibfnamefont {S.}~\bibnamefont {Howard}},
  \bibinfo {author} {\bibfnamefont {S.}~\bibnamefont {Ran}}, \bibinfo {author}
  {\bibfnamefont {Z.}~\bibnamefont {Wang}}, \bibinfo {author} {\bibfnamefont
  {J.~O.}\ \bibnamefont {Rodriguez}}, \bibinfo {author} {\bibfnamefont
  {M.}~\bibnamefont {Sigrist}}, \bibinfo {author} {\bibfnamefont
  {Z.}~\bibnamefont {Wang}}, \bibinfo {author} {\bibfnamefont {N.~P.}\
  \bibnamefont {Butch}}, \ and\ \bibinfo {author} {\bibfnamefont
  {V.}~\bibnamefont {Madhavan}},\ }\href@noop {} {\bibfield  {journal}
  {\bibinfo  {journal} {Nature}\ }\textbf {\bibinfo {volume} {579}},\ \bibinfo
  {pages} {523} (\bibinfo {year} {2020})}\BibitemShut {NoStop}%
\bibitem [{\citenamefont {Yamakage}\ and\ \citenamefont
  {Sato}(2014)}]{yamakage_2014}%
  \BibitemOpen
  \bibfield  {author} {\bibinfo {author} {\bibfnamefont {A.}~\bibnamefont
  {Yamakage}}\ and\ \bibinfo {author} {\bibfnamefont {M.}~\bibnamefont
  {Sato}},\ }\href {\doibase 10.1016/j.physe.2013.08.030} {\bibfield  {journal}
  {\bibinfo  {journal} {Physica E: Low-dimensional Systems and Nanostructures}\
  }\textbf {\bibinfo {volume} {55}},\ \bibinfo {pages} {13} (\bibinfo {year}
  {2014})}\BibitemShut {NoStop}%
\bibitem [{\citenamefont {Shiozaki}\ and\ \citenamefont
  {Sato}(2014)}]{shiozaki_2014}%
  \BibitemOpen
  \bibfield  {author} {\bibinfo {author} {\bibfnamefont {K.}~\bibnamefont
  {Shiozaki}}\ and\ \bibinfo {author} {\bibfnamefont {M.}~\bibnamefont
  {Sato}},\ }\href {\doibase 10.1103/PhysRevB.90.165114} {\bibfield  {journal}
  {\bibinfo  {journal} {Phys. Rev. B}\ }\textbf {\bibinfo {volume} {90}},\
  \bibinfo {pages} {165114} (\bibinfo {year} {2014})}\BibitemShut {NoStop}%
\bibitem [{\citenamefont {Kobayashi}\ \emph {et~al.}(2014)\citenamefont
  {Kobayashi}, \citenamefont {Shiozaki}, \citenamefont {Tanaka},\ and\
  \citenamefont {Sato}}]{kobayashi_2014}%
  \BibitemOpen
  \bibfield  {author} {\bibinfo {author} {\bibfnamefont {S.}~\bibnamefont
  {Kobayashi}}, \bibinfo {author} {\bibfnamefont {K.}~\bibnamefont {Shiozaki}},
  \bibinfo {author} {\bibfnamefont {Y.}~\bibnamefont {Tanaka}}, \ and\ \bibinfo
  {author} {\bibfnamefont {M.}~\bibnamefont {Sato}},\ }\href {\doibase
  10.1103/PhysRevB.90.024516} {\bibfield  {journal} {\bibinfo  {journal} {Phys.
  Rev. B}\ }\textbf {\bibinfo {volume} {90}},\ \bibinfo {pages} {024516}
  (\bibinfo {year} {2014})}\BibitemShut {NoStop}%
\bibitem [{\citenamefont {Zhao}\ \emph {et~al.}(2016)\citenamefont {Zhao},
  \citenamefont {Schnyder},\ and\ \citenamefont {Wang}}]{zhao_2016}%
  \BibitemOpen
  \bibfield  {author} {\bibinfo {author} {\bibfnamefont {Y.~X.}\ \bibnamefont
  {Zhao}}, \bibinfo {author} {\bibfnamefont {A.~P.}\ \bibnamefont {Schnyder}},
  \ and\ \bibinfo {author} {\bibfnamefont {Z.~D.}\ \bibnamefont {Wang}},\
  }\href {\doibase 10.1103/PhysRevLett.116.156402} {\bibfield  {journal}
  {\bibinfo  {journal} {Phys. Rev. Lett.}\ }\textbf {\bibinfo {volume} {116}},\
  \bibinfo {pages} {156402} (\bibinfo {year} {2016})}\BibitemShut {NoStop}%
\bibitem [{\citenamefont {Agterberg}\ \emph {et~al.}(2017)\citenamefont
  {Agterberg}, \citenamefont {Brydon},\ and\ \citenamefont
  {Timm}}]{agterberg_2017}%
  \BibitemOpen
  \bibfield  {author} {\bibinfo {author} {\bibfnamefont {D.~F.}\ \bibnamefont
  {Agterberg}}, \bibinfo {author} {\bibfnamefont {P.~M.~R.}\ \bibnamefont
  {Brydon}}, \ and\ \bibinfo {author} {\bibfnamefont {C.}~\bibnamefont
  {Timm}},\ }\href {\doibase 10.1103/PhysRevLett.118.127001} {\bibfield
  {journal} {\bibinfo  {journal} {Phys. Rev. Lett.}\ }\textbf {\bibinfo
  {volume} {118}},\ \bibinfo {pages} {127001} (\bibinfo {year}
  {2017})}\BibitemShut {NoStop}%
\bibitem [{\citenamefont {Bzdu\ifmmode~\check{s}\else \v{s}\fi{}ek}\ and\
  \citenamefont {Sigrist}(2017)}]{bzdusek_2017}%
  \BibitemOpen
  \bibfield  {author} {\bibinfo {author} {\bibfnamefont {T.}~\bibnamefont
  {Bzdu\ifmmode~\check{s}\else \v{s}\fi{}ek}}\ and\ \bibinfo {author}
  {\bibfnamefont {M.}~\bibnamefont {Sigrist}},\ }\href {\doibase
  10.1103/PhysRevB.96.155105} {\bibfield  {journal} {\bibinfo  {journal} {Phys.
  Rev. B}\ }\textbf {\bibinfo {volume} {96}},\ \bibinfo {pages} {155105}
  (\bibinfo {year} {2017})}\BibitemShut {NoStop}%
\bibitem [{\citenamefont {Timm}\ \emph {et~al.}(2017)\citenamefont {Timm},
  \citenamefont {Schnyder}, \citenamefont {Agterberg},\ and\ \citenamefont
  {Brydon}}]{timm_2017}%
  \BibitemOpen
  \bibfield  {author} {\bibinfo {author} {\bibfnamefont {C.}~\bibnamefont
  {Timm}}, \bibinfo {author} {\bibfnamefont {A.~P.}\ \bibnamefont {Schnyder}},
  \bibinfo {author} {\bibfnamefont {D.~F.}\ \bibnamefont {Agterberg}}, \ and\
  \bibinfo {author} {\bibfnamefont {P.~M.~R.}\ \bibnamefont {Brydon}},\ }\href
  {\doibase 10.1103/PhysRevB.96.094526} {\bibfield  {journal} {\bibinfo
  {journal} {Phys. Rev. B}\ }\textbf {\bibinfo {volume} {96}},\ \bibinfo
  {pages} {094526} (\bibinfo {year} {2017})}\BibitemShut {NoStop}%
\bibitem [{\citenamefont {Brydon}\ \emph {et~al.}(2018)\citenamefont {Brydon},
  \citenamefont {Agterberg}, \citenamefont {Menke},\ and\ \citenamefont
  {Timm}}]{brydon_2018}%
  \BibitemOpen
  \bibfield  {author} {\bibinfo {author} {\bibfnamefont {P.~M.~R.}\
  \bibnamefont {Brydon}}, \bibinfo {author} {\bibfnamefont {D.~F.}\
  \bibnamefont {Agterberg}}, \bibinfo {author} {\bibfnamefont {H.}~\bibnamefont
  {Menke}}, \ and\ \bibinfo {author} {\bibfnamefont {C.}~\bibnamefont {Timm}},\
  }\href {\doibase 10.1103/PhysRevB.98.224509} {\bibfield  {journal} {\bibinfo
  {journal} {Phys. Rev. B}\ }\textbf {\bibinfo {volume} {98}},\ \bibinfo
  {pages} {224509} (\bibinfo {year} {2018})}\BibitemShut {NoStop}%
\bibitem [{\citenamefont {Umerski}(1997)}]{umerski_1997}%
  \BibitemOpen
  \bibfield  {author} {\bibinfo {author} {\bibfnamefont {A.}~\bibnamefont
  {Umerski}},\ }\href {\doibase 10.1103/PhysRevB.55.5266} {\bibfield  {journal}
  {\bibinfo  {journal} {Phys. Rev. B}\ }\textbf {\bibinfo {volume} {55}},\
  \bibinfo {pages} {5266} (\bibinfo {year} {1997})}\BibitemShut {NoStop}%
\bibitem [{\citenamefont {Fukui}\ \emph {et~al.}(2005)\citenamefont {Fukui},
  \citenamefont {Hatsugai},\ and\ \citenamefont {Suzuki}}]{fukui_2005a}%
  \BibitemOpen
  \bibfield  {author} {\bibinfo {author} {\bibfnamefont {T.}~\bibnamefont
  {Fukui}}, \bibinfo {author} {\bibfnamefont {Y.}~\bibnamefont {Hatsugai}}, \
  and\ \bibinfo {author} {\bibfnamefont {H.}~\bibnamefont {Suzuki}},\ }\href
  {\doibase 10.1143/JPSJ.74.1674} {\bibfield  {journal} {\bibinfo  {journal}
  {J. Phys. Soc. Jpn.}\ }\textbf {\bibinfo {volume} {74}},\ \bibinfo {pages}
  {1674} (\bibinfo {year} {2005})}\BibitemShut {NoStop}%
\bibitem [{\citenamefont {Sato}\ \emph {et~al.}(2011)\citenamefont {Sato},
  \citenamefont {Tanaka}, \citenamefont {Yada},\ and\ \citenamefont
  {Yokoyama}}]{sato_2011}%
  \BibitemOpen
  \bibfield  {author} {\bibinfo {author} {\bibfnamefont {M.}~\bibnamefont
  {Sato}}, \bibinfo {author} {\bibfnamefont {Y.}~\bibnamefont {Tanaka}},
  \bibinfo {author} {\bibfnamefont {K.}~\bibnamefont {Yada}}, \ and\ \bibinfo
  {author} {\bibfnamefont {T.}~\bibnamefont {Yokoyama}},\ }\href {\doibase
  10.1103/PhysRevB.83.224511} {\bibfield  {journal} {\bibinfo  {journal} {Phys.
  Rev. B}\ }\textbf {\bibinfo {volume} {83}},\ \bibinfo {pages} {224511}
  (\bibinfo {year} {2011})}\BibitemShut {NoStop}%
\bibitem [{\citenamefont {Sato}\ and\ \citenamefont
  {Fujimoto}(2010)}]{sato_2010}%
  \BibitemOpen
  \bibfield  {author} {\bibinfo {author} {\bibfnamefont {M.}~\bibnamefont
  {Sato}}\ and\ \bibinfo {author} {\bibfnamefont {S.}~\bibnamefont
  {Fujimoto}},\ }\href {\doibase 10.1103/PhysRevLett.105.217001} {\bibfield
  {journal} {\bibinfo  {journal} {Phys. Rev. Lett.}\ }\textbf {\bibinfo
  {volume} {105}},\ \bibinfo {pages} {217001} (\bibinfo {year}
  {2010})}\BibitemShut {NoStop}%
\bibitem [{\citenamefont {Hu}(1994)}]{CRHu1994}%
  \BibitemOpen
  \bibfield  {author} {\bibinfo {author} {\bibfnamefont {C.-R.}\ \bibnamefont
  {Hu}},\ }\href {\doibase 10.1103/PhysRevLett.72.1526} {\bibfield  {journal}
  {\bibinfo  {journal} {Phys. Rev. Lett.}\ }\textbf {\bibinfo {volume} {72}},\
  \bibinfo {pages} {1526} (\bibinfo {year} {1994})}\BibitemShut {NoStop}%
\bibitem [{\citenamefont {Kashiwaya}\ and\ \citenamefont
  {Tanaka}(2000)}]{kashiwaya_2000}%
  \BibitemOpen
  \bibfield  {author} {\bibinfo {author} {\bibfnamefont {S.}~\bibnamefont
  {Kashiwaya}}\ and\ \bibinfo {author} {\bibfnamefont {Y.}~\bibnamefont
  {Tanaka}},\ }\href {\doibase 10.1088/0034-4885/63/10/202} {\bibfield
  {journal} {\bibinfo  {journal} {Rep. Prog. Phys.}\ }\textbf {\bibinfo
  {volume} {63}},\ \bibinfo {pages} {1641} (\bibinfo {year}
  {2000})}\BibitemShut {NoStop}%
\bibitem [{\citenamefont {Fu}\ and\ \citenamefont {Kane}(2008)}]{fu_2008}%
  \BibitemOpen
  \bibfield  {author} {\bibinfo {author} {\bibfnamefont {L.}~\bibnamefont
  {Fu}}\ and\ \bibinfo {author} {\bibfnamefont {C.~L.}\ \bibnamefont {Kane}},\
  }\href {\doibase 10.1103/PhysRevLett.100.096407} {\bibfield  {journal}
  {\bibinfo  {journal} {Phys. Rev. Lett.}\ }\textbf {\bibinfo {volume} {100}},\
  \bibinfo {pages} {096407} (\bibinfo {year} {2008})}\BibitemShut {NoStop}%
\bibitem [{\citenamefont {Tanaka}\ \emph {et~al.}(2009)\citenamefont {Tanaka},
  \citenamefont {Yokoyama},\ and\ \citenamefont {Nagaosa}}]{Tanaka2009}%
  \BibitemOpen
  \bibfield  {author} {\bibinfo {author} {\bibfnamefont {Y.}~\bibnamefont
  {Tanaka}}, \bibinfo {author} {\bibfnamefont {T.}~\bibnamefont {Yokoyama}}, \
  and\ \bibinfo {author} {\bibfnamefont {N.}~\bibnamefont {Nagaosa}},\ }\href
  {\doibase 10.1103/PhysRevLett.103.107002} {\bibfield  {journal} {\bibinfo
  {journal} {Phys. Rev. Lett.}\ }\textbf {\bibinfo {volume} {103}},\ \bibinfo
  {pages} {107002} (\bibinfo {year} {2009})}\BibitemShut {NoStop}%
\bibitem [{\citenamefont {Qi}\ \emph {et~al.}(2009)\citenamefont {Qi},
  \citenamefont {Hughes}, \citenamefont {Raghu},\ and\ \citenamefont
  {Zhang}}]{qi_2009}%
  \BibitemOpen
  \bibfield  {author} {\bibinfo {author} {\bibfnamefont {X.-L.}\ \bibnamefont
  {Qi}}, \bibinfo {author} {\bibfnamefont {T.~L.}\ \bibnamefont {Hughes}},
  \bibinfo {author} {\bibfnamefont {S.}~\bibnamefont {Raghu}}, \ and\ \bibinfo
  {author} {\bibfnamefont {S.-C.}\ \bibnamefont {Zhang}},\ }\href {\doibase
  10.1103/PhysRevLett.102.187001} {\bibfield  {journal} {\bibinfo  {journal}
  {Phys. Rev. Lett.}\ }\textbf {\bibinfo {volume} {102}},\ \bibinfo {pages}
  {187001} (\bibinfo {year} {2009})}\BibitemShut {NoStop}%
\bibitem [{\citenamefont {Ohashi}\ \emph {et~al.}(2020)\citenamefont {Ohashi},
  \citenamefont {Tanaka},\ and\ \citenamefont {Yada}}]{ohashi_2020}%
  \BibitemOpen
  \bibfield  {author} {\bibinfo {author} {\bibfnamefont {R.}~\bibnamefont
  {Ohashi}}, \bibinfo {author} {\bibfnamefont {Y.}~\bibnamefont {Tanaka}}, \
  and\ \bibinfo {author} {\bibfnamefont {K.}~\bibnamefont {Yada}},\ }\href
  {\doibase 10.7566/JPSJ.89.054701} {\bibfield  {journal} {\bibinfo  {journal}
  {J. Phys. Soc. Jpn.}\ }\textbf {\bibinfo {volume} {89}},\ \bibinfo {pages}
  {054701} (\bibinfo {year} {2020})}\BibitemShut {NoStop}%
\bibitem [{\citenamefont {Matsumoto}\ and\ \citenamefont
  {Shiba}(1995)}]{matsumoto_1995}%
  \BibitemOpen
  \bibfield  {author} {\bibinfo {author} {\bibfnamefont {M.}~\bibnamefont
  {Matsumoto}}\ and\ \bibinfo {author} {\bibfnamefont {H.}~\bibnamefont
  {Shiba}},\ }\href {\doibase 10.1143/JPSJ.64.1703} {\bibfield  {journal}
  {\bibinfo  {journal} {Journal of the Physical Society of Japan}\ }\textbf
  {\bibinfo {volume} {64}},\ \bibinfo {pages} {1703} (\bibinfo {year}
  {1995})}\BibitemShut {NoStop}%
\bibitem [{\citenamefont {Tanaka}\ and\ \citenamefont
  {Kashiwaya}(1995)}]{tanaka_1995}%
  \BibitemOpen
  \bibfield  {author} {\bibinfo {author} {\bibfnamefont {Y.}~\bibnamefont
  {Tanaka}}\ and\ \bibinfo {author} {\bibfnamefont {S.}~\bibnamefont
  {Kashiwaya}},\ }\href {\doibase 10.1103/PhysRevLett.74.3451} {\bibfield
  {journal} {\bibinfo  {journal} {Phys. Rev. Lett.}\ }\textbf {\bibinfo
  {volume} {74}},\ \bibinfo {pages} {3451} (\bibinfo {year}
  {1995})}\BibitemShut {NoStop}%
\bibitem [{\citenamefont {Tanuma}\ \emph {et~al.}(2001)\citenamefont {Tanuma},
  \citenamefont {Tanaka},\ and\ \citenamefont {Kashiwaya}}]{tanuma_2001}%
  \BibitemOpen
  \bibfield  {author} {\bibinfo {author} {\bibfnamefont {Y.}~\bibnamefont
  {Tanuma}}, \bibinfo {author} {\bibfnamefont {Y.}~\bibnamefont {Tanaka}}, \
  and\ \bibinfo {author} {\bibfnamefont {S.}~\bibnamefont {Kashiwaya}},\ }\href
  {\doibase 10.1103/PhysRevB.64.214519} {\bibfield  {journal} {\bibinfo
  {journal} {Phys. Rev. B}\ }\textbf {\bibinfo {volume} {64}},\ \bibinfo
  {pages} {214519} (\bibinfo {year} {2001})}\BibitemShut {NoStop}%
\bibitem [{\citenamefont {Yada}\ \emph {et~al.}(2011)\citenamefont {Yada},
  \citenamefont {Sato}, \citenamefont {Tanaka},\ and\ \citenamefont
  {Yokoyama}}]{yada_2011}%
  \BibitemOpen
  \bibfield  {author} {\bibinfo {author} {\bibfnamefont {K.}~\bibnamefont
  {Yada}}, \bibinfo {author} {\bibfnamefont {M.}~\bibnamefont {Sato}}, \bibinfo
  {author} {\bibfnamefont {Y.}~\bibnamefont {Tanaka}}, \ and\ \bibinfo {author}
  {\bibfnamefont {T.}~\bibnamefont {Yokoyama}},\ }\href {\doibase
  10.1103/PhysRevB.83.064505} {\bibfield  {journal} {\bibinfo  {journal} {Phys.
  Rev. B}\ }\textbf {\bibinfo {volume} {83}},\ \bibinfo {pages} {064505}
  (\bibinfo {year} {2011})}\BibitemShut {NoStop}%
\bibitem [{\citenamefont {Kobayashi}\ \emph {et~al.}(2015)\citenamefont
  {Kobayashi}, \citenamefont {Tanaka},\ and\ \citenamefont
  {Sato}}]{kobayashi_2015}%
  \BibitemOpen
  \bibfield  {author} {\bibinfo {author} {\bibfnamefont {S.}~\bibnamefont
  {Kobayashi}}, \bibinfo {author} {\bibfnamefont {Y.}~\bibnamefont {Tanaka}}, \
  and\ \bibinfo {author} {\bibfnamefont {M.}~\bibnamefont {Sato}},\ }\href
  {\doibase 10.1103/PhysRevB.92.214514} {\bibfield  {journal} {\bibinfo
  {journal} {Phys. Rev. B}\ }\textbf {\bibinfo {volume} {92}},\ \bibinfo
  {pages} {214514} (\bibinfo {year} {2015})}\BibitemShut {NoStop}%
\bibitem [{\citenamefont {Tamura}\ \emph {et~al.}(2017)\citenamefont {Tamura},
  \citenamefont {Kobayashi}, \citenamefont {Bo},\ and\ \citenamefont
  {Tanaka}}]{tamura_2017}%
  \BibitemOpen
  \bibfield  {author} {\bibinfo {author} {\bibfnamefont {S.}~\bibnamefont
  {Tamura}}, \bibinfo {author} {\bibfnamefont {S.}~\bibnamefont {Kobayashi}},
  \bibinfo {author} {\bibfnamefont {L.}~\bibnamefont {Bo}}, \ and\ \bibinfo
  {author} {\bibfnamefont {Y.}~\bibnamefont {Tanaka}},\ }\href {\doibase
  10.1103/PhysRevB.95.104511} {\bibfield  {journal} {\bibinfo  {journal} {Phys.
  Rev. B}\ }\textbf {\bibinfo {volume} {95}},\ \bibinfo {pages} {104511}
  (\bibinfo {year} {2017})}\BibitemShut {NoStop}%
\bibitem [{\citenamefont {Lapp}\ \emph {et~al.}(2020)\citenamefont {Lapp},
  \citenamefont {B\"orner},\ and\ \citenamefont {Timm}}]{Lapp2020}%
  \BibitemOpen
  \bibfield  {author} {\bibinfo {author} {\bibfnamefont {C.~J.}\ \bibnamefont
  {Lapp}}, \bibinfo {author} {\bibfnamefont {G.}~\bibnamefont {B\"orner}}, \
  and\ \bibinfo {author} {\bibfnamefont {C.}~\bibnamefont {Timm}},\ }\href
  {\doibase 10.1103/PhysRevB.101.024505} {\bibfield  {journal} {\bibinfo
  {journal} {Phys. Rev. B}\ }\textbf {\bibinfo {volume} {101}},\ \bibinfo
  {pages} {024505} (\bibinfo {year} {2020})}\BibitemShut {NoStop}%
\bibitem [{\citenamefont {Lee}\ and\ \citenamefont {Fisher}(1981)}]{lee_1981}%
  \BibitemOpen
  \bibfield  {author} {\bibinfo {author} {\bibfnamefont {P.~A.}\ \bibnamefont
  {Lee}}\ and\ \bibinfo {author} {\bibfnamefont {D.~S.}\ \bibnamefont
  {Fisher}},\ }\href {\doibase 10.1103/PhysRevLett.47.882} {\bibfield
  {journal} {\bibinfo  {journal} {Phys. Rev. Lett.}\ }\textbf {\bibinfo
  {volume} {47}},\ \bibinfo {pages} {882} (\bibinfo {year} {1981})}\BibitemShut
  {NoStop}%
\end{thebibliography}%

\end{document}